%% file: ANA-SUSY-2018-08-PAPER.tex
\pdfoutput=1
\documentclass[cernpreprint,atlasdraft=false,texlive=2016,UKenglish,texmf]{atlasdoc}
 
\usepackage[subfigure]{atlaspackage}
\usepackage[block=space,backref=false]{atlasbiblatex}
\usepackage{pifont}
\usepackage{xcolor}

\usepackage{atlasphysics}
\usepackage{atlasbsm}

\graphicspath{{logos/}}
 
\usepackage{ANA-SUSY-2018-08-PAPER-defs}
\usepackage{multirow}
\usepackage{graphicx}
\usepackage{epstopdf}
\usepackage{adjustbox}
\usepackage{rotating}
\usepackage{footnote}
\usepackage[utf8]{inputenc}
\usepackage[T1]{fontenc}
\makesavenoteenv{tabular}
\makesavenoteenv{table}

\AtlasTitle{Search for new phenomena in events with two opposite-charge leptons, jets and missing transverse momentum  in $pp$ collisions at $\sqrt{s} = 13$ TeV\ with the ATLAS detector}

\AtlasAbstract{
The results of a search for direct pair production of top squarks and for dark matter in events with two opposite-charge leptons (electrons or muons), jets and missing transverse momentum are reported, using 139~\mbox{fb\(^{-1}\)} of integrated luminosity from proton--proton collisions at $\sqrt{s}=13$ \ensuremath{\text{Te\kern -0.1em V}}, collected by the ATLAS detector at the Large Hadron Collider during Run~2 (2015--2018).
This search considers the pair production of top squarks and is sensitive across a wide range of mass differences between the top squark and the lightest neutralino.
Additionally, spin-0 mediator dark-matter models are considered, in which the mediator is produced in association with a pair of top quarks. The mediator subsequently decays to a pair of dark-matter particles.
No significant excess of events is observed above the Standard Model background, and limits are set at $95\%$\ confidence level.
The results exclude  top squark masses up to about $1$ \ensuremath{\text{Te\kern -0.1em V}}, and masses of the lightest neutralino up to about $500$ \ensuremath{\text{Ge\kern -0.1em V}}.
Limits on dark-matter production are set for scalar (pseudoscalar) mediator  masses up to about $250~(300)$ \ensuremath{\text{Ge\kern -0.1em V}}.}

\author{The ATLAS Collaboration}
\AtlasJournal{JHEP}
\AtlasJournalRef{JHEP 04 (2021) 165}
\PreprintIdNumber{CERN-EP-2020-243}
\AtlasDOI{10.1007/JHEP04(2021)165}

\hypersetup{pdftitle={ATLAS document},pdfauthor={The ATLAS Collaboration}}
 
\begin{document}
 
\maketitle

\section{Introduction}
\label{sec:intro}
 
The Standard Model (SM) of particle physics is extremely successful in describing the phenomena of elementary particles and their interactions. Its predictive power has been proven with high precision by a wide range of experiments. However, despite its success, several important questions remain unanswered within the SM. One particularly striking omission is that it does not provide any explanation for dark matter (DM)~\cite{Zwicky, Bertone_2005}. This is a non-baryonic, non-luminous matter component of the universe, for which there is strong evidence from a range of astrophysical observations.
A weakly interacting dark-matter candidate particle can be produced at the Large Hadron Collider (LHC)~\cite{Evans:2008zzb} in a variety of ways, as described, for example, by supersymmetry (SUSY)~\cite{Golfand:1971iw,Volkov:1973ix,Wess:1974tw,Wess:1974jb,Ferrara:1974pu,Salam:1974ig} or DM models. At the LHC, one of the most promising modes is the production of DM particle pairs in association with on- or off-shell top quarks. 
Previous searches for DM candidates in association with a top quark pair have been performed by the ATLAS~\cite{SUSY-2018-12,SUSY-2014-07,SUSY-2016-15,SUSY-2016-16,SUSY-2016-26,EXOT-2016-27,TOPQ-2016-10} and CMS~\cite{CMS-SUS-15-004,CMS-SUS-15-005,CMS-SUS-16-008,CMS-SUS-16-049,CMS-SUS-16-051,CMS-SUS-17-001,CMS-SUS-19-005,CMS-SUS-19-006,CMS-SUS-19-009,CMS-EXO-16-049} collaborations. However, those previous searches were statistically limited, or sensitive only up to limited particle masses.
They also suffered from significant regions in which no limit could be placed because the kinematics of the decays made the signal events particularly difficult to identify.
This paper aims to extend the sensitivity beyond that of the previous searches to higher masses, and to cover the regions in which the previous ATLAS results had no sensitivity~\cite{SUSY-2016-17,SUSY-2016-18}. It achieves this in part by exploiting a larger dataset, corresponding to $139~\ifb$ of proton--proton collision data collected by the ATLAS experiment during Run~2 of the LHC (2015--2018) at a centre-of-mass energy $\sqrt{s}=13$~\TeV. Further improvements in sensitivity are obtained by using a new discriminating variable, the `object-based \MET significance' \cite{ATLAS-CONF-2018-038}, lowering the lepton \pT thresholds, and optimising a dedicated selection to target signal models in the most difficult kinematic regions.

\subsection*{Signal models and kinematic regions}
For DM production, the simplified benchmark models~\cite{goodman2011lhc,Alwall:2008ag,Alves:2011wf} assume the existence of a mediator particle which couples both to the SM and to the dark sector~\cite{Abercrombie_2019, Buckley_2015, haisch2015simplified}. The couplings of the mediator to the SM fermions are then severely restricted by precision flavour measurements. An ansatz that automatically relaxes these constraints is Minimal Flavour Violation \cite{D_Ambrosio_2002}. This assumption implies that the interaction between any new neutral spin-0 state and SM matter is proportional to the fermion masses via Yukawa-type couplings.\footnote{Following Ref.~\cite{Buckley_2015}, couplings to $W$ and $Z$ bosons, as well as explicit dimension-4 $\phi$--$h$ or $a$--$h$ couplings, are set to zero in this simplified model. In addition, the coupling of the mediator to the dark sector is not taken to be proportional to the mass of the DM candidates.} It follows that colour-neutral mediators would be produced mainly through loop-induced gluon fusion or in association with heavy-flavour quarks.
Here, the DM particles $\chi$ are assumed to be pair produced through the exchange of a spin-0 mediator, which can be a colour-neutral scalar or pseudoscalar particle (denoted by $\phi$ or $a$, respectively), in association with a top quark pair: $pp\rightarrow \chi \bar{\chi} t \bar t$ (Figure~\ref{subfig:DM}).
 
Alternatively, dark-matter particles are also predicted in supersymmetry, a space-time symmetry that for each SM particle postulates the existence of a partner particle whose spin differs by one-half unit. To avoid violation of baryon number ($B$) and lepton number ($L$) conservation, a multiplicative quantum number $R$-parity~\cite{Farrar:1978xj}, defined as $R~=~(-1)^{3(B-L)+2S}$, is assumed to be conserved. SUSY particles are then produced in pairs, and the lightest supersymmetric particle (LSP) is stable and, if only weakly interacting, a candidate for dark matter~\cite{Goldberg:1983nd,Ellis:1983ew}.  In the framework of a generic $R$-parity-conserving Minimal Supersymmetric Standard Model (MSSM)~\cite{Fayet:1976et,Fayet:1977yc}, the supersymmetric scalar partners of right-handed and left-handed quarks (squarks), \squarkR and \squarkL, can mix to form two mass eigenstates, $\tilde q_1$ and $\tilde q_2$, with $\tilde q_1$ defined to be the lighter one. In the case of the supersymmetric partner of the top quark, \stop, large mixing effects can lead to one of the top squark mass eigenstates, \stopone, being significantly lighter than the other squarks. The charginos and neutralinos are mixtures of the bino, winos and Higgsinos that are superpartners of the U(1) and SU(2) gauge bosons and the Higgs bosons, respectively. Their mass eigenstates are referred to as $\tilde{\chi}_{i}^{\pm}$~$(i=1,2)$ and $\tilde{\chi}_{j}^{0}$~$(j=1,2,3,4)$ in order of increasing mass. In a large variety of models, the LSP, which is the DM candidate, is the lightest neutralino \ninoone. Searches for direct pair production of the top squark and DM particles can be performed in final states with two leptons (electrons or muons) of opposite electric charge, jets and missing transverse momentum (Figures~\ref{subfig:stoptB}--\ref{subfig:stoptN1}).
Depending on the mass difference between the top squark and the lighter SUSY particles, different decay modes are relevant. For $m(W)+m(b) < m(\stopone)-m(\ninoone) < m(t)$, the three-body decay $\stopone \rightarrow b W \ninoone$ occurs through an off-shell top quark (Figure~\ref{subfig:stoptB}). For smaller mass differences, i.e.  $m(\stopone) - m(\ninoone) < m(W)+m(b)$, the four-body decay channel $\stop \rightarrow b f f' \ninoone$, where $f$ and $f'$ are two fermions from the off-shell $(\Wboson^{*})$ decay, is assumed to occur (Figure~\ref{subfig:stopfB}). In this search, $f$ and $f'$ are a charged lepton and its associated anti-neutrino (or vice versa). For each of these two decay modes a dedicated event selection is performed to maximise the sensitivity. These selections are referred to as \textit{three-body} and \textit{four-body} selections in this paper.
Direct pair production of top squarks which decay into an on-shell top quark and the lightest neutralino $\stopone \rightarrow t \ninoone$, will occur when $m(\stopone) - m(\ninoone) > m(t)$ (Figure~\ref{subfig:stoptN1}).
The signature of the $\ttbar$+DM process is similar to that of the simplified model shown in Figure~\ref{subfig:DM}, so the same selection is also used to constrain the $\stopone \rightarrow t \ninoone$ model and it is referred to as the \textit{two-body} selection.
 
The paper proceeds as follows; after a description of the ATLAS detector in Section~\ref{sec:detector}, the data and simulated Monte Carlo (MC) samples used in the analysis are detailed in Section~\ref{sec:samples} and the object identification is documented in Section~\ref{sec:objects}. The search strategy, the SM background estimations, and the systematic uncertainties  are discussed in Sections~\ref{sec:selection},~\ref{sec:backgrounds} and~\ref{sec:systs}.   The results and their statistical interpretations are presented in  Sections~\ref{sec:results}~and~\ref{sec:interpretations}. Finally, Section~\ref{sec:conclusion} presents the conclusions.
 
\begin{figure}[!t]
\begin{center}
\subfigure[]{
\label{subfig:DM}\includegraphics[width=0.2\textwidth]{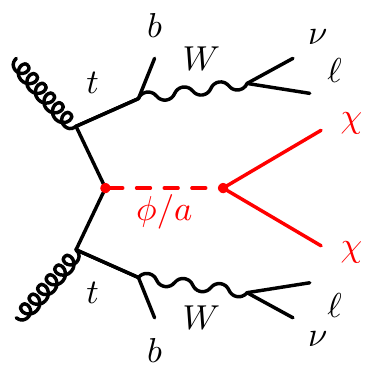}
}
\subfigure[]{
\label{subfig:stoptB}\includegraphics[width=0.24\textwidth]{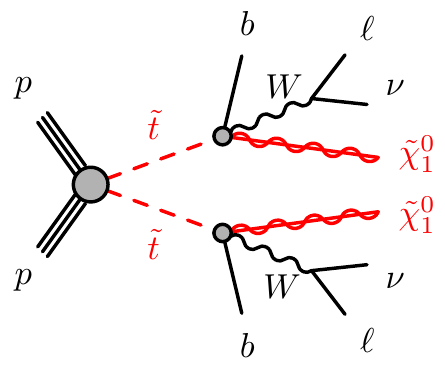}
}
\subfigure[]{
\label{subfig:stopfB}\includegraphics[width=0.24\textwidth]{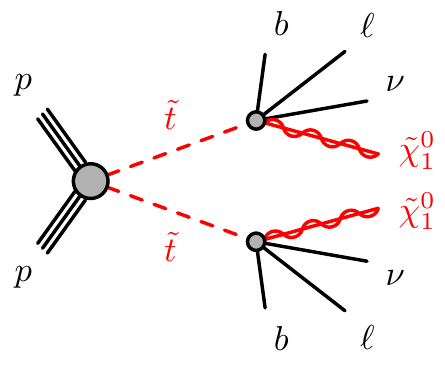}
}
\subfigure[]{
\label{subfig:stoptN1}\includegraphics[width=0.24\textwidth]{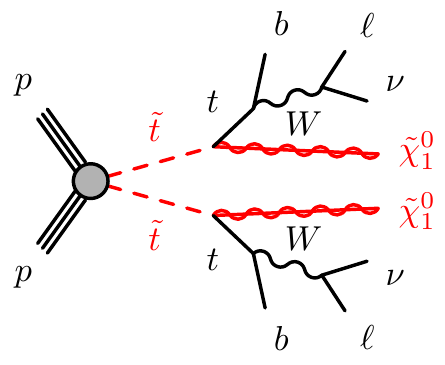}
}
\end{center}
\caption{Diagrams representing the signal models targeted by the searches: (a) the spin-0 mediator models, where the mediator decays into a pair of dark-matter particles and is produced in association with a pair of top quarks ($pp\rightarrow \chi \bar{\chi} t \bar t$), (b) the three-body \stopone decay mode into an on-shell $W$ boson, a $b$-quark and the lightest neutralino ($\stopone \rightarrow b W \ninoone$), (c) the four-body \stopone decay mode ($\stopone \rightarrow b \bar{\ell} \nu \ninoone$) where $\bar{\ell}$ and $\nu$ are a anti-lepton with its neutrino and (d) the two-body \stopone decay into an on-shell top quark and the lightest neutralino ($\stopone \rightarrow t \ninoone$). For all the diagrams (a-d) the distinction between particle and anti-particle is omitted.}
\label{fig:diagram_stop}
\end{figure}

\FloatBarrier

\section{ATLAS detector}
\label{sec:detector}

The ATLAS detector~\cite{PERF-2007-01} at the LHC covers nearly the entire solid angle around the collision point.\footnote{\AtlasCoordFootnote \, A vector energy $\vec{E}$ is defined by combining the energy deposited in the calorimeter with its deposit direction.}
It consists of an inner tracking detector surrounded by a thin superconducting solenoid, electromagnetic and hadronic calorimeters,
and a muon spectrometer with three large superconducting toroidal magnets.
 
The inner-detector system (ID) is immersed in a \SI{2}{\tesla} axial magnetic field
and provides charged-particle tracking in the range \(|\eta| < 2.5\).
The high-granularity silicon pixel detector covers the vertex region and typically provides four measurements per track,
the first hit normally being in the insertable B-layer installed before Run~2~\cite{ATLAS-TDR-19,Abbott:2018ikt}.
It is followed by the silicon microstrip tracker, which usually provides eight measurements per track.
These silicon detectors are complemented by the transition radiation tracker (TRT),
which enables radially extended track reconstruction up to \(|\eta| = 2.0\).
The TRT also provides electron identification information
based on the fraction of hits (typically 30 in total) above a higher energy-deposit threshold corresponding to transition radiation.
 
The calorimeter system covers the pseudorapidity range \(|\eta| < 4.9\).
Within the region \(|\eta|< 3.2\), electromagnetic calorimetry is provided by barrel and
endcap high-granularity lead/liquid-argon (LAr) calorimeters,
with an additional thin LAr presampler covering \(|\eta| < 1.8\)
to correct for energy loss in material upstream of the calorimeters.
Hadronic calorimetry is provided by the steel/scintillating-tile calorimeter,
segmented into three barrel structures within \(|\eta| < 1.7\), and two copper/LAr hadronic endcap calorimeters.
The solid angle coverage is completed with forward copper/LAr and tungsten/LAr calorimeter modules
optimised for electromagnetic and hadronic measurements respectively.
 
The muon spectrometer (MS) comprises separate trigger and
high-precision tracking chambers measuring the deflection of muons in a magnetic field generated by the superconducting air-core toroids.
The field integral of the toroids ranges between \num{2.0} and \SI{6.0}{\tesla\metre}
across most of the detector.
A set of precision chambers covers the region \(|\eta| < 2.7\) with three layers of monitored drift tubes,
complemented by cathode-strip chambers in the forward region, where the background is highest.
The muon trigger system covers the range \(|\eta| < 2.4\) with resistive-plate chambers in the barrel, and thin-gap chambers in the endcap regions.
 
Interesting events are selected to be recorded by the first-level trigger system implemented in custom hardware,
followed by selections made by algorithms implemented in software in the high-level trigger~\cite{TRIG-2016-01}.
The first-level trigger accepts events from the \SI{40}{\MHz} bunch crossings at a rate below \SI{100}{\kHz},  which the
high-level trigger reduces in order to record events to disk at about \SI{1}{\kHz}.
 
\section{Data and simulated event samples}
\label{sec:samples}
The data used in this analysis were collected by the ATLAS detector during $pp$ collisions at a centre-of-mass energy of $\sqrt{s} = 13$~\TeV\ from 2015 to 2018. The average number $\langle \mu \rangle$ of $pp$ interactions per bunch crossing (pile-up) varies from 14  during 2015 to 38 during  2017--2018. Only events taken in stable beam conditions, and for which all relevant detector systems were operational, are considered in this analysis. After data-quality requirements  the data sample amounts to a total integrated luminosity of 139~\ifb. The uncertainty in the combined 2015--2018 integrated luminosity is $1.7\%$ \cite{ATLAS-CONF-2019-021}, obtained using the LUCID-2 detector \cite{LUCID2}.

The two-body and three-body selections use events accepted by a trigger that requires a minimum of two electrons, two muons, or an electron and a muon~\cite{TRIG-2016-01}.
Different  trigger-level thresholds for the transverse momentum of the leptons were used  in different  data-taking periods, ranging between 8 and $22~\GeV$. Tighter thresholds are applied in the lepton offline selection, to ensure that the  trigger efficiency is `on plateau' in  all of the relevant kinematic region.
Missing transverse momentum triggers~\cite{TRIG-2019-01}  are used in the four-body selection to increase the acceptance of  low-\pT leptons. The missing transverse momentum trigger threshold varied depending on data-taking conditions in the four years: 70~\GeV\ for data collected during  2015;  in the range 90--110~\GeV\ for data collected during  2016,  and  110 \GeV\ for data collected during  2017 and 2018.  Tighter offline requirements on the missing transverse momentum are defined accordingly  to ensure event selection on the plateau region of the trigger efficiency curve.
 
Simulated event samples are used for SM background estimations and to model the signal samples. Standard Model MC samples were processed through a full $\GEANT4$~\cite{Agostinelli:2002hh} simulation of the ATLAS detector, while a fast simulation based on parameterisation of the calorimeter response and $\GEANT4$ simulation for all the other detector components~\cite{SOFT-2010-01} is used for the SUSY and DM signal samples. MC events are reconstructed using the same algorithms  used for the data. To compensate for small  residual  differences between data and simulation in the lepton reconstruction efficiency, energy scale, energy resolution, trigger modelling, and $b$-tagging efficiency, the simulated events are reweighted  using correction factors derived from data~\cite{EGAM-2018-01,PERF-2015-10,FTAG-2018-01}.
 
The events targeted by this analysis are characterised by two leptons with opposite electric charge, jets and missing transverse momentum.
The main SM background contributions are expected to come from top quark pair production (\ttbar), associated production of a $Z$ boson and a top quark pair (\ttZ), single-top decay in the $Wt$ production channel ($Wt$), \zjets production and diboson processes ($VV$ with $V=W,Z$).
 
Matrix element and showering generators  used for the SM backgrounds and signals are listed in Table~\ref{tab:mc_backgr} along with the relevant  parton distribution function (PDF) sets, the configuration of underlying-event and hadronisation parameters (tunes), and the cross-section order in $\alpha_\mathrm{s}$ used to normalise the event yields.
Additional MC samples are used to  estimate systematic uncertainties, as detailed in Section~\ref{sec:systs}.
 
The SUSY top squark pair signal samples were generated from leading-order (LO) matrix elements with up to two extra partons using \MGMCatNLOV{2.6.2}~\cite{Alwall:2014hca}. \MGMCatNLO was interfaced to \PYTHIAV{8.212}+~\textsc{MadSpin}~\cite{Sjostrand:2014zea,Artoisenet:2012st} for the signal samples used in the three-body and four-body selections, while it was interfaced to \PYTHIAV{8.212} for the SUSY signal samples used for the interpretation of the two-body selection results. Signal cross-sections were calculated to next-to-next-to-leading order (NNLO) in $\alpha_\mathrm{s}$, adding the resummation of soft gluon emission at next-to-next-to-leading-logarithm accuracy (NNLO+NNLL)~\cite{Beenakker:2016lwe,Beenakker:2014sma,Beenakker:2013mva,Beenakker:2011sf,Beenakker:2009ha,Kulesza:2009kq,Kulesza:2008jb,Beenakker:1996ch}. The nominal cross section and the uncertainty are derived using the PDF4LHC15 PDF set, following the recommendations presented in Ref.~\cite{Butterworth:2015oua}. Jet--parton matching was performed following the CKKW-L prescription~\cite{Lonnblad:2012ix}. The A14 tune~\cite{ATL-PHYS-PUB-2014-021} was used for the modelling of parton showering, hadronisation and the underlying event. Parton luminosities were provided by the NNPDF2.3LO PDF set~\cite{Ball:2012cx}.
 
The dark-matter signal samples were also generated from leading-order matrix elements, with up to one extra parton, using \MGMCatNLOV{2.6.2} interfaced to \PYTHIAV{8.212}. In the DM samples generation the couplings of the scalar and pseudoscalar mediators to the SM and DM particles ($g_q$ and $g_{\chi}$) are set to one. The kinematics of the mediator decay are not strongly dependent on the values of the couplings; however, the particle kinematic distributions are sensitive to the nature of the mediator and to the mediator and DM particle masses. The cross-sections were computed at NLO~\cite{Mattelaer:2015haa,Backovic:2015soa}.
 
Inelastic $pp$ interactions were generated and overlaid onto the hard-scattering process to simulate the effect of multiple proton--proton interactions occurring during the same (in-time) or a nearby (out-of-time) bunch crossing. These were produced using \PYTHIAV{8.186}~\cite{Sjostrand:2008} and EvtGen~\cite{Lange:2001uf} with the NNPDF2.3LO set of PDFs~\cite{Ball:2012cx} and the A3 tune~\cite{ATL-PHYS-PUB-2016-017}. The MC samples were reweighted so that the distribution of the average number of interactions per bunch crossing reproduces the observed distribution in the data.

\begin{table}[ht]
\footnotesize
\centering
\caption{\label{tab:mc_backgr}Simulated signal and  background event samples with the corresponding matrix element and parton shower (PS) generators, cross-section order in $\alpha_\mathrm{s}$ used to normalise the event yield, and the generator and PS  PDF sets used.}
\begin{adjustbox}{width=1.\textwidth}
\begin{tabular}{llllll}
\noalign{\smallskip}\hline\noalign{\smallskip}
Physics process                          & Generator                                                                                    & Parton shower                                                                                 & Normalisation                                                                                                                    & PDF (generator)                               & PDF (PS)\\
\noalign{\smallskip}\hline\noalign{\smallskip}
SUSY Signals                             & \multirow{2}{*}{\MGMCatNLO~\cite{Alwall:2014hca}.}                                           & \multirow{2}{*}{\PYTHIAV{8.212}+~\textsc{MadSpin}~\cite{Sjostrand:2014zea,Artoisenet:2012st}} & \multirow{2}{*}{NNLO+NNLL~\cite{Beenakker:2016lwe,Beenakker:2014sma,Beenakker:2013mva,Beenakker:2011sf,Beenakker:2009ha,Kulesza:2009kq,Kulesza:2008jb,Beenakker:1996ch} }& \multirow{2}{*}{NNPDF2.3LO~\cite{Ball:2012cx}}& \multirow{2}{*}{NNPDF2.3LO~\cite{Ball:2012cx}} \\
(three-body, four-body)                  & \\
SUSY Signals (two-body)                  & \MGMCatNLO                                                                                   & \PYTHIAV{8.212}                                                                               & NNLO+NNLL~\cite{Beenakker:2016lwe,Beenakker:2014sma,Beenakker:2013mva,Beenakker:2011sf,Beenakker:2009ha,Kulesza:2009kq,Kulesza:2008jb,Beenakker:1996ch}                  & NNPDF2.3LO                                    & NNPDF2.3LO\\
DM Signals (two-body)                    & \MGMCatNLO                                                                                   & \PYTHIAV{8.212}                                                                               & NLO~\cite{Mattelaer:2015haa,Backovic:2015soa}                                                                                    & NNPDF2.3LO                                    & NNPDF2.3LO\\
\noalign{\smallskip}\hline\noalign{\smallskip}
$t\bar{t}$                               & \POWHEGBOXV{v2}~\cite{Frixione:2007vw,Alioli:2010,Nason:2004rx}                              & \PYTHIAV{8.230}                                                                               & NNLO+NNLL~\cite{Czakon:2011xx}                                                                                                   & NNPDF3.0NLO~\cite{Ball:2014uwa}               & NNPDF2.3LO\\
$t\bar{t}+V\ (V=W,Z)$                    & \MGMCatNLO                                                                                   & \PYTHIAV{8.210}                                                                               & NLO~\cite{ATL-PHYS-PUB-2016-005,Alwall:2014hca}                                                                                  & NNPDF3.0NLO                                   & NNPDF2.3LO    \\
Single top                               & \POWHEGBOXV{v2}~\cite{Frixione:2007vw,Alioli:2010,Nason:2004rx,Frederix:2012,Alioli:2009NLO} & \PYTHIAV{8.230}                                                                               &NLO+NNLL~\cite{Aliev:2010zk,Kant:2014oha,Kidonakis1,Kidonakis2,Kidonakis3}                                                        & NNPDF3.0NLO                                   & NNPDF2.3LO            \\
$Z/\gamma^*(\rightarrow \ell\ell )$+jets & \SHERPAV{2.2.1}~\cite{Gleisberg:2008ta,ATL-PHYS-PUB-2016-003}                                & \SHERPAV{2.2.1}~~                                                                             &NNLO~\cite{Gavin_2011}                                                                                                            & NNPDF3.0NNLO~\cite{Ball:2014uwa}              &NNPDF3.0NNLO~\cite{Ball:2014uwa} \\
Diboson $VV\ (V=W,Z)$                    & \SHERPAV{2.2.1} or 2.2.2~\cite{Gleisberg:2008ta}                                             & \SHERPAV{2.2.1} or 2.2.2~~                                                                    &NLO~\cite{ATL-PHYS-PUB-2017-005}                                                                                                  & NNPDF3.0NNLO                                  & NNPDF3.0NNLO         \\
Triboson $VVV\ (V=W,Z)$                  & \SHERPAV{2.2.2}                                                                              & \SHERPAV{2.2.2}  	~~                                                                          &NLO~\cite{Gleisberg:2008ta,ATL-PHYS-PUB-2017-005}                                                                                 & NNPDF3.0NNLO                                  & NNPDF3.0NNLO         \\
$\ttbar H$                               & \POWHEGBOXV{v2}~\cite{Frixione:2007vw,Alioli:2010,Hartanto}                                  & \PYTHIAV{8.230}                                                                               &NLO~\cite{ATL-PHYS-PUB-2016-005,Alwall:2014hca}                                                                                  & NNPDF3.0NLO                                   & NNPDF2.3LO \\
$t\bar{t}WW$                             & \MGMCatNLO                                                                                   & \PYTHIAV{8.186}~\cite{Sjostrand:2008}                                                         &NLO~\cite{Alwall:2014hca}                                                                                                         & NNPDF2.3LO                                    & NNPDF2.3LO           \\
$t\bar{t}WZ$                             & \MGMCatNLO                                                                                   & \PYTHIAV{8.212}                                                                               &NLO~\cite{Alwall:2014hca}                                                                                                         & NNPDF3.0NLO                                   & NNPDF2.3LO           \\
$tZ,t\bar{t}t\bar{t},t\bar{t}t$          & \MGMCatNLO                                                                                   & \PYTHIAV{8.230}                                                                               &NLO~\cite{Alwall:2014hca}                                                                                                         & NNPDF3.0NLO                                   & NNPDF2.3LO       \\
\noalign{\smallskip}\hline\noalign{\smallskip}
\end{tabular}
\end{adjustbox}
\end{table}

\section{Object identification}
\label{sec:objects}
Candidate events are required to have a reconstructed vertex with at least two associated tracks, each with $\pT > 500$~\MeV\ and originating from the beam collision region in the $x$--$y$ plane. The primary  vertex in the event is the
vertex with the highest scalar sum of the squared transverse momenta of associated tracks.
 
The leptons selected for analysis are classified  as baseline or signal leptons depending  on  an increasingly stringent set of reconstruction quality  criteria  and kinematic selections,
so that signal leptons are a subset of the baseline leptons.
Baseline leptons are used in the calculation of missing transverse momentum ($\mathbf p^\mathrm{miss}_\mathrm{T}$), to resolve ambiguities between the analysis objects in the event, as described later, and  for the fake/non-prompt (FNP) lepton background estimation described in Section~\ref{sec:backgrounds}.
Signal leptons are used for the final event selection.
 
Baseline electron candidates are reconstructed from three-dimensional  clusters of energy deposition in the electromagnetic calorimeter matched to ID tracks.  These  electron candidates  are required to have pseudorapidity $|\eta|<2.47$, $\ET>\SI{4.5}{\GeV}$, and to pass a \textit{Loose} likelihood-based identification requirement~\cite{EGAM-2018-01} with an additional condition on  the number of hits in the B-layer.  The  tracks associated  with  electron candidates are  required to have a longitudinal impact parameter\footnote{The transverse impact parameter is defined as the distance of closest approach in the transverse plane between a track and the beam-line. The longitudinal impact parameter corresponds to the $z$-coordinate distance between the point along the track at which the transverse impact parameter is defined and the primary vertex.}  relative to the primary vertex  $|z_0 \sin\theta|  < 0.5$~mm, where $\theta$ is the track's polar angle.
 
Baseline muon candidates are reconstructed by matching ID tracks,  in the pseudorapidity region $|\eta|<2.4$ for the two-body and three-body selections and $|\eta|<2.7$ for the four-body selection, with   MS tracks or  energy deposits in the calorimeter compatible with a minimum-ionising particle (calo-tagged muon).
The resulting tracks are  required to have a  $\pT>\SI{4}{\GeV}$ and  a $|z_0 \sin\theta| < 0.5$~mm from  the primary vertex.
Muon candidates are required to satisfy the \textit{Medium} identification requirement, defined in Ref.~\cite{PERF-2015-10},   based on  the numbers of hits in the different ID and MS subsystems, and on the significance of the charge-to-momentum ratio $q/p$.
 
Additional tighter selections are applied to the baseline lepton candidates to select the signal electrons or muons. Signal electrons are required to satisfy a \textit{Medium} likelihood-based identification requirement~\cite{EGAM-2018-01} and the track associated with a signal electron is required to have a significance $\vert d_0\vert/\sigma(d_0) < 5$, where $d_0$ is the transverse impact parameter relative to the reconstructed primary vertex and $\sigma(d_0)$ is its uncertainty.
Isolation criteria are applied to electrons by placing an upper limit on the sum of the transverse energy of the calorimeter energy clusters in a cone of size $\Delta R_\eta = \sqrt{(\Delta \eta)^2+(\Delta\phi)^2} = 0.2$ around the electron (excluding the deposit from the electron itself) and the scalar sum of the \pT of tracks within a cone of $\Delta R_\eta = 0.2$ around the electron (excluding its own track).
The isolation criteria are optimised such that the isolation selection efficiency is uniform across $\eta$. This varies from 90\% for $\pT = 25\ \GeV$ to 99\% for $\pT = 60\ \GeV$ in events with a $Z$ boson decaying into pair of electrons~\cite{EGAM-2018-01}.
 
For signal muons a significance in the transverse impact parameter  $\vert d_0\vert/\sigma(d_0) < 3$ is required. Isolation criteria applied to muons require the scalar sum of the \pT of tracks inside a cone of $\Delta R_\eta = 0.3$ around the muon (excluding its own track) to be less than 15\% of the muon \pT. In addition, the sum of the transverse energy of the calorimeter energy clusters in a cone of $\Delta R_\eta = 0.2$ around the muon (excluding the energy from the lepton itself) must be less than 30\% of the muon \pT~\cite{PERF-2015-10}.
 
Jets are reconstructed from  three-dimensional  clusters of energy in the calorimeter~\cite{PERF-2014-07} using the anti-$k_t$ jet clustering algorithm~\cite{Cacciari:2008} as implemented in the FastJet package~\cite{Fastjet}, with a radius parameter $R=0.4$. The reconstructed jets are then calibrated by the application of a jet energy scale derived from 13~\TeV\ data and simulation~\cite{PERF-2016-04}. Only jet candidates with $\pT>\SI{20}{\GeV}$ and $|\eta|<2.8$ are considered.\footnote{Hadronic $\tau$-lepton decay products are treated as jets.}
 
To reduce the effects of pile-up, for jets with $|\eta|\leq 2.5$ and $\pt<\SI{120}{\GeV}$ a significant fraction of the tracks associated with each jet are required to have an origin compatible with the primary vertex, as defined by the jet vertex tagger (JVT)~\cite{ATLAS-CONF-2014-018}. This requirement reduces the fraction of jets from pile-up to 1\%, with an efficiency for pure hard-scatter jets of about 90\%. Finally, in order to remove events impacted by detector noise and non-collision backgrounds, specific jet-quality requirements~\cite{ATLAS-CONF-2015-029,PERF-2014-03} are applied, designed to provide an efficiency of selecting jets from proton--proton collisions above 99.5\% (99.9\%) for $\pT > 20\ (100)$~\GeV.
 
The MV2C10 boosted decision tree algorithm~\cite{FTAG-2018-01} identifies jets containing $b$-hadrons (`$b$-jets') by using quantities such as the impact parameters of associated tracks, and well-reconstructed secondary vertices. A selection that provides 77\% efficiency for tagging $b$-jets in simulated $t\bar{t}$ events is used. The corresponding rejection factors against jets originating from $c$-quarks, from $\tau$-leptons, and from light quarks and gluons in the same sample at this working point are 4.9, 15 and 110, respectively.
 
To avoid reconstruction ambiguities and double counting of analysis objects, an overlap removal procedure  is applied to the baseline leptons and jets in the order which follows.
First, the calo-tagged muons are removed if sharing the track with electrons and, next, all electrons sharing an ID track with a muon are removed.
Jets which are not $b$-tagged (with the tagging parameters corresponding to an efficiency of 85\%) and which lie within a cone of $\Delta R = \sqrt{(\Delta y)^2+(\Delta\phi)^2} = 0.2$ around an electron candidate are removed. All jets lying within $\Delta R =0.2$ of an electron are  removed if the electron has $\pT > 100$ GeV.
Finally,  any  lepton candidate is removed  in  favour of a jet candidate  if it lies a distance  $\Delta R < \mathrm{min }(0.4, 0.04 + 10/{p_{\mathrm T}(\ell)})$ from the jet, where  ${p_{\mathrm T}(\ell)}$ is the \pT of the lepton.
 
The missing transverse momentum ($\mathbf p^\mathrm{miss}_\mathrm{T}$), with magnitude \MET, is defined as the negative vector sum of the transverse momenta for all  baseline electrons, photons, muons and jets. Low-momentum tracks from the primary vertex that are not associated with reconstructed analysis objects  are also included in the calculation. The \MET value is adjusted for the calibration of the selected physics objects~\cite{PERF-2016-07}.
Linked to the \MET value is the `object-based \MET significance',  called  simply `\METsig' in this paper.
This quantity measures the  significance of \MET based upon the transverse momentum  resolution of
all objects used in the calculation of the $\mathbf p^\mathrm{miss}_\mathrm{T}$.
It is defined as
\begin{equation}
\METsig = \frac{|\mathbf p^\mathrm{miss}_\mathrm{T}|}{\sqrt{\sigma_{\textup{L}}^2(1- \rho_{\textup{LT}}^2)}}  \nonumber
\label{eq.metSig}
\end{equation}
where  $\sigma_{\textup{L}}$  is the (longitudinal) component parallel to the $\mathbf p^\mathrm{miss}_\mathrm{T}$ of the total transverse momentum resolution for all objects in the event and the quantity
$\rho_{\textup{LT}}$ is the  correlation factor between  the parallel and perpendicular components of the transverse momentum  resolution for each object. On an event-by-event basis, given the full event composition, \METsig evaluates the $p$-value that the observed \MET is consistent with the null hypothesis of zero real \MET, as further detailed in Ref.~\cite{ATLAS-CONF-2018-038}. In this way \METsig  helps to separate events with true \MET, arising from weakly interacting particles such as dark matter or neutralinos, from those where \MET  is consistent with particle mismeasurement, resolution or identification inefficiencies, thus providing better background rejection.
 
\FloatBarrier
\section{Event selection}
\label{sec:selection}
Different event selections are inspired by previous published strategies~\cite{SUSY-2016-17,SUSY-2016-18} reoptimised to fully exploit the larger available dataset.
For all selections, an improvement in the sensitivity is obtained with the introduction of the \METsig variable, which enables further optimisation of the selection variables.
The four-body  sensitivity also benefits from a reduction in the lepton  \pT threshold in the region with  small mass differences \DeltaM ~between \stopone and \ninoone.
The threshold for the muon (electron) \pT was lowered from $7$~\GeV\ to $4$~\GeV\ ($4.5$~\GeV).
 
Events are required to have exactly two signal leptons (two electrons, two muons, or one electron and one muon) with opposite electric charge. In the two-body  and three-body  selections, an invariant mass \mll greater than $20~\GeV$ condition is applied to remove leptons from Drell--Yan and low-mass resonances, while in the four-body selection, given the softer \pT spectrum of the leptons,  \mll is required to be  higher than $10~\GeV$.
Events with same flavour (SF) lepton pairs ($e^{\pm} e^{\mp}$ and $\mu^{\pm} \mu^{\mp}$) with \mll\ between 71.2 and 111.2~\GeV\ are rejected to reduce the $Z$ boson background, except for the four-body selection.
No additional \mll\ selection is imposed on the different flavour (DF) lepton pairs ($e^{\pm} \mu^{\mp}$).
Different jet ($b$-jet) multiplicities, labelled as \njet (\nbjet),  are required in the three selections, as detailed below.
\subsection{Discriminators and kinematic variables}
\label{sec:variables}
Final event selections are obtained by separating signal from SM background using different kinematic variables.
Two variables are constructed from the \MET and the \pT of the leading leptons and jets:
\begin{equation}
\Rll = \MET / \left( \pT(\ell_1) + \pT(\ell_2) \right) \ \ \ \ \ \mathrm{and}\ \ \ \ \ \Rlj = {\MET/\left(\MET + \pT(\ell_1) + \pT(\ell_2) + \sum_{i=1,...,N\leq 4}\pT(j_i)\right)}    \nonumber
\end{equation}
where $\pT(\ell_1)$ and $\pT(\ell_2)$ are the leading and sub-leading lepton transverse momenta respectively and $\pT(j_{i =1,...,N\leq 4})$ are the transverse momenta of the up to four leading jets, in decreasing order. For some backgrounds, e.g.\ \zjets, the variable \Rll has a distribution that peaks at lower values than the signal, and it is thus used to reject those backgrounds. Similarly, $R_{2\ell 4j}$ is employed for its high rejection power against multi-jet events.
 
Another variable employed is \pbll, which is defined as the vectorial sum of $\mathbf p^{\mathrm{miss}}_{\mathrm{T}}$ and the leptons' transverse momentum vectors $\mathbf \pT(\ell_1)$  and  $\mathbf \pT(\ell_2)$.
Its magnitude, \ptllboost , can be interpreted as the magnitude of the vector sum of all the transverse hadronic activity in the event.
The azimuthal angle between the $ \mathbf p^{\mathrm{miss}}_{\mathrm{T}} $ vector and the \pbll\ vector  is defined as  \dphib. This variable is useful for selecting events where the non hadronic component ($e$, $\mu$, $\nu$ and $\chi$ or $\ninoone$) is collimated.
 
The lepton-based stransverse mass~\cite{Lester:1999tx,Barr:2003rg} is a kinematic variable used to bound the masses of a pair of identical particles which have each decayed into a visible and an invisible particle. This quantity is defined as
\begin{equation}
\mttwo
( \mathbf p_{\mathrm{T,1}}, \mathbf p_{\mathrm{T,2}}, \mathbf p_{\mathrm{T}}^{\mathrm{miss}}) = \min_{\mathbf q_{\mathrm{T,1}} + \mathbf q_{\mathrm{T,2}} =
\mathbf p_{\mathrm{T}}^{\mathrm{miss}} }
\left\{ \max [\;
m_{\mathrm{T}}( \mathbf p_{\mathrm{T,1}}, \mathbf q_{\mathrm{T,1}} ),
m_{\mathrm{T}}( \mathbf p_{\mathrm{T,2}}, \mathbf q_{\mathrm{T,2}}
)
\;]
\right\} ,  \nonumber
\label{mt2eq}
\end{equation}
\noindent where $m_{\mathrm T}$ indicates the transverse mass,\footnote{The transverse mass is defined by the equation $m_\textup{T}(\mathbf p_{\mathrm{T}},\mathbf q_{\mathrm{T}} )=\sqrt{2|\mathbf p_{\mathrm{T}}|| \mathbf q_{\mathrm{T}} |(1-\mathrm{cos}(\Delta\phi))}$, where $\Delta\phi$ is the angle between particles of negligible mass with transverse momenta $\mathbf p_{\mathrm{T}}$ and $ \mathbf q_{\mathrm{T}}$.} $\mathbf p_{\mathrm{T,1}}$ and $\mathbf p_{\mathrm{T,2}}$ are the transverse momentum vectors of two visible particles, and $\mathbf q_{\mathrm{T,1}}$ and $\mathbf q_{\mathrm{T,2}}$ are transverse momentum vectors with $\mathbf p_{\mathrm{T}}^{\mathrm{miss}}  = \mathbf q_{\mathrm{T,1}} + \mathbf q_{\mathrm{T,2}}$.
The minimisation is performed over all the possible decompositions of $\mathbf p_{\mathrm{T}}^{\mathrm{miss}} $.
In this paper, $\mathbf p_{\mathrm{T,1}}$ and $\mathbf p_{\mathrm{T,2}}$ are the transverse momentum vectors of the two leptons and $\mttwo(\mathbf \pT(\ell_1), \mathbf \pT(\ell_2), \mathbf p^{\mathrm{miss}}_{\mathrm{T}})$ is referred to simply as \mttwoL. For the \mttwoL calculation, the invisible particles are assumed to be massless.
The \mttwoL\ distribution is expected to have an endpoint corresponding to the $W$ mass for backgrounds such as \ttbar while it is expected to reach higher values in the case of SUSY events, due to the presence of the neutralinos~\cite{Cho:2007dh,Burns:2008va}.

The three-body selection uses a number of `super-razor' variables~\cite{Buckley:2013kua},  which are derived with a series of assumptions made in order to approximate the centre-of-mass energy frame (Razor Frame) of two parent particles (i.e.\ top squarks) and the decay frames. Each parent particle is assumed to decay into a set of visible (only leptons are considered in this case) and invisible particles (i.e.\ neutrinos and neutralinos).
These variables are \RPT, the Lorentz factor \GAMr, the azimuthal angle \DPB\ and \MDR. The first variable is
$\RPT = |\vec{J}_{\mathrm T}|/ (|\vec{J}_{\mathrm T}| + \sqrt{\hat{s}}_{\mathrm R}/4)$ with $\vec{J}_{\mathrm T}$ as the vector sum of the transverse momenta of the visible particles and the missing transverse momentum, and $\sqrt{\hat{s}_{\mathrm R}}$ as an estimate of the system's energy in the razor frame $R$, defined  as the frame in which the two visible leptons have equal and opposite longitudinal momentum ($p_{\mathrm z}$).
The value of $|\vec{J}_{\mathrm T}|$ vanishes for events where leptons are the only visible particles, such as diboson events, leading to \RPT values that tend toward zero. Instead, in events that contain additional activity, such as \ttbar, this variable tends towards unity.
The Lorentz factor, \GAMr, is associated with the boost from the razor frame $R$ to the approximation of the two decay frames of the parent particles and is expected to have values tending towards unity for back-to-back visible particles or when they have different momenta.
Lower values of \GAMr are otherwise expected when the two visible particles are collinear and have comparable momentum.
The azimuthal angle \DPB\ is defined between the razor boost from the laboratory to the $R$ frame and the sum of the visible momenta as evaluated in the $R$ frame.
It is a good discriminator when used in searches for signals from models with small mass differences between the massive pair-produced particle and the invisible particle produced in the decay.
Finally, the last variable is $\MDR = \sqrt{\hat{s}_{\mathrm R}} / \gamma_{{\mathrm R}+1} $, which is particularly powerful in discriminating between signal events and \ttbar and diboson background, since it has a kinematic end-point that is proportional to the mass-splitting between the parent particle and the invisible particle.

\FloatBarrier
\subsection{Two-body event selection}
\label{sec:twobody_selection}
 
This selection targets the dark-matter signal model that assumes the production of a pair of dark-matter particles  through the exchange of a spin-0 mediator, in association with a pair of top quarks (Figure~\ref{subfig:DM}).
It is also used for a search for top squarks decaying into an on-shell top and neutralino (Figure~\ref{subfig:stoptN1}).
 
For each event, the leading lepton, $\ell_1$, is required to have $\ptleplead>25$~\GeV, while for the sub-leading lepton, $\ell_2$, the requirement is $\ptlepsubl>20$~\GeV. The event selection also requires at least one reconstructed $b$-jet, \dphib\ lower than 1.5 and \METsig\ greater than 12, and finally \mttwoL\ greater than $110~\GeV$.
Following the classification of the events, two sets of signal regions (SRs) are defined: a set of exclusive SRs \textit{binned} in the \mttwoL\ variable, to maximise model-dependent search sensitivity, and a set of \textit{inclusive} SRs, to be used for model-independent results.
For the binned SRs, events are separated according to the lepton flavours, different flavour or same flavour, and by the range $[x,\ y)$ of the \mttwoL\ interval: $\SRtwoBDF_{[x,y)}$ or $\SRtwoBSF_{[x,y)}$.
For the inclusive signal regions, referred to as $\SRtwoB_{[x,\infty)}$ with $x$ being the lower bound placed on the \mttwoL\ variable, DF and SF events are combined.
The common definition of these two sets of signal regions is shown in Table~\ref{tab:SRtwoB_defs}.
 
\begin{table}[!htb]
\caption{\label{tab:SRtwoB_defs} Two-body selection. Common definition of the binned and the inclusive sets of signal regions.}
\begin{center}
\begin{tabular}{l|cc}
\noalign{\smallskip}\hline\noalign{\smallskip}
& \multicolumn{2}{c}{\SRtwoB} \\
\noalign{\smallskip}\hline\noalign{\smallskip}
Leptons flavour                  & ~~~~DF~~~~  & ~~~~SF~~~~ \\
\ptleplead [\GeV]                 & \multicolumn{2}{c}{$>25$}\\
\ptlepsubl [\GeV]                 & \multicolumn{2}{c}{$>20$}\\
$\mll$ [\GeV]                    & \multicolumn{2}{c}{$>20$}\\
\noalign{\smallskip}\hline\noalign{\smallskip}
$|\mll-m_Z|$ [\GeV]  & --  & $>20$ \\
\nbjet                           & \multicolumn{2}{c}{$\geq1$}\\
\dphib [rad]                           &\multicolumn{2}{c}{$<1.5$}\\
\METsig                          &\multicolumn{2}{c}{$>12$}\\
\noalign{\smallskip}\hline\noalign{\smallskip}
\mttwoL\ [\GeV]                   & \multicolumn{2}{c}{$>110$}\\
\noalign{\smallskip}\hline\noalign{\smallskip}
\end{tabular}
\end{center}
\end{table}
 
\FloatBarrier

\subsection{Three-body event selection}
\label{sec:threebody_selection}
 
The three-body decay mode of the top squark shown in Figure~\ref{subfig:stoptB} is dominant in the region where  $ m(\stopone) > m(\ninoone)+m(W)+m(b) $  and $ m(\stopone) < m(\ninoone) + m(t)$.
The signal kinematics  in this region  resemble that of $WW$ production when \mwsel\ and that of $t\bar{t}$ production when \mtsel.
The signal selection was optimised to reject these dominant backgrounds while not degrading signal efficiency.
The $b$-jet multiplicity is highly dependent on the mass-splitting between the top squark and the neutralino, $\DeltaM=m(\stopone)-m(\ninoone)$, since for  lower $\DeltaM$ the $b$-jets have lower momentum and cannot be reconstructed efficiently.
Accordingly, two orthogonal signal regions were defined: \SRTBW\ targeting \mwsel, applying a $b$-jet veto, and \SRTBT\ targeting $\mtsel$, allowing for $b$-jets.
Separation between same-flavour and different-flavour events is also kept to optimise model-dependent search sensitivity, thus defining four different SRs: \SRTBWDF, \SRTBWSF, \SRTBTDF\ and \SRTBTSF.
The signal regions make use of a common set of requirements on the \pt\ of the two leptons, \METsig\ and \GAMr.
The definitions of these regions are summarised in Table~\ref{tab:threebodysrdefinitions}.
\begin{table}[!htb]
\centering
\caption{Three-body selection. Signal regions definition.}
\label{tab:threebodysrdefinitions}
\begin{center}
\begin{tabular}{l| cc | cc}
\noalign{\smallskip}\hline\noalign{\smallskip}
& \multicolumn{2}{c|}{\SRTBW}  & \multicolumn{2}{c}{\SRTBT} \\
\noalign{\smallskip}\hline\noalign{\smallskip}
Leptons flavour                 & ~~~~DF~~~~                           & ~~~~SF~~~~                                   & ~~~~DF~~~~  & ~~~~SF~~~~ \\
\ptleplead [\GeV]                & \multicolumn{2}{c|}{$>25$}   & \multicolumn{2}{c}{$>25$} \\
\ptlepsubl [\GeV]                & \multicolumn{2}{c|}{$>20$}   & \multicolumn{2}{c}{$>20$} \\
\mll [GeV]                      & \multicolumn{2}{c|}{$>20$}   & \multicolumn{2}{c}{$>20$} \\
\noalign{\smallskip}\hline\noalign{\smallskip}
 
$|\mll-m_Z|$ [\GeV]             & --                           & $ > 20$                             & -- & $ > 20$   \\
\nbjet                          & \multicolumn{2}{c|}{$=0$}    & \multicolumn{2}{c}{$\geq1$}\\
\DPB [rad]                            & \multicolumn{2}{c|}{$>2.3$}  & \multicolumn{2}{c}{$>2.3$}      \\
\METsig                         & \multicolumn{2}{c|}{$>12$}   & \multicolumn{2}{c}{$>12$} \\
\GAM                            & \multicolumn{2}{c|}{$>0.7$}  & \multicolumn{2}{c}{$>0.7$}   \\
\RPT                            & \multicolumn{2}{c|}{$>0.78$} & \multicolumn{2}{c}{$>0.70$}       \\
\MDR [\GeV]                      & \multicolumn{2}{c|}{$>105$}  & \multicolumn{2}{c}{$>120$}      \\
\noalign{\smallskip}\hline\noalign{\smallskip}
\end{tabular}
\end{center}
\end{table}
\FloatBarrier
\subsection{Four-body event selection}
\label{sec:fourbody_selection}
In the kinematic region defined by $m(\stopone) < m(\ninoone) + m(b) + m(W)$ and $m(\stopone) > m(\ninoone) + m(b)$, the top squarks are assumed to decay via a four-body process through an off-shell top quark and $W$ boson as shown in Figure~\ref{subfig:stopfB}.
In this region the final-state leptons from the virtual \Wboson\ boson decay  are expected to have lower momentum and can be efficiently selected when imposing both a lower and upper bound on the \pT\ of the leptons. A transverse momentum lower bound of 4.5~\GeV\ (4~\GeV) is applied for electrons (muons), together with an upper bound, which is optimised separately for the leading and the sub-leading leptons.
Two separate signal regions  are defined to cover different \DeltaM\ ranges: the first one, \SRFBA, targets small values of \DeltaM\ and requires $\ptleplead < 25$~\GeV\ and $\ptlepsubl < 10$~\GeV; the second one, \SRFBB, targets larger values of \DeltaM\  and instead requires $\ptlepsubl>10$~\GeV. This condition also ensures orthogonality between the two SRs.
The presence of an energetic initial-state radiation (ISR) jet recoiling against the system of the two top squarks is required, introducing an imbalance in the event kinematics with an enhanced value of \MET that allows signal events to be distinguished from SM processes. For this reason, for each event, the leading jet $j_1$ is considered to be a jet from ISR and required to have $\pT>150$~\GeV.
A further reduction of the SM background is achieved with selections on \METsig, \ptllboost, \Rll and \Rlj variables.
An additional requirement is applied to improve the sub-leading lepton isolation, using the following isolation variable:
\begin{equation}
\mindRSubLJ = \min_{j_i \in \mathrm{[jets]}} \Delta R_{\eta} (\ell_{2},j_{i}) \nonumber
\end{equation}
where `[jets]' contains all the jets in the event. This reduces the probability of lepton misidentification or selecting a lepton originating from heavy-flavour or $\pi/K$ decays in jets.
The definitions of these regions are  summarised in Table~\ref{tab:fourbodysrdefinitions}.

\begin{table}[!h]
\centering
\caption{\label{tab:fourbodysrdefinitions} Four-body selection. Signal regions definition. }
\begin{tabular}{l |c |  c}
\noalign{\smallskip}\hline\noalign{\smallskip}
& \SRFBA                      & \SRFBB \\
\noalign{\smallskip}\hline\noalign{\smallskip}
\ptleplead [GeV] & $~~~~~~~~<25~~~~~~~~$ & $ ~~~~~~~~<100$~~~~~~~~       \\
\ptlepsubl [GeV] & $<10$  & $[10,\ 50] $ \\
\mll [GeV]       & \multicolumn{2}{c}{$>10$}\\
\noalign{\smallskip}\hline\noalign{\smallskip}
$\pT(j_1)$ [\GeV] & \multicolumn{2}{c}{$>150$}\\
\mindRSubLJ      & \multicolumn{2}{c}{$>1$}\\
\METsig          & \multicolumn{2}{c}{$>10$}\\
\ptllboost [\GeV]      & \multicolumn{2}{c}{$>280$}\\
\MET [\GeV]       & \multicolumn{2}{c}{$>400$}\\
\noalign{\smallskip}\hline\noalign{\smallskip}
$\Rll$           & $> 25$                      & $>13$\\
$\Rlj$           & $> 0.44$                    & $>0.38$\\
\noalign{\smallskip}\hline\noalign{\smallskip}
 
\end{tabular}
\end{table}

\FloatBarrier
\section{Background estimation}
\label{sec:backgrounds}

The MC predictions for the dominant SM background processes are improved using  a data-driven normalisation procedure, while non-dominant processes are estimated directly using  MC simulation.
A simultaneous profile likelihood fit~\cite{Cowan:2010js}  is used to constrain the MC yields   with the observed data  in dedicated background control regions (CRs).
The fit is performed using  standard minimisation software~\cite{HistFitter, pyhf} where the normalisations of the targeted backgrounds are allowed to float, while the MC simulation is used to describe the shape of kinematic variables.
Systematic uncertainties that could affect the expected yields in the different regions are taken into account in the fit  through nuisance parameters.
Each uncertainty source is described by a single nuisance parameter, and  correlations between  nuisance parameters,  background processes and selections are taken into account.
A list of the systematic uncertainties considered in the fits is provided in Section~\ref{sec:systs}.
The SM background thus modelled is validated in dedicated validation regions (VRs) which are disjoint from both the control and signal regions.
 
Important sources of reducible background are events with jets which are misidentified as leptons.
The fake/non-prompt (FNP) lepton  background comes    from   $\pi/K$ and  heavy-flavour hadron decays  and  photon conversions. This  is particularly important  for   the low-\pT  leptons  targeted by  the   four-body selection.
The FNP background is mainly  suppressed by the lepton isolation requirements described in Section~\ref{sec:objects}, but a non-negligible residual contribution is expected.
This is estimated from data using the `fake factor' method~\cite{ATLAS-2011-1104.5225, EXOT-2012-25, CMS:2018ect, CMS-PAS-EXO-17-029} which uses two orthogonal lepton definitions, labelled  as `Id' and `anti-Id', to define a control data sample enriched in fake leptons.
The Id lepton corresponds to the signal lepton identification criteria used in this analysis. Anti-Id electrons fail either the signal identification or isolation requirement, while anti-Id muons fail the isolation requirement.
The sample used for the fake-factor computation is enriched in $Z$+jets events.
Events with three leptons are selected, with the two same-flavour leptons of opposite electric charge (SFOS leptons) identified as the $Z$ boson decay products ($\ell^Z_1$ and $\ell^Z_2$, in order of decreasing \pT) satisfying the Id requirements, and  the  third unpaired lepton, called the \textit{probe} lepton ($\ell^{\mathrm{probe}}$), satisfying either the Id or anti-Id criteria.
The fake factor  is defined as the ratio of the Id lepton yield to the anti-Id probe lepton yield. Residual contributions from processes producing  prompt leptons are subtracted using the MC predictions.
Fake factors are measured separately for electrons and muons  and as a function of the lepton \pT\ and $\eta$. These  are derived in the \CRFNP\ region whose selection is summarised in Table~\ref{tab:zjets_cr_def}.
The FNP estimates in each analysis region are derived by applying  the fake factors  to events satisfying that region's criteria but replacing  at least one of the signal leptons by an anti-Id one.
 
\begin{table}[h!]
\caption{FNP selection. Detailed definition of the CR$^{\textrm{FNP}}$ region. \label{tab:zjets_cr_def}}
\centering
\begin{tabular}{l|c}
\noalign{\smallskip}\hline\noalign{\smallskip}
& CR$^{\mathrm{FNP}}$   \\
\noalign{\smallskip}\hline\noalign{\smallskip}
Lepton multiplicity                        & 3 \\
$|\mll-m_Z|$ [\GeV]                        & $<10$ for SFOS pair \\
$\pT(\ell^Z_1)$ [\GeV]                     & $>25$ \\
$\pT(\ell^Z_2)$ [\GeV]                     & $>20$ \\
$\pT(\ell^{\mathrm{probe}})$ [\GeV]                 & $>4.5\,(4.0) ~e\,(\mu)$ \\
\noalign{\smallskip}\hline\noalign{\smallskip}
$\Delta R_\eta(\ell^\mathrm{probe}, \ell_i)$ & $>0.2$ \\
$m_{\mathrm{T}}(\ell^\mathrm{probe},\MET)$ [\GeV]          & $<40$         \\
\noalign{\smallskip}\hline\noalign{\smallskip}
\multirow{3}{*}{Additional requirements}           &$\pT(\ell^{\mathrm{probe}})<16$~\GeV\\
& or   \\
&$\MET < 50~\GeV$\\
\noalign{\smallskip}\hline\noalign{\smallskip}
\end{tabular}
\end{table}
 
The three selections in this paper use different sets of CRs and VRs, specifically designed to be   kinematically similar to the respective SRs. The definitions of the regions used in each analysis and the results of the fits are described in the following subsections.
 
\FloatBarrier
\subsection{Estimation of the backgrounds in the two-body selection}
\label{sec:twobody_backgrounds}
 
The main background sources for the two-body selection are \ttbar\ and \ttZ\ with invisible decay of the $Z$ boson.
These processes are normalised to data in dedicated CRs: \CRtwoT\ and \CRtwoTZ.
The \ttbar normalisation factor is extracted from different-flavour dilepton events.
In order to test the reliability of the \ttbar background prediction, two validation regions \VRtwoTDF\ and \VRtwoTSF\ are defined.
The \ttZ\ production events with invisible decay of the $Z$ boson are expected to dominate the tail of the \mttwoL\ distribution in the SRs and are normalised in the dedicated control region \CRtwoTZ. Given the difficulty in achieving sufficient purity for this SM process because of the high contamination from \ttbar events, a strategy based on a three-lepton final state is adopted. Events are selected if characterised by three charged leptons including at least one pair of SFOS leptons having invariant mass consistent with that of the $Z$~boson ($|\mll-m_{Z}|<20$~\GeV). If more than one pair is identified, the one with $\mll$ closest to the $Z$~boson mass is chosen. Events are further required to have a jet multiplicity, \njet, greater than or equal to three with at least two $b$-tagged jets.
These selections target \ttZ\ production with the $Z$ boson decaying into two leptons and \ttbar\ decaying in the semileptonic channel.
In order to select \ttZ\ events whose kinematics, regardless of subsequent \ttbar\ and $Z$ decays, emulate the kinematics of this background in the SRs, the momenta of the two leptons of the SFOS pair $(\mathbf p(\ell^\mathrm{Z}_1),\mathbf p(\ell^\mathrm{Z}_2))$ are vectorially added to the $\mathbf p^\mathrm{miss}_\mathrm{T}$, effectively treating them like the neutrino pair from the $Z$ boson decay. A variable  called $\METcorr=\left|\left(\mathbf p^\mathrm{miss}_\mathrm{T}+\mathbf p(\ell^\mathrm{Z}_1)+\mathbf p(\ell^\mathrm{Z}_2)\right)_{\mathrm{T}}\right|$ is constructed. Events characterised by high \mttwoL\ in the SRs are emulated by requiring high \METcorr\ values in \CRtwoTZ. In order to check the $\ttZ$ background estimation, the validation region \VRtwoTZ\ was defined. For this region, events with four leptons are selected and required to have at least one pair of SFOS leptons compatible with the $Z$ boson decay.
A variant of the \mttwo\ variable called $\mttwo^{4\ell}$ is defined from the $\pTcorr=\left(\mathbf p^\mathrm{miss}_\mathrm{T}+\mathbf p(\ell^\mathrm{Z}_1)+\mathbf p(\ell^\mathrm{Z}_2)\right)_\mathrm{T}$\ and the momenta of the remaining two leptons.
The definition of the control and validation regions used in the two-body selection is summarised in Table~\ref{tab:twobody_CRdef}.
The expected signal contamination in the CRs is generally below $\sim 1\%$. The signal contamination in the VRs is less than $15\%$ ($7\%$) for a DM signal model with scalar (pseudoscalar) mediator mass of $100~\GeV$ and DM mass of $1~\GeV$.

\begin{table}[!htb]
\caption{Two-body selection. Control and validation regions definition. The common selection defined in Section~\ref{sec:selection} also applies to all regions.}
\label{tab:twobody_CRdef}
\begin{center}
\begin{adjustbox}{width=1.\textwidth}
\begin{tabular}{l | c c | c c | c }
\noalign{\smallskip}\hline\noalign{\smallskip}
& \CRtwoT    & \CRtwoTZ                        & \VRtwoTDF                     & \VRtwoTSF & \VRtwoTZ \\
\noalign{\smallskip}\hline\noalign{\smallskip}
Lepton multiplicity    & 2          & 3                               & \multicolumn{2}{c}{$2$}       & 4 \\
Lepton flavour         & DF         & at least one SFOS pair          & DF                            & SF        & at least one SFOS pair \\
\ptleplead [\GeV]       & $>25$      & $>25$                           & \multicolumn{2}{c}{$>25$}     & $>25$  \\
\ptlepsubl [\GeV]       & $>20$      & $>20$                           & \multicolumn{2}{c}{$>20$}     & $>20$ \\
$\pT(\ell_3)$ [\GeV]    & --         & $>20$                           & \multicolumn{2}{c}{--}        & $>20$  \\
$\pT(\ell_4)$ [\GeV]    & --         & --                              & \multicolumn{2}{c}{--}        & $>20$  \\
\mll                   & $>20$      & --                              & \multicolumn{2}{c}{$>20$}     & --  \\
\noalign{\smallskip}\hline\noalign{\smallskip}
$|\mll-m_Z|$ [\GeV]    & --         &$<20$ for at least one SFOS pair & --                            & $>20$     & $<20$ for the SFOS pair  \\
\nbjet                 & $\geq1$    &$\geq2$ with $\njet\geq3$        &\multicolumn{2}{c}{$\geq$ 1}   & $>0$\\
\dphib [rad]           & $\geq 1.5$ & --                              &\multicolumn{2}{c}{$< 1.5$}    & --\\
\METsig                & $> 8$      &--                               &\multicolumn{2}{c}{$> 12$}     & --\\
\METcorr [\GeV]         & --         &$>140$                           &\multicolumn{2}{c}{--}         & --\\
\noalign{\smallskip}\hline\noalign{\smallskip}
\mttwoL\ [\GeV]         & [100, 120] &--                               &\multicolumn{2}{c}{[100, 110]} & -- \\
$\mttwo^{4\ell}$ [\GeV] & --         &--                               & \multicolumn{2}{c}{--}        & >110 \\
\noalign{\smallskip}\hline\noalign{\smallskip}
\end{tabular}
\end{adjustbox}
\end{center}
\end{table}
 
Figure~\ref{fig:var_CRbC} illustrates the modelling of the shape of two important variables after the background fit:  (a)  shows the \dphib distribution  with the \CRtwoT\  selection, and (b) shows the \mll distribution of the SFOS leptons in the \CRtwoTZ\ selection. Good agreement is found between the data and the background model for all of the selection variables.
 
\begin{figure}[htb!]
\begin{center}
\subfigure[]{
\includegraphics[width=0.48\textwidth]{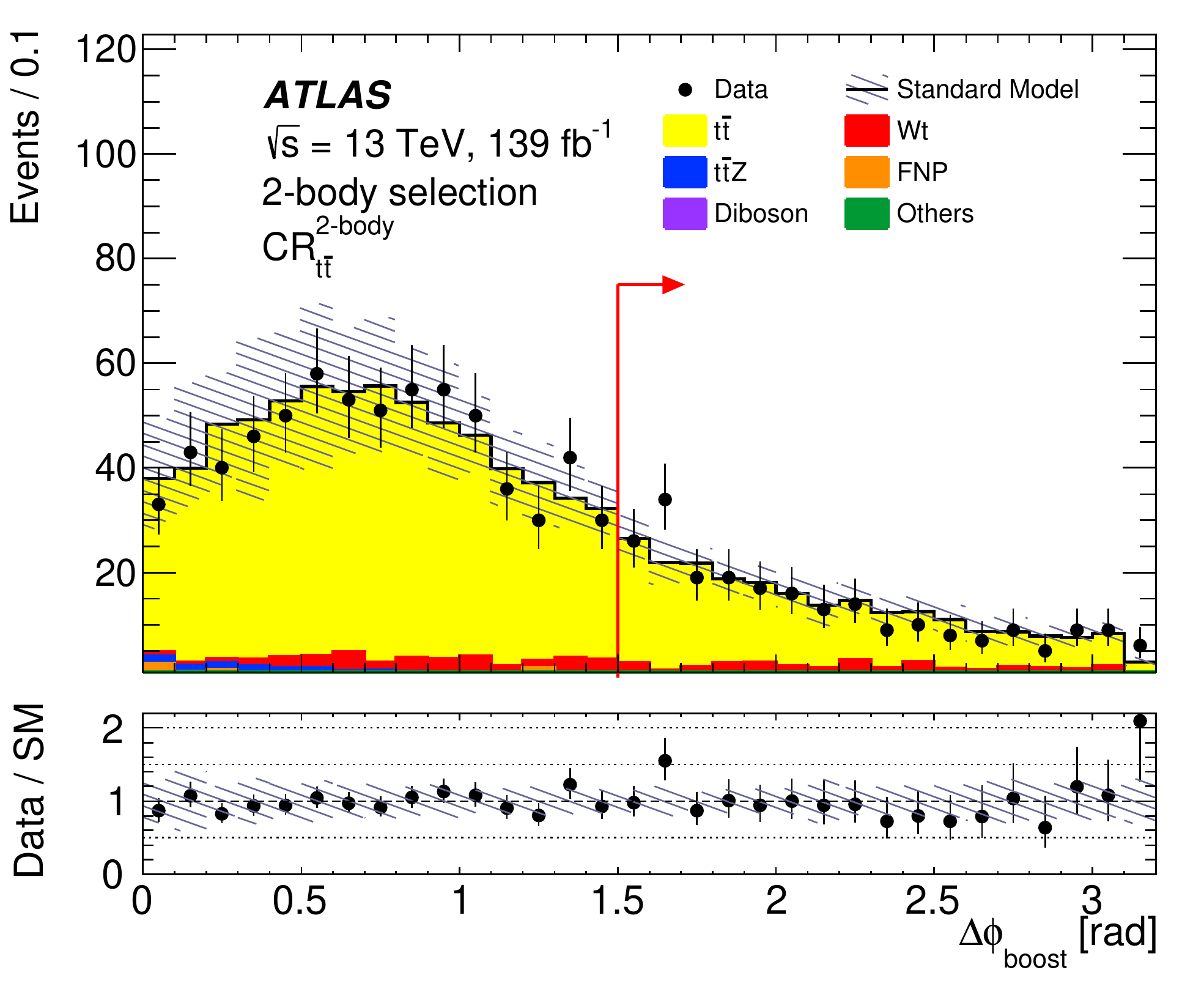}
}
\subfigure[]{
\includegraphics[width=0.48\textwidth]{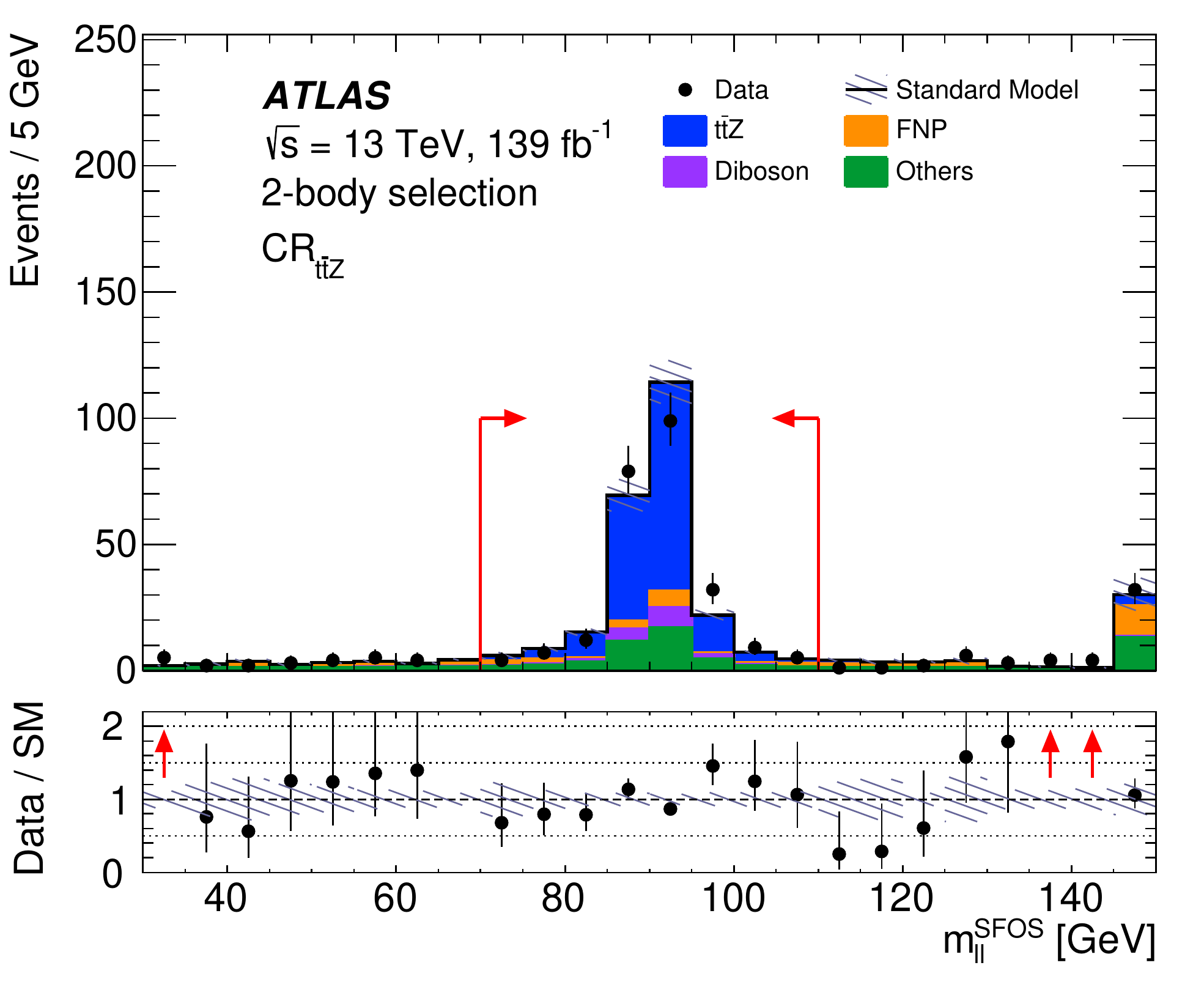}
}
\end{center}
\caption{Two-body selection. Distributions of (a) \dphib in \CRtwoT\ and (b) \mll of the two same-flavour and opposite-charge leptons candidate in \CRtwoTZ, each after the background fit. The contributions from all SM backgrounds are shown as a histogram stack. ``Others'' includes  the contributions from $VVV$, $\ttbar t$, $\ttbar\ttbar$, $\ttbar W$, $\ttbar WW$, $\ttbar WZ$, $\ttbar H$, and $tZ$. The hatched bands represent the total statistical and detector-related systematic uncertainty. The rightmost bin of (b) includes overflow events. In the upper panels, red arrows indicate the control region selection criteria. The bottom panels show the ratio of the observed data to the total SM background prediction, with hatched bands representing the total uncertainty in the background prediction; red arrows show data outside the vertical-axis range. }
\label{fig:var_CRbC}
\end{figure}
 
\begin{figure}[htb!]
\begin{center}
\subfigure[]{
\includegraphics[width=0.48\textwidth]{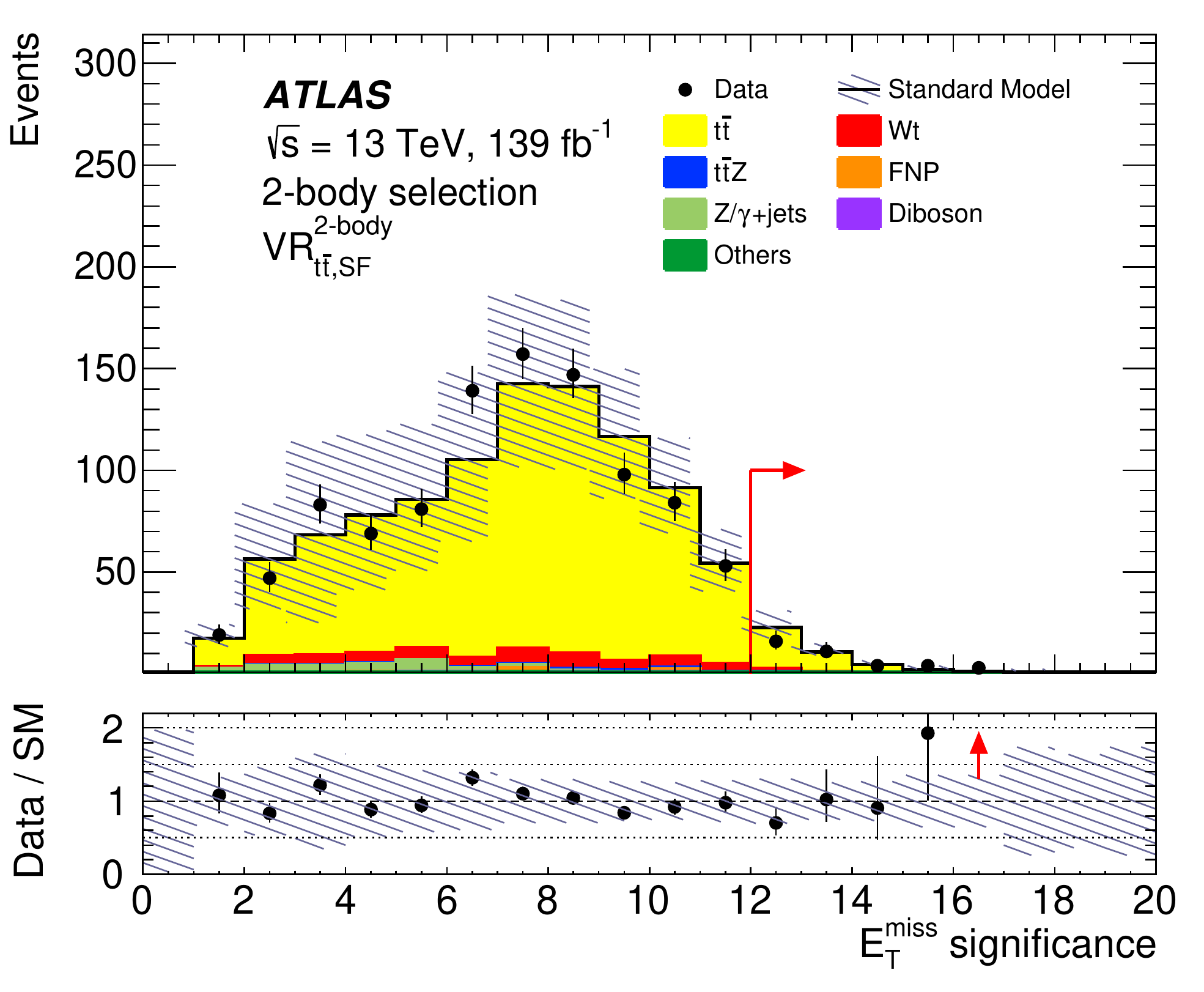}
}
\subfigure[]{
\includegraphics[width=0.48\textwidth]{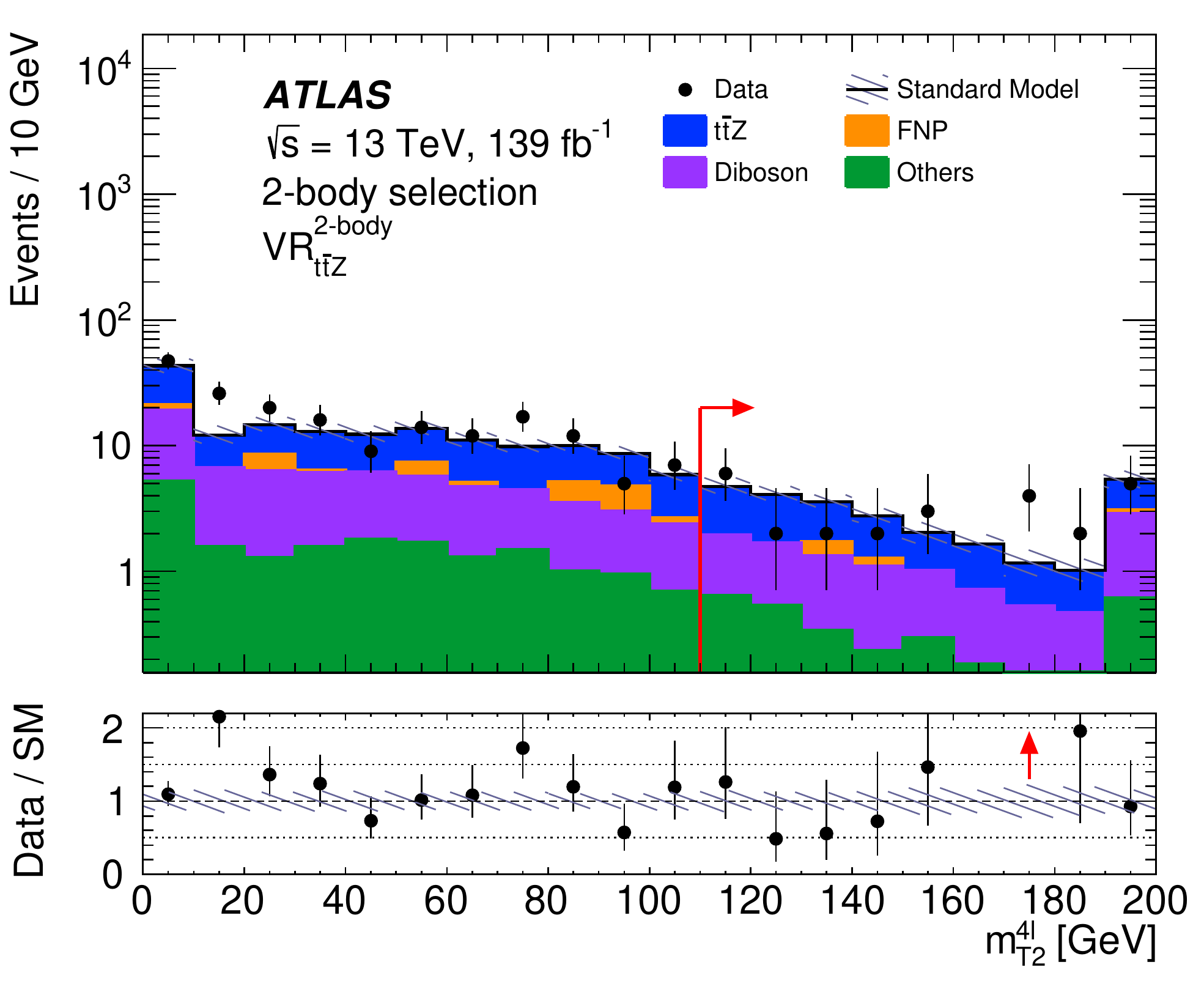}
}
\end{center}
\caption{Two-body selection. Distributions of the \METsig in (a) \VRtwoTSF\ and (b) $\mttwo^{4\ell}$ in \VRtwoTZ, each after the background fit. The contributions from all SM backgrounds are shown as a histogram stack. ``Others'' includes contributions from $VVV$, $\ttbar t$, $\ttbar\ttbar$, $\ttbar W$, $\ttbar WW$, $\ttbar WZ$, $\ttbar H$, and $tZ$ processes.  The hatched bands represent the total statistical and detector-related systematic uncertainty. The rightmost bin of each plot includes overflow events. In the upper panels, red arrows indicate the validation region selection criteria. The bottom panels show the ratio of the observed data to the total SM background prediction, with hatched bands representing the total uncertainty in the background prediction; red arrows show data outside the vertical-axis range. }
\label{fig:var_CRbD}
\end{figure}

The results of the  fit  are reported in Table~\ref{tab:twoBODYfitRES} for the two-body CRs and VRs. The normalisations for fitted backgrounds are found to be consistent with the theoretical predictions when uncertainties are considered: the normalisation factors obtained from the fit for \ttbar\ and \ttZ\ are $0.88 \pm 0.08$  and $1.07 \pm 0.14$ respectively.
Good agreement, within one standard deviation of the SM background prediction, is observed in the VRs (see Figure~\ref{fig:var_CRbD}).
 
\begin{table}[h!]
\caption{Two-body selection. Background fit results for \CRtwoT, \CRtwoTZ, \VRtwoTDF, \VRtwoTSF\ and \VRtwoTZ. ``Others'' includes contributions from $VVV$, $\ttbar t$, $\ttbar\ttbar$, $\ttbar W$, $\ttbar WW$, $\ttbar WZ$, $\ttbar H$, and $tZ$ processes. Combined statistical and systematic uncertainties are given. Entries marked `--' indicate a negligible background contribution (less than 0.001 events). The individual uncertainties can be correlated, and do not necessarily add up in quadrature to the total background uncertainty.
\label{tab:twoBODYfitRES}
}
\begin{center}
\small
\begin{tabular}{l|cc|ccc}
\noalign{\smallskip}\hline\noalign{\smallskip}
& \CRtwoT              & \CRtwoTZ       & \VRtwoTDF            & \VRtwoTSF            & \VRtwoTZ       \\[-0.05cm]
\noalign{\smallskip}\hline\noalign{\smallskip}
Observed events               & $230$                & $247$          & $45$                 & $38$                 & $26$       \\
\noalign{\smallskip}\hline\noalign{\smallskip}
Total (post-fit) SM events    & $230 \pm 15$~~         & $246 \pm 16$~~   & $50 \pm 15$          & $42 \pm 11$          & $25.7 \pm 3.4$~~ \\
\noalign{\smallskip}\hline\noalign{\smallskip}
Post-fit, \ttbar              & $196 \pm 17$~~         & --             & $44 \pm 15$          & $36 \pm 11$          & --  \\
Post-fit, \ttZ                & $0.49 \pm 0.23$      & $170 \pm 22$~~   & $1.7 \pm 0.6$        & $1.9 \pm 0.6$        & $14.0 \pm 2.1$~~ \\
$Wt$                          & $31 \pm 7$~~           & --             & $2.7 \pm 1.2$        & $2.6 \pm 1.2$        & --  \\
Diboson                       & $1.0 \pm 0.6$        & $17 \pm 4$~~     & $0.50 \pm 0.25$      & $0.59 \pm 0.32$      & $8.7 \pm 3.0$  \\
Others                        & $1.1 \pm 0.5$        & $44 \pm 12$    & $1.0 \pm 0.6 $       & $0.8 \pm 0.5 $       & $3.01 \pm 0.87$\\
Fake and non-prompt           & $0.0_{-0.0}^{+0.5}$~ & $16\pm 8$~~      & $0.0_{-0.0}^{+0.5}$~  & $0.0_{-0.0}^{+0.5}$~  & $0.0_{-0.0}^{+0.5}$~        \\
\noalign{\smallskip}\hline\noalign{\smallskip}
\end{tabular}
\end{center}
\end{table}

\FloatBarrier
\subsection{Estimation of the backgrounds in the three-body selection}
\label{sec:threebody_backgrounds}
 
The dominant SM backgrounds in the  three-body signal regions  are diboson, \ttbar and $\ttZ$ production.
Dedicated CRs were defined, labelled as \CRthreeV\ and \CRthreeT, which are kinematically close to the SRs and which have good purity in diboson and \ttbar events respectively. The orthogonality between CRs and SRs is mainly ensured by the inversion of the \DPB cut.
The normalisation of the $\ttZ$ background is extracted using the same control region \CRtwoTZ\  defined for the two-body selection in Section~\ref{sec:twobody_backgrounds}.
Dedicated validation regions were defined to test the modelling of these processes: \VRthreeV\ for the diboson background,  and \VRthreeONET\ and \VRthreeTWOT\ for the validation of the \ttbar background, where \VRthreeONET\ is characterised by a $b$-jet veto while at least one $b$-jet is required in \VRthreeTWOT.
The definition of the control and validation regions is summarised in Table~\ref{tab:threebody_CRdef}.
The expected signal contamination is below $2\%$ in the CRs and reaches a maximum of $10\%$ in the VRs for a top squark mass of $\sim 430$~\GeV.

\begin{table}[!htb]
\caption{Three-body selection. Control and validation regions definitions. The common selection defined in Section~\ref{sec:selection} also applies to all regions. A further control region \CRtwoTZ\ was defined previously in Table~\ref{tab:twoBODYfitRES}.}
\label{tab:threebody_CRdef}
\begin{center}
\small
\begin{tabular}{l | c c | c c c}
\noalign{\smallskip}\hline\noalign{\smallskip}
& \CRthreeT & \CRthreeV       & \VRthreeONET\ & \VRthreeTWOT & \VRthreeV\       \\ 
\noalign{\smallskip}\hline\noalign{\smallskip}
Lepton flavour                        & DF        & DF+SF           & DF            & DF           & DF+SF              \\
\ptleplead [\GeV]                      & $> 25$    & $> 25$          & $> 25$        & $> 25$       & $> 25$              \\
\ptlepsubl [\GeV]                      & $> 20$    & $> 20$          & $> 20$        & $> 20$       & $> 20$              \\
\mll [GeV]                            & $> 20$    & $> 20$          & $> 20$        & $> 20$       & $> 20$              \\
\noalign{\smallskip}\hline\noalign{\smallskip}
$|\mll-m_{Z}|$ [\GeV]                  & --        & $>20$ (SF only) & --            & --           & $>20$ (SF only)    \\
\nbjet                                & $\geq 2$  & $=0$            & $=0$          & $\geq 1$     & $=0$               \\
\MDR\ [\GeV]                           & $>80$     & $>100$          & $[80,105]$    & $[80,120]$   & $>100$         \\
\RPT                                  & --        & $> 0.3$         & $> 0.7$       & $> 0.7$      & $> 0.7$             \\
\GAM                                  & > 0.7     & $> 0.7$         & $> 0.7$       & $> 0.7$      & $[0.45,0.7]$       \\
\METsig                               & $> 10$    & $> 10$          & $> 12$        & $>12$        & $>12$              \\
\DPB [rad]                            & $< 2.3$   & $< 2.3$         & $> 2.3$       & $> 2.3$      & $ >2.3$             \\
\noalign{\smallskip}\hline\noalign{\smallskip}
\end{tabular}
\end{center}
\end{table}
 
Table~\ref{tab:threebodybkgonly_CRVR} shows the expected and observed numbers of events in each of the control and validation regions after the background fit. The normalisation factors extracted from the fit of the backgrounds for the diboson, \ttbar\ and \ttZ\ production processes are $0.92 \pm 0.28$, $0.96 \pm 0.09$ and $1.06 \pm 0.15$ respectively. The total number of fitted background events in the validation regions is in agreement with the observed number of data events. Figure~\ref{fig:threebodykinCRPlots} shows the distributions of \DPB\ for the  \CRthreeV\ and \CRthreeT\ selections after the background fit, illustrating  the MC modelling of the shape for this  variable.
Figure~\ref{fig:threebodykinVRPlots} shows  distributions of $\RPT$ in \VRthreeONET\ and \VRthreeTWOT, and  of  $\DPB $ in \VRthreeV,  after the background fit.  Good agreement, within one standard deviation of the SM background prediction, is observed in the validation regions.
 
\begin{table}[!htb]
\caption{Three-body selection. Background fit results for \CRthreeV, \CRthreeT, \CRtwoTZ, \VRthreeV, \VRthreeONET\ and \VRthreeTWOT. ``Others'' includes contributions from $VVV$, $\ttbar t$, $\ttbar\ttbar$, $\ttbar W$, $\ttbar WW$, $\ttbar WZ$, $\ttbar H$, and $tZ$ processes. Combined statistical and systematic uncertainties are given. Entries marked `--' indicate a negligible background contribution (less than 0.001 events). The individual uncertainties can be correlated, and do not necessarily add up in quadrature to the total background uncertainty.}
\label{tab:threebodybkgonly_CRVR}
\begin{center}
\small
\begin{tabular}{ l | c c c | c c c }
\noalign{\smallskip}\hline\noalign{\smallskip}
& \CRthreeT              & \CRthreeV\      & \CRtwoTZ\    & \VRthreeONET\          & \VRthreeTWOT           & \VRthreeV\  \\
\noalign{\smallskip}\hline\noalign{\smallskip}
Observed events            & $192$                  & $169$           & $247$        & $41$                   & $137$                  & $84$        \\
\noalign{\smallskip}\hline\noalign{\smallskip}
Total (post-fit) SM events & $192 \pm 14$~~           & $169 \pm 13$~~    & $247 \pm 16$~~ & $38.3 \pm 5.9$~~         & $142 \pm 25$~~           & $97 \pm 15$              \\
\noalign{\smallskip}\hline\noalign{\smallskip}
Post-fit, $\ttbar$         & $180 \pm 14$~~           & $65 \pm 14$     & $-$          & $25 \pm 5$~~             & $130 \pm 24$~~           & $44 \pm 11$              \\
Post-fit,  \ttZ            & $1.57 \pm 0.33$        & $1.36 \pm 0.31$ & $172 \pm 23$~~ & $0.07_{-0.07}^{+0.12}$~\, & $1.6 \pm 0.7$          & $1.0  \pm 0.4$              \\
Post-fit,  diboson         & $0.063 \pm 0.035$      & $74 \pm 21$     & $16 \pm 7$~~   & $11 \pm 4$~~             & $0.9 \pm 0.5$          & $41 \pm 14$              \\
$Wt$                       & $9.0 \pm 1.4$          & $7.6 \pm 2.3$   & $-$          & $1.9 \pm 0.6$          & $8.1 \pm 2.0$          & $8.1 \pm 1.1$              \\
\zjets          & $-$                    & $13 \pm 5$~~      & $-$          & $-$                    & $-$                    & $0.04_{-0.04}^{+0.05}$~\,              \\
Others                     & $1.39 \pm 0.21$        & $3.57 \pm 0.24$ & $43 \pm 12$  & $0.27\pm0.06$          & $1.11 \pm 0.18$        & $1.15 \pm 0.11$              \\
Fake and non-prompt        & $0.00_{-0.00}^{+0.22}$~\, & $5.0 \pm 1.9$   &$16 \pm 8$~~    & $0.00_{-0.00}^{+0.27}$~\, & $0.00_{-0.00}^{+0.27}$~\, & $1.8 \pm 1.5$              \\
\noalign{\smallskip}\hline\noalign{\smallskip}
\end{tabular}
\end{center}
\end{table}
 
\begin{figure}[!htb]
\begin{center}
\subfigure[]{
\includegraphics[width=0.48\textwidth]{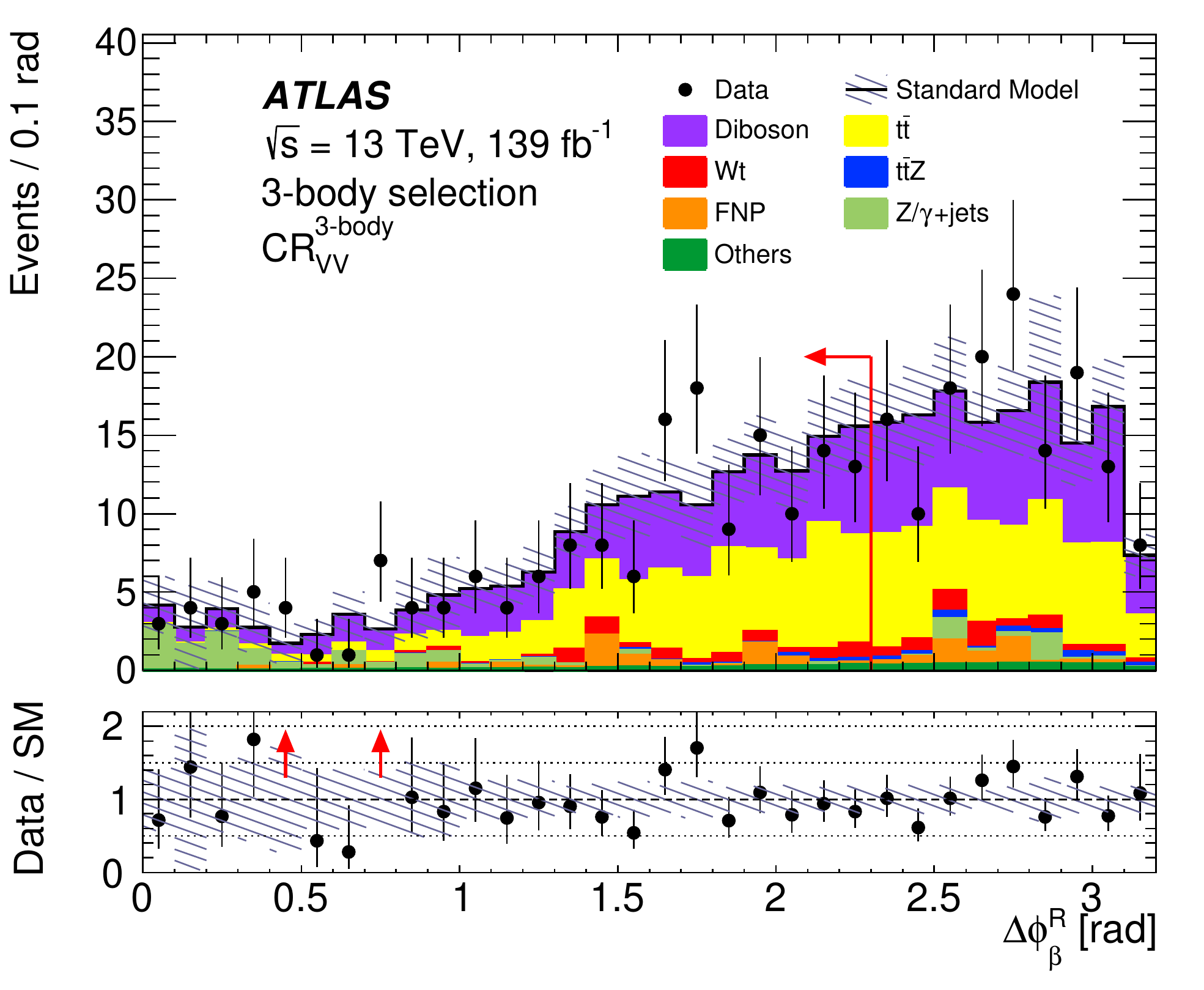}
}
\subfigure[]{
\includegraphics[width=0.48\textwidth]{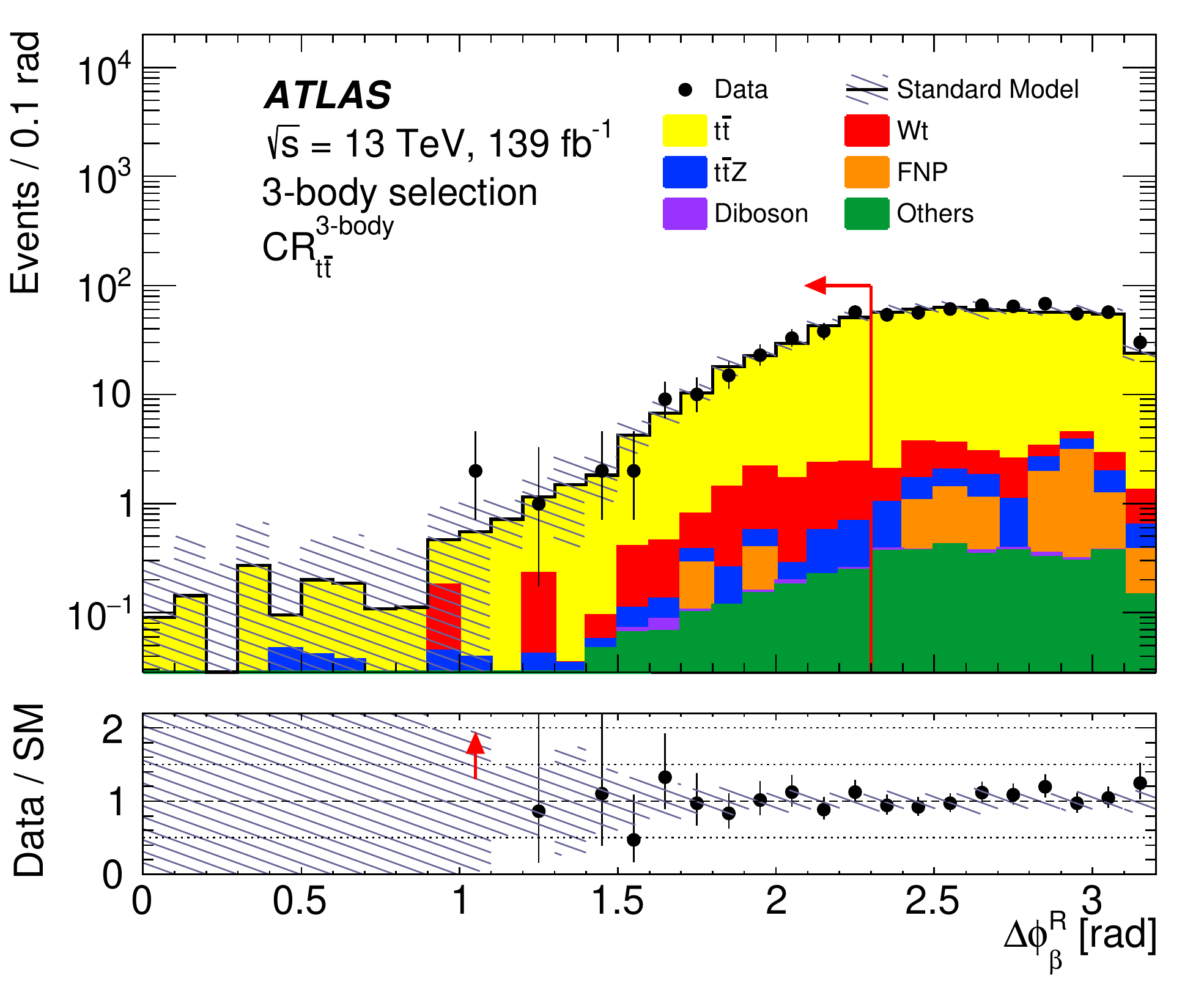}
}
\end{center}
\caption{Three-body selection. Distributions of (a) $\DPB$ in the \CRthreeV\ selection, and (b) in the \CRthreeT\ selection, after the background fit. The contributions from all SM backgrounds are shown as a histogram stack. ``Others'' includes contributions from $VVV$, $\ttbar t$, $\ttbar\ttbar$, $\ttbar W$, $\ttbar WW$, $\ttbar WZ$, $\ttbar H$, and $tZ$ processes.  The hatched bands represent the total statistical and detector-related systematic uncertainty. In the upper panels, red arrows indicate the control region selection criteria. The bottom panels show the ratio of the observed data to the total SM background prediction, with hatched bands representing the total uncertainty in the background prediction; red arrows show data outside the vertical-axis range. }
\label{fig:threebodykinCRPlots}
\end{figure}

\begin{figure}[!htb]
\begin{center}
\subfigure[]{
\includegraphics[width=0.48\textwidth]{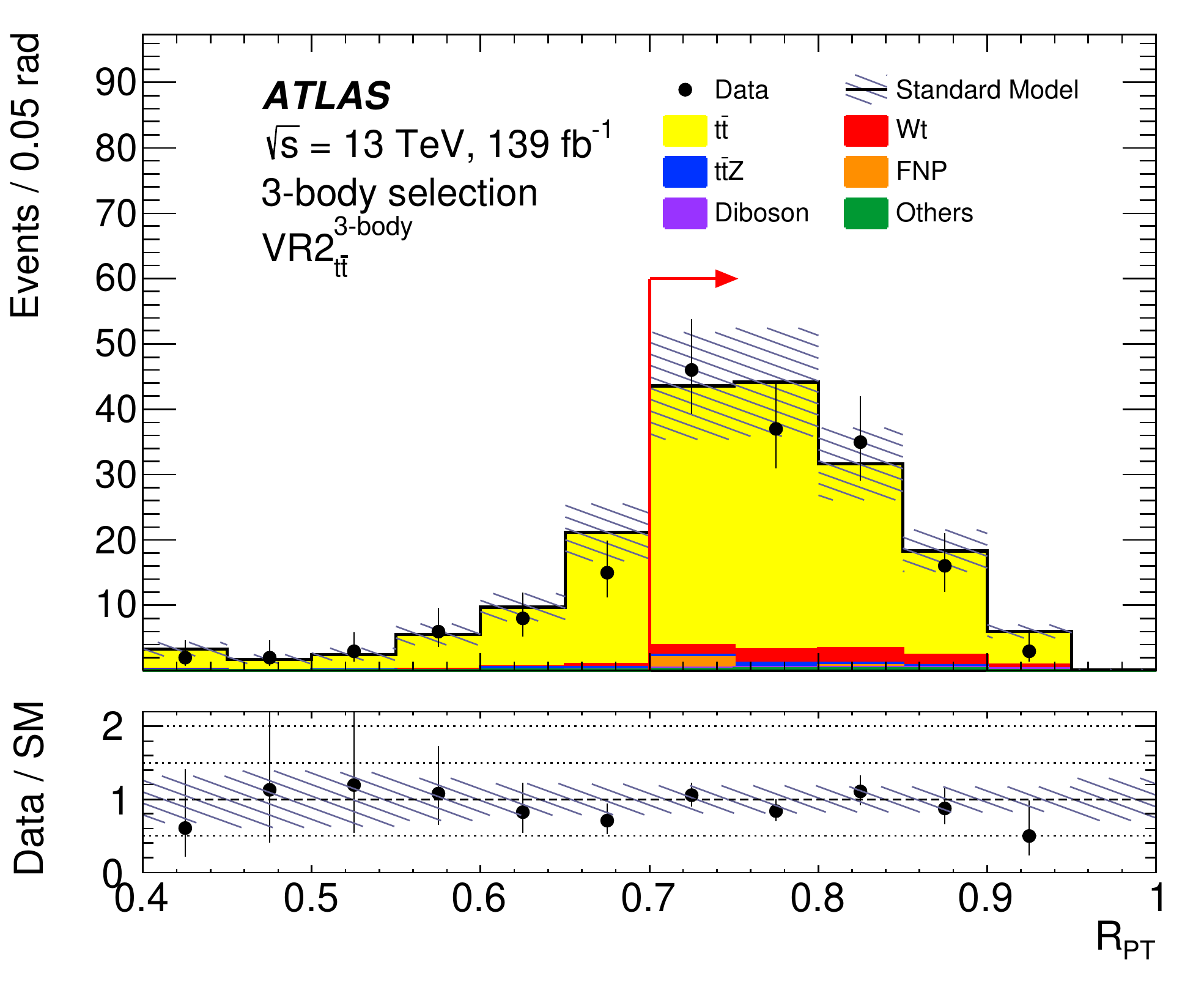}
}
\subfigure[]{
\includegraphics[width=0.48\textwidth]{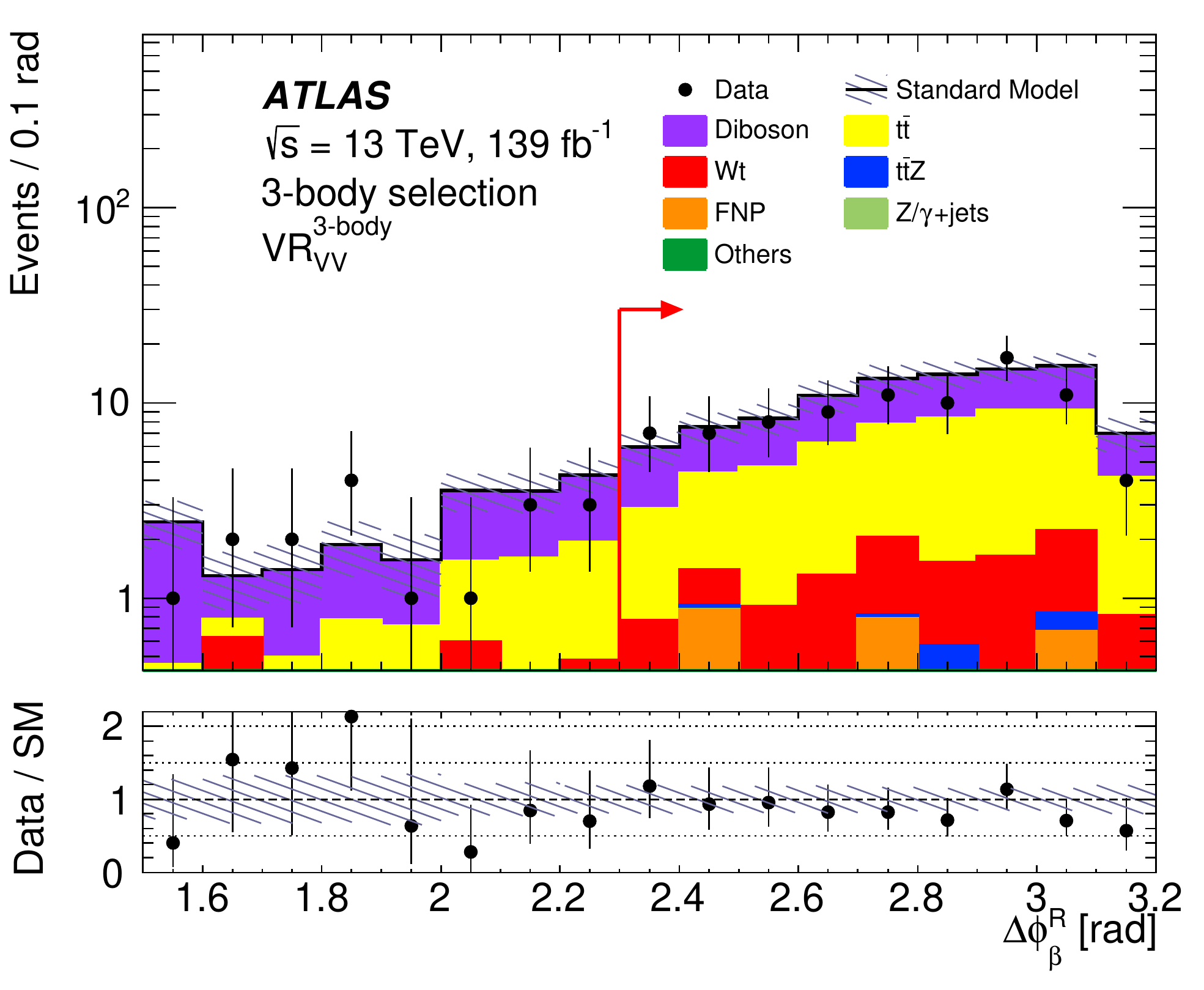}
}
\end{center}
\caption{Three-body selection. Distributions of (a) $\RPT$ in  the validation region \VRthreeTWOT\ and  (b) $\DPB$ in the validation region \VRthreeV,  after the background fit. The contributions from all SM backgrounds are shown as a histogram stack. ``Others'' includes contributions from $VVV$, $\ttbar t$, $\ttbar\ttbar$, $\ttbar W$, $\ttbar WW$, $\ttbar WZ$, $\ttbar H$, and $tZ$ processes. The hatched bands represent the total statistical and detector-related systematic uncertainty. The bottom panels show the ratio of the observed data to the total SM background prediction, with hatched bands representing the total uncertainty in the background prediction.
}
\label{fig:threebodykinVRPlots}
\end{figure}
 
\FloatBarrier
\subsection{Estimation of the backgrounds in the four-body selection}
\label{sec:fourbody_backgrounds}
 
The dominant irreducible SM background sources for the four-body selection are \ttbar\ and diboson: these backgrounds are normalised in two dedicated background-enriched  control regions labelled as \CRfourT\ and \CRfourV.
Some of the requirements defining the kinematics of the SRs are relaxed in order to allow the selection of \ttbar events in \CRfourT, while the \Rll selection is adjusted to maintain complete orthogonality with the SRs.
The diboson contribution in \CRfourV\ is enhanced by  limiting  the number of jets in the event  and the sub-leading jet \pT, and by the additional  veto on $b$-jets.
The background predictions are tested in validation regions: \VRfourT\ for \ttbar validation and \VRfourV\ and \VRfourVlll\ for diboson validation, with the latter two selecting, respectively, events with two and three leptons in the final state.
For \VRfourVlll\  a new set of variables is defined in order to mimic the dibosons' kinematics in the signal regions.
The two SFOS leptons with an invariant mass closest to $m_Z$ are considered as the two leptons coming from the decay of the $Z$ boson.
The momentum of the lepton ($\mathbf p(\ell^\mathrm{Z}_{\mathrm{paired}})$) of the selected pair having the same electric charge as the non-paired lepton is added to the $\mathbf p^\mathrm{miss}_{\mathrm{T}}$ in order to define $E_{T,1\ell,\mathrm{corr}}^{\mathrm{miss}}=\left|\left(\mathbf p^\mathrm{miss}_{\mathrm{T}}+\mathbf p(\ell^\mathrm{Z}_{\mathrm{paired}})\right)_{\mathrm{T}}\right|$ and $R_{\ell,\mathrm{corr}}$ is defined as the ratio of $E_{T,1\ell,\mathrm{corr}}^{\mathrm{miss}}$ to the sum of the transverse momenta of two remaining OS leptons. The invariant mass of the remaining two leptons, called $m_{\ell\ell,\mathrm{corr}}$, is also used.
The definition of the control and validation regions used in the four-body selection is summarised in Table~\ref{tab:fourbodycrvrdef}.
In the \ttbar control region the signal contamination is $\sim 1\%$ or less. In  \CRfourV, the typical  signal contamination is about $\sim \mathrm{1--2}\%$, but reaches a maximum value of $\sim 5$\% for a top squark mass of $\sim 400$~\GeV\ and  lightest-neutralino mass of $\sim 310$~\GeV\ at the boundary of the region excluded by the previous analysis. Signal contamination in the validation regions is below 10\%.
 
\begin{table}[!htb]
\caption{Four-body selection. Control and validation regions definition. The common selection defined in Section~\ref{sec:selection} also applies to all regions.}
\label{tab:fourbodycrvrdef}
\begin{center}
\small
\begin{tabular}{l | c c | c c c}
\noalign{\smallskip}\hline\noalign{\smallskip}
& \CRfourT & \CRfourV              &\VRfourT        & \VRfourV          & \VRfourVlll\\[-0.05cm]
\noalign{\smallskip}\hline\noalign{\smallskip}
Lepton multiplicity                            & $2 $     & $2$                   & $2$            & $2$               & $3$ \\
Lepton flavour                                 & DF+SF    & DF+SF                 & DF+SF          & DF+SF             & at least one SFOS pair\\
\ptleplead [\GeV]                              & $<100$   & $<100$                & $<100$         &$<100$             & $<100$\\
\ptlepsubl [\GeV]                              & ~~$<50$    & ~~$<50$                 & ~~$<50$          & ~~$<50$              & $<100$\\
$\pT(\ell_3)$ [\GeV]                           & --       & --                    & --             &--                 & $<100$\\
\mll [\GeV]                                    & ~~$>10$    & ~~$>45$                 & ~~$>10$          & ~~$>45$             & ~~$>10$ \\
\noalign{\smallskip}\hline\noalign{\smallskip}
$|\mll-m_{Z}|$ [\GeV]                          & --       & $>10$ for SF only     & --             & $>10$ for SF only & -\\
\MET [\GeV]                                    &$>350$    &$>250$                 &$>250$          &$>250$             & $>250$         \\
$\pT(j_1)$ [\GeV]                              & $>150$   & $>150$                & $>150$         & $>150$            & $>150$ \\
\mindRSubLJ                                    & $>1$     & $>1$                  & $>1$           & $>1$              & $>1$   \\
\njet                                          & --       & $\leq 2$              & --             & $\leq 4$          & $<5$     \\
\nbjet                                         & $\geq 2$ & $=0$                  & $\geq 1$       & $=0$              & $=0$     \\
$b$-tagged $j_1$                               & --       & --                    & True           & --                & --     \\
$\pT(j_2)$ [\GeV]                              & --       & $<40$ if $j_2$ exists & --             & --                & -- \\
\METsig                                        & $>10$    & $>10$                 & $>10$          & $>10$             & $>5$  \\
\ptllboost [\GeV]                              & $>280$   & $>280$                & $>280$         & $>280$            & --\\
\Rll                                           & $< 5$    &$< 4$                  & $> 5$          & $[4,\ 5]$         & --    \\
\Rlj                                           & --       & --                    & $[0.3,\ 0.38]$ & --                & --     \\
\noalign{\smallskip}\hline\noalign{\smallskip}
$E_{\mathrm{T},1\ell,\mathrm{corr}}^{\mathrm{miss}}$ [\GeV] &--        &--                     &--              &--                 & $>300$\\
$R_{2\ell,\mathrm{corr}}$                        &--        &--                     &--              &--                 & $>5$\\
$m_{\ell\ell,\mathrm{corr}}$ [\GeV]              &--        &--                     &--              &--                 & ~~$>10$\\
\noalign{\smallskip}\hline\noalign{\smallskip}
\end{tabular}
\end{center}
\end{table}
 
Table~\ref{tab:fourbodybkgonly_CRVR} shows the expected and observed numbers of events in each of the control and validation regions after the background fit.
The normalisation factors extracted by the fit for the diboson and \ttbar production processes are $1.00 \pm 0.25$ and $0.90 \pm 0.12$ respectively.
The distributions of \MET  in \CRfourT\ and \Rll  in \CRfourV, after the background fit, are shown in Figure~\ref{fig:fourbodykinCRPlots}.
The distributions of \ptlepsubl in \VRfourT, \njet in \VRfourV\ and $E_{T,1\ell,\mathrm{corr}}^{\mathrm{miss}}$  in \VRfourVlll, after the background fit, are shown in Figure~\ref{fig:fourbodykinVRPlots}. Good agreement between data and the SM predictions is observed.
 
\begin{table}[!htb]
\caption{Four-body selection. Background fit results for \CRfourT, \CRfourV, \VRfourT, \VRfourV\ and \VRfourVlll. The `Others' category contains the contributions from $VVV$, $\ttbar t$, $\ttbar\ttbar$, $\ttbar W$, $\ttbar WW$, $\ttbar WZ$, $\ttbar H$, and $tZ$. Combined statistical and systematic uncertainties are given.  Entries marked `--' indicate a negligible background contribution (less than 0.001 events). The individual uncertainties can be correlated, and do not necessarily add up in quadrature to the total background uncertainty.}
\label{tab:fourbodybkgonly_CRVR}
\begin{center}
\small
\begin{tabular}{l|cc|ccc}
\noalign{\smallskip}\hline\noalign{\smallskip}
& \CRfourT        & \CRfourV        &\VRfourT         & \VRfourV         & \VRfourVlll\\[-0.05cm]
\noalign{\smallskip}\hline\noalign{\smallskip}
Observed events                       & $149$           & $163$           & $86$            & $168$            & $25$                    \\
\noalign{\smallskip}\hline\noalign{\smallskip}
Total (post-fit) SM events            & $149 \pm 12$~~    & $162 \pm 13$~~    & $86 \pm 20$     & $173 \pm 14$~~     & $27 \pm 5$~~              \\
\noalign{\smallskip}\hline\noalign{\smallskip}
Post-fit, $\ttbar$                    & $115 \pm 13$~~    & $39 \pm 13$     & $41 \pm 19$     & $57 \pm 14$      & --              \\
Post-fit, diboson                     & $0.7 \pm 0.5$   & $89 \pm 18$     & $1.5 \pm 0.6$   & $75 \pm 18$      & $19 \pm 6$~~              \\
$Wt$                                  & $27 \pm 4$~~      & $11.9 \pm 1.8$~~  & $18 \pm 5$~~      & $10.3 \pm 0.8$~~   & --              \\
\zjets                     & $0.18 \pm 0.07$ & $2.1 \pm 1.1$   & $2.1 \pm 0.5$   & $0.81 \pm 0.35$  & --              \\
\ttZ                                  & $1.32 \pm 0.34$ & $0.18 \pm 0.09$ & $0.52 \pm 0.17$ & $0.41 \pm 0.16$  & $0.120 \pm 0.029$              \\
Others                                & $2.41 \pm 0.17$ & $0.30 \pm 0.26$ & $1.34 \pm 0.20$ & $1.2 \pm 0.2$    & $0.095 \pm 0.028$              \\
Fake and non-prompt                   & $2.3 \pm 2.1$   & $20 \pm 4$~~      & $20.7 \pm 3.4$~~  & $28 \pm 5$~~       & $7.9 \pm 1.1$              \\
\noalign{\smallskip}\hline\noalign{\smallskip}
\end{tabular}
\end{center}
\end{table}

\begin{figure}[!htb]
\begin{center}
\subfigure[]{
\includegraphics[width=0.48\textwidth]{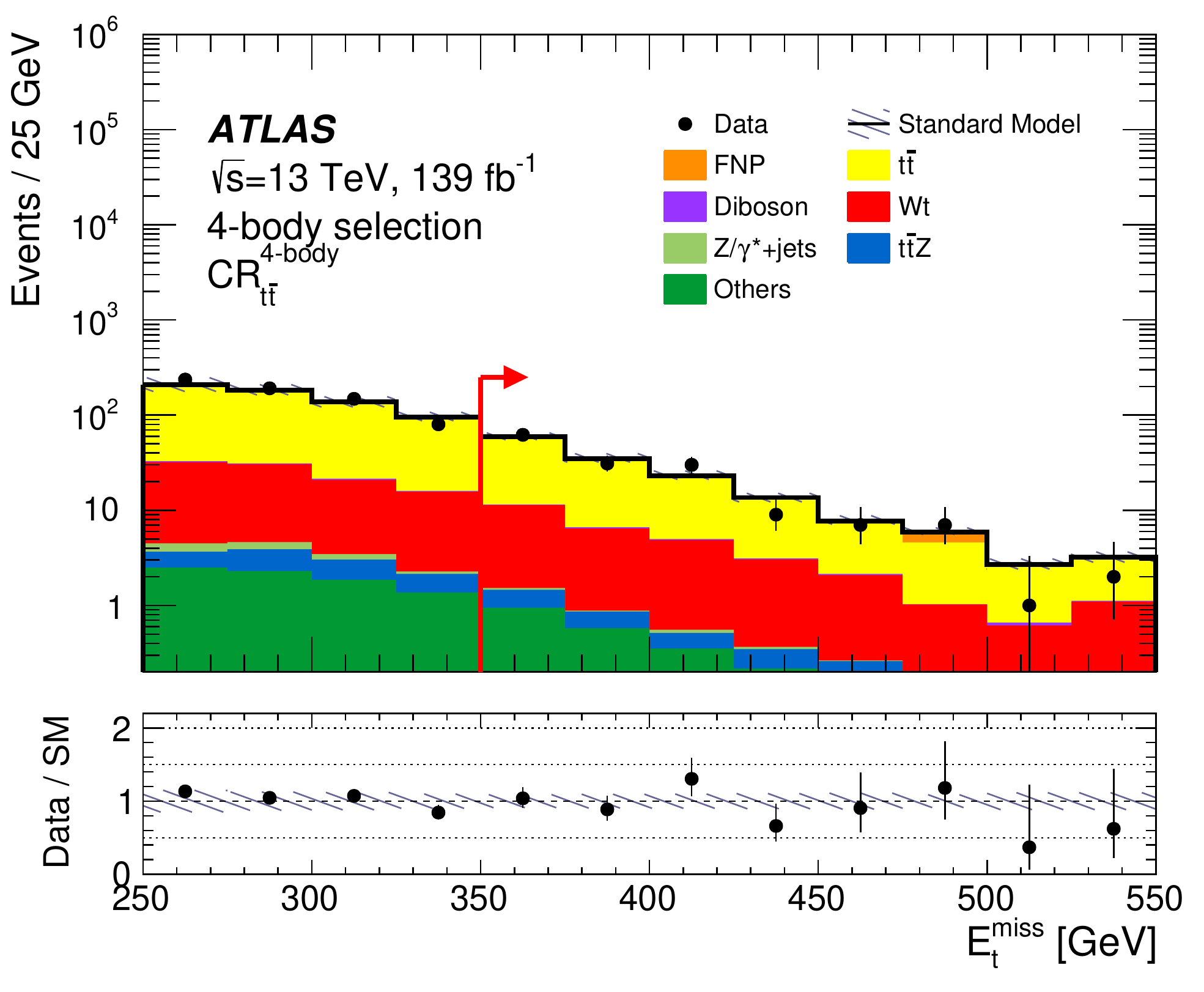}
}
\subfigure[]{
\includegraphics[width=0.48\textwidth]{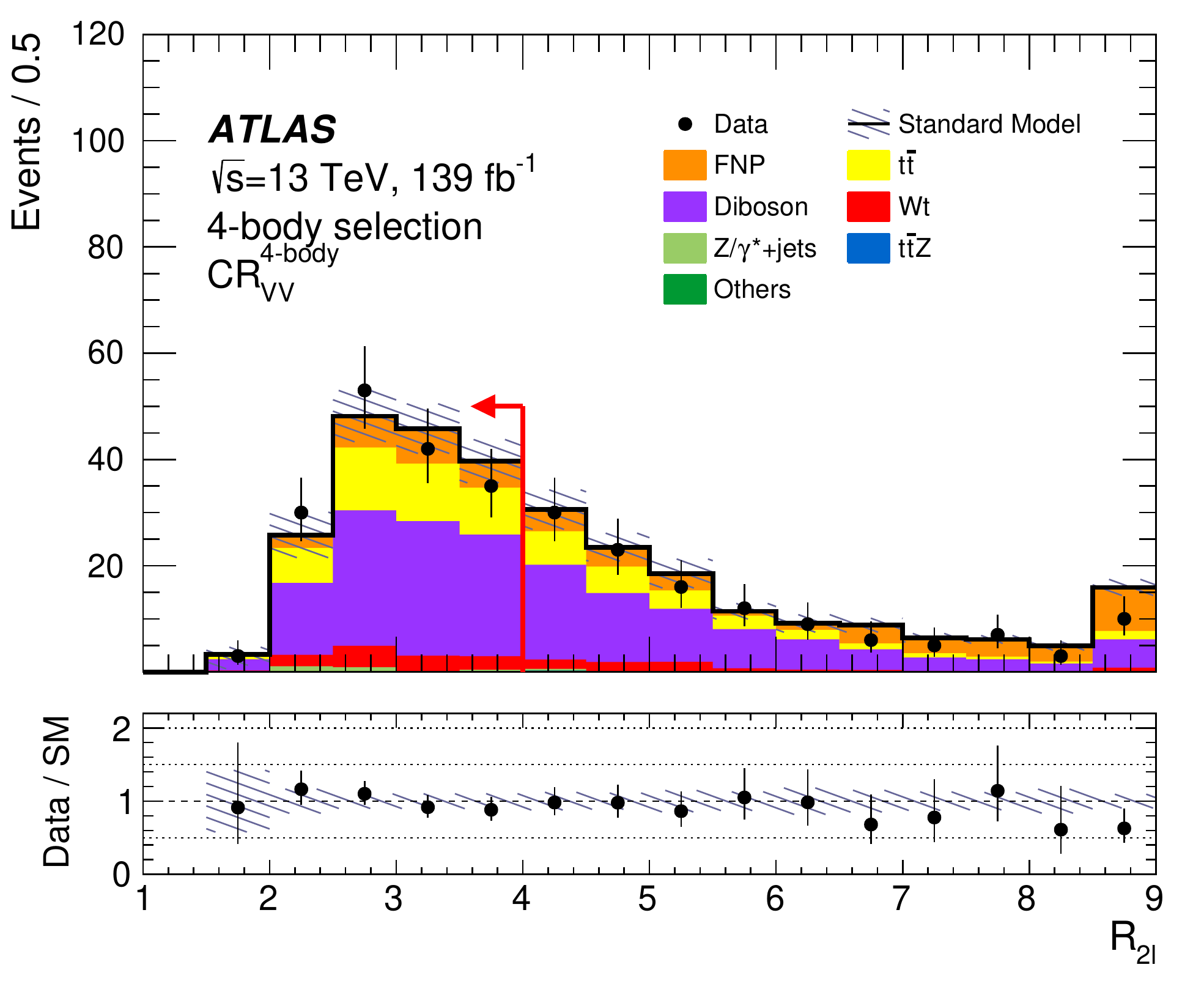}
}
\end{center}
\caption{Four-body selection. Distributions of (a) \MET in \CRfourT\ and (b)  \Rll  in \CRfourV\ after the background fit. The contributions from all SM backgrounds are shown as a histogram stack. ``Others'' includes contributions from $VVV$, $\ttbar t$, $\ttbar\ttbar$, $\ttbar W$, $\ttbar WW$, $\ttbar WZ$, $\ttbar H$, and $tZ$ processes.  The hatched bands represent the total statistical and detector-related systematic uncertainty. The rightmost bin of each plot includes overflow events. In the upper panels, red arrows indicate the control region selection criteria. The bottom panels show the ratio of the observed data to the total SM background prediction, with hatched bands representing the total uncertainty in the background prediction.}
\label{fig:fourbodykinCRPlots}
\end{figure}
 
\begin{figure}[!htb]
\begin{center}
\subfigure[]{
\includegraphics[width=0.48\textwidth]{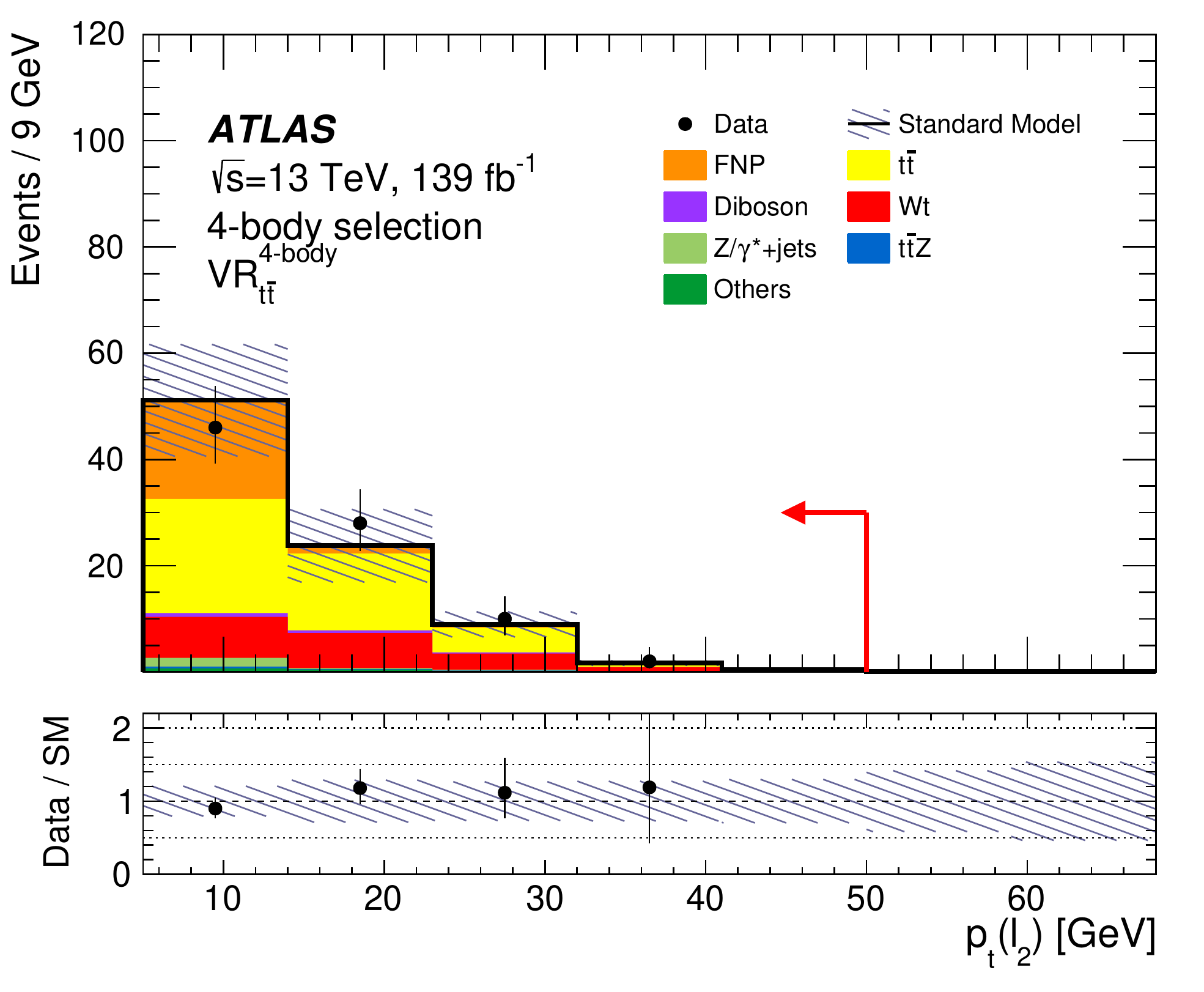}
}
\subfigure[]{
\includegraphics[width=0.48\textwidth]{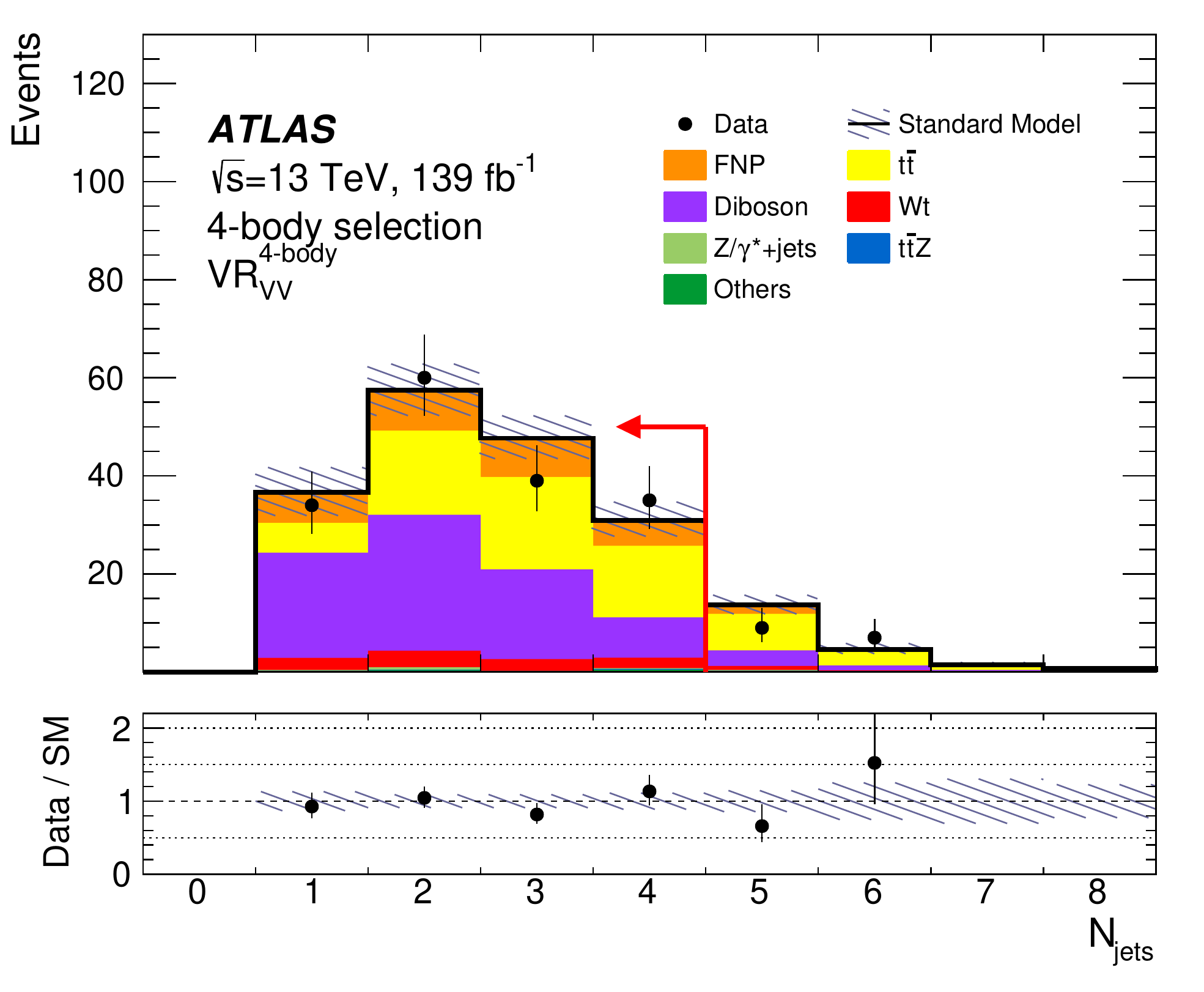}
}
\subfigure[]{
\includegraphics[width=0.48\textwidth]{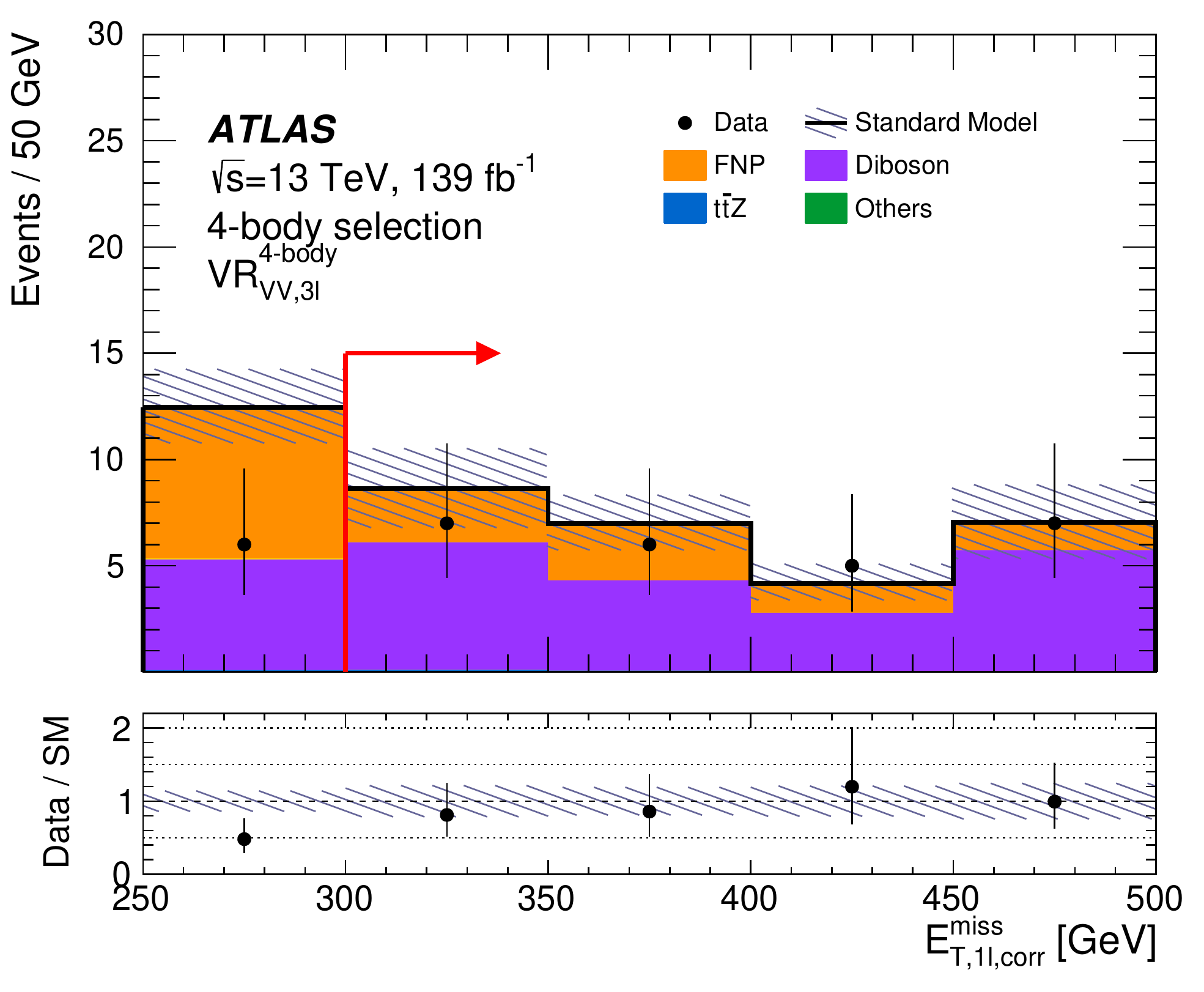}
}
\end{center}
\caption{Four-body selection. Distributions of (a) \ptlepsubl in \VRfourT, (b) \njet in \VRfourV\ and (c) $E_{\mathrm{T},1\ell,\mathrm{corr}}^{\mathrm{miss}}$ in \VRfourVlll\ after the background fit. The contributions from all SM backgrounds are shown as a histogram stack. ``Others'' includes contributions from $VVV$, $\ttbar t$, $\ttbar\ttbar$, $\ttbar W$, $\ttbar WW$, $\ttbar WZ$, $\ttbar H$, and $tZ$ processes.  The hatched bands represent the total statistical and detector-related systematic uncertainty. The rightmost bin of each plot includes overflow events.
In the upper panels, red arrows indicate the validation region selection criteria. The bottom panels show the ratio of the observed data to the total SM background prediction, with hatched bands representing the total uncertainty in the background prediction.}
\label{fig:fourbodykinVRPlots}
\end{figure}
 
\FloatBarrier

\section{Systematic uncertainties}
\label{sec:systs}
 
Systematic uncertainties are evaluated for the signal and for the background predictions. The main experimental uncertainties in the yields of the reconstructed objects, the theoretical uncertainties in the processes' yields, and  the uncertainties related to the MC modelling of the SM backgrounds are described in this section.  The statistical uncertainties in the simulated event samples are also taken into account.
 
The main sources of experimental uncertainty are related to the jet energy scale (JES) and the jet energy resolution (JER).
The JES and JER uncertainties are derived as a function of the \pT and $\eta$ of the jet, as well as of the pile-up conditions and the jet-flavour composition of the selected jet sample~\cite{ATL-PHYS-PUB-2015-015}. Uncertainties associated with the modelling of the $b$-tagging efficiencies for $b$-jets, $c$-jets and light-flavour jets~\cite{ATLAS-CONF-2014-004,ATLAS-CONF-2014-046} are also considered.
The systematic uncertainties related to the modelling of \MET in the simulation are estimated by propagating the uncertainties in the energy and momentum scales of electrons, muons and jets, as well as the uncertainties in the resolution and scale of the soft term~\cite{ATL-PHYS-PUB-2015-023}.
Other detector-related systematic uncertainties, including those arising from lepton reconstruction efficiency, energy scale, energy resolution and in the modelling of the trigger efficiency~\cite{PERF-2015-10,EGAM-2018-01,TRIG-2016-01,TRIG-2018-05,2020arXiv200413447A}, or the ones due to the pile-up reweighting and JVT are found to have a small impact on the results. 
 
Systematic uncertainties in the theoretical modelling of the observed final states can be broadly divided  into uncertainties in the description of the parton-level final states (uncertainties in the proton PDF, cross-section, and strong coupling constant) and further uncertainties arising from the parton showering and hadronisation processes that convert partons into the hadronic final states.
The uncertainties in the modelling of the $\ttbar$ background are estimated by varying the renormalisation and factorisation scales, as well as the amount of initial- and final-state radiation produced when generating the samples~\cite{ATL-PHYS-PUB-2016-004, ATL-PHYS-PUB-2018-009}.
Comparison between the yields obtained with \POWHEG and \MGMCatNLO~\cite{ATL-PHYS-PUB-2016-004} is used to estimate uncertainties from the  event generator choice.
For $\ttZ$ production, in the two-body and three-body selections, the effects of QCD scale uncertainties are evaluated using seven-point variations of the factorisation and renormalisation scales~\cite{pubnote_mc_ttv}.
Uncertainties for additional radiation contributions (ISR, FSR) are evaluated by comparing the nominal sample with one obtained with a \PYTHIA tune enhancing the radiation~\cite{Sjostrand:2014zea}.
In the four-body selection, since the $\ttZ$ background contribution is minor, a total theoretical error of 14\%, coming from the cross-section uncertainty~\cite{TOPQ-2016-11}, is applied instead.
For \ttbar and \ttZ production, the parton showering and hadronisation uncertainties are covered by the  difference between samples obtained using the two different showering models implemented in \PYTHIA and in \HERWIG.
Single top quark production via the $Wt$-channel is a minor background in all the selections. An uncertainty in the acceptance due to the interference between \ttbar and  $Wt$ production is assigned by comparing dedicated  samples produced with \POWHEG and \PYTHIA using the diagram removal (DR) and the diagram subtraction (DS) approaches~\cite{Frixione:2008yi}.
The modelling uncertainties for the  diboson background are estimated using the seven-point variations of the renormalisation and factorisation scales.
Additional uncertainties  in the resummation (QSF) and matching (CKKM)   scales between the matrix element generator and  parton shower are computed by varying the scale  parameters in \SHERPA~\cite{ATL-PHYS-PUB-2017-005}.
For the other background processes which make minor contributions a conservative uncertainty is applied.
These minor backgrounds are mainly $\ttbar WZ$ and $ttW$ processes. A 30\% uncertainty, driven by the DR versus DS difference for the $\ttbar WZ$~\cite{Bylund:2016qau} process, is applied in the two-body and three-body selections. For the four-body selection a 22\% uncertainty is applied for the uncertainty in the $\ttbar W$ cross-section~\cite{TOPQ-2016-11}.
For all the processes mentioned above the PDF uncertainties~\cite{Butterworth:2015oua} were evaluated and found to be negligible.
 
Systematic uncertainties in the data-driven FNP background estimate are expected due to potential differences in the FNP composition (heavy flavour, light flavour or photon conversions) between the  regions defined in Section~\ref{sec:backgrounds} and the \CRFNP\ used to extract the fake factor. A FNP systematic error is evaluated in each of the regions by varying the FNP composition in the \CRFNP\ to match that of the considered analysis region. The statistical error is also included by propagating the statistical uncertainty in the ratio used to compute the fake factor.
For the four-body selection, where the  FNP lepton background is dominant, a FNP closure uncertainty is also evaluated from the full difference between the data and the FNP predictions as observed in a validation region with two same-charge leptons with kinematics similar to the four-body selection. The closure uncertainty ranges between 13\% and 33\% in the regions where the FNP background is important.
 
A 1.7\% uncertainty in the luminosity measurement is considered for all signal and background estimates that are derived directly from MC simulations~\cite{ATLAS-CONF-2019-021}.
 
Tables~\ref{tab:SR2bSF_syst},~\ref{tab:SR2bDF_syst} and~\ref{tab:SR3b4b_syst} summarise the contributions from the different sources of systematic uncertainty in the total SM background predictions for the two-body, three-body and four-body signal regions.
The total systematic uncertainty ranges between 14\% and 26\%, with the dominant sources being the MC statistical error, the JES and JER, the uncertainty in the background normalisation and the theoretical uncertainties.
 
The SUSY signal cross-section uncertainty is evaluated from an envelope of the cross-section predictions using different PDF sets and factorisation and renormalisation scales as described in Ref.~\cite{Borschensky:2014cia}.
The uncertainty in the DM production cross-section is derived from the scale variations and the PDF choices.
The SUSY and DM theory signal uncertainties are computed from the variation of the radiation, renormalisation, factorisation and merging scales.
These uncertainties are most relevant for the four-body selection, where the largest theory uncertainties are those resulting from radiation and are in the range $10\%$ to $24\%$ depending on the mass difference $m(\stopone)-m(\ninoone)$.
For the DM signals the total systematic uncertainty is between $5\%$ and $20\%$.

\begin{table}[h!]
\caption{Two-body selection. Sources of systematic uncertainty in the SM background estimates, after the background fits, for the SF selection.  The values are given as relative uncertainties in the total expected background event yields in the SRs. Entries marked `--' indicate  a contribution smaller than 1\%.  `MC statistical uncertainty' refers to the statistical uncertainty from the simulated event samples. `Other theoretical uncertainties' represent the theoretical uncertainty coming from $VVV$, $\ttbar t$, $\ttbar\ttbar$, $\ttbar W$, $\ttbar WW$, $\ttbar WZ$, $\ttbar H$, and $tZ$ contributions. The individual components can be correlated and therefore do not necessarily add up in quadrature to the total systematic uncertainty.
\label{tab:SR2bSF_syst}}
\begin{center}
\footnotesize
\setlength{\tabcolsep}{0.2pc}
\begin{tabular}{l|cccccc}
\noalign{\smallskip}\hline\noalign{\smallskip}
Signal Region                     & \SRSFzero & \SRSFone & \SRSFtwo & \SRSFthree & \SRSFfour & \SRSFfive              \\
\noalign{\smallskip}\hline\noalign{\smallskip}
Total SM background uncertainty   & $19\%$    & $20\%$   & $17\%$   & $15\%$     & $15\%$    & $20\%$             \\
\noalign{\smallskip}\hline\noalign{\smallskip}
$VV$~theoretical uncertainties      & $-$       & $2.4\%$  & $3.5\%$  & $4.9\%$    & $4.4\%$   & $7.1\%$       \\
\ttbar ~theoretical uncertainties & $10\%$    & $11\%$   & $6.2\%$  & $-$        & $1.7\%$   & $2.7\%$       \\
\ttZ~theoretical uncertainties    & $1.0\%$   & $2.2\%$  & $4.2\%$  & $5.2\%$    & $5.0\%$   & $11\%$       \\
\ttbar--\(Wt\)~interference        & $-$       & $-$      & $-$      & $-$        & $1.0\%$   & $5.7\%$       \\
Other theoretical uncertainties   & $1.0\%$   & $1.4\%$  & $2.7\%$  & $2.5\%$    & $2.6\%$   & $1.9\%$       \\
\noalign{\smallskip}\hline\noalign{\smallskip}
MC  statistical uncertainty       & $5.1\%$   & $5.4\%$  & $7.0\%$  & $7.7\%$    & $9.9\%$   & $8.7\%$       \\
\noalign{\smallskip}\hline\noalign{\smallskip}
\ttbar~normalisation              & $7.6\%$   & $4.8\%$  & $1.0\%$  & $-$        & $-$       & $-$       \\
\ttZ~normalisation               & $1.1\%$   & $3.2\%$  & $5.6\%$  & $7.2\%$    & $6.4\%$   & $4.8\%$       \\
\noalign{\smallskip}\hline\noalign{\smallskip}
Jet energy scale                  & $11\%$    & $6.7\%$  & $9.6\%$  & $2.0\%$    & $3.4\%$   & $2.0\%$       \\
Jet energy resolution             & $3.6\%$   & $13\%$   & $7.0\%$  & $6.1\%$    & $3.6\%$   & $7.7\%$       \\
\MET modelling                    & $2.9\%$   & $3.6\%$  & $1.0\%$  & $4.1\%$    & $2.7\%$   & $1.2\%$       \\
Lepton modelling                  & $3.6\%$   & $1.8\%$  & $1.8\%$  & $3.8\%$    & $3.7\%$   & $6.4\%$       \\
Flavour tagging                    & $1.0\%$   & $1.0\%$  & $1.0\%$  & $2.6\%$    & $3.0\%$   & $2.4\%$       \\
Pile-up reweighting and JVT       & $-$       & $1.4\%$  & $1.0\%$  & $1.0\%$    & $1.7\%$   & $-$       \\
Fake and non-prompt leptons       & $-$       & $-$      & $1.1\%$  & $-$        & $2.8\%$   & $4.3\%$       \\
\noalign{\smallskip}\hline\noalign{\smallskip}
\end{tabular}
\end{center}
\end{table}
 
\begin{table}[h!]
\caption{Two-body selection. Sources of systematic uncertainty in the SM background estimates, after the background fits, for the DF selection. The values are given as relative uncertainties in the total expected background event yields in the SRs. Entries marked `--' indicate a contribution smaller than 1\%. `MC statistical uncertainty' refers to the statistical uncertainty from the simulated event samples. `Other theoretical uncertainties' represent the theoretical uncertainty coming from $VVV$, $\ttbar t$, $\ttbar\ttbar$, $\ttbar W$, $\ttbar WW$, $\ttbar WZ$, $\ttbar H$, and $tZ$ contributions. The individual components can be correlated and therefore do not necessarily add up in quadrature to the total systematic uncertainty.
\label{tab:SR2bDF_syst}}
\begin{center}
\footnotesize
\setlength{\tabcolsep}{0.2pc}
\begin{tabular}{l|cccccc}
\noalign{\smallskip}\hline\noalign{\smallskip}
Signal Region                     & \SRDFzero & \SRDFone & \SRDFtwo & \SRDFthree & \SRDFfour & \SRDFfive              \\
\noalign{\smallskip}\hline\noalign{\smallskip}
Total SM background uncertainty   & $20\%$    & $20\%$   & $15\%$   & $16\%$     & $14\%$    & $21\%$             \\
\noalign{\smallskip}\hline\noalign{\smallskip}
$VV$ theoretical uncertainties      & $1.0\%$   & $1.3\%$  & $2.6\%$  & $1.0\%$    & $2.0\%$   & $1.8\%$       \\
\ttbar ~theoretical uncertainties & $9.6\%$   & $12\%$   & $7.6\%$  & $-$        & $3.1\%$   & $-$       \\
\ttZ ~theoretical uncertainties   & $1.2\%$   & $2.0\%$  & $5.3\%$  & $6.6\%$    & $5.7\%$   & $16\%$       \\
\ttbar--\(Wt\)~interference       & $-$       & $-$      & $-$      & $-$        & $-$       & $-$       \\
Other theoretical uncertainties   & $1.0\%$   & $1.2\%$  & $2.8\%$  & $3.2\%$    & $2.7\%$   & $3.3\%$       \\
\noalign{\smallskip}\hline\noalign{\smallskip}
MC statistical uncertainty        & $4.7\%$   & $5.0\%$  & $6.9\%$  & $8.2\%$    & $7.7\%$   & $6.6\%$       \\
\noalign{\smallskip}\hline\noalign{\smallskip}
\ttbar  ~normalisation            & $7.2\%$   & $5.6\%$  & $1.2\%$  & $-$        & $-$       & $-$       \\
\ttZ  ~normalisation              & $1.4\%$   & $2.8\%$  & $6.9\%$  & $9.1\%$    & $7.3\%$   & $7.2\%$       \\
\noalign{\smallskip}\hline\noalign{\smallskip}
Jet energy scale                  & $8.5\%$   & $10\%$   & $2.5\%$  & $6.1\%$    & $1.0\%$   & $2.6\%$       \\
Jet energy resolution             & $13\%$    & $6.6\%$  & $6.2\%$  & $4.3\%$    & $5.3\%$   & $2.0\%$       \\
\MET modelling                    & $3.5\%$   & $6.1\%$  & $1.0\%$  & $2.2\%$    & $2.2\%$   & $1.0\%$       \\
Lepton modelling                  & $1.5\%$   & $1.1\%$  & $1.6\%$  & $1.3\%$    & $1.3\%$   & $1.0\%$       \\
Flavour tagging                    & $1.0\%$   & $1.0\%$  & $1.3\%$  & $2.0\%$    & $1.0\%$   & $1.0\%$       \\
Pile-up reweighting and JVT       & $-$       & $1.6\%$  & $1.0\%$  & $-$        & $1.0\%$   & $-$       \\
Fake and non-prompt leptons       & $-$       & $3.5\%$  & $-$      & $-$        & $7.1\%$   & $13\%$       \\
\noalign{\smallskip}\hline\noalign{\smallskip}
\end{tabular}
\end{center}
\end{table}
 
\begin{table}[h!]
\caption{Three-body and four-body selections. Sources of systematic uncertainty in the SM background estimates, after the background fits. The values are given as relative uncertainties in the total expected background event yields in the SRs. Entries marked `--' indicate  a  contribution  smaller than 1\%.  `MC statistical uncertainty' refers to the statistical uncertainty from the simulated event samples. `Other theoretical uncertainties' represent the theoretical uncertainty coming from $VVV$, $\ttbar t$, $\ttbar\ttbar$, $\ttbar W$, $\ttbar WW$, $\ttbar WZ$, $\ttbar H$, and $tZ$ contributions. The individual components can be correlated and therefore do not necessarily add up in quadrature to the total systematic uncertainty.
\label{tab:SR3b4b_syst}}
\begin{center}
\footnotesize
\setlength{\tabcolsep}{0.2pc}
\begin{tabular}{l|cccc|cc}
\noalign{\smallskip}\hline\noalign{\smallskip}
Signal Region                     & \SRTBWDF & \SRTBWSF & \SRTBTDF & \SRTBTSF & \SRFBA  &\SRFBB \\
\noalign{\smallskip}\hline\noalign{\smallskip}
Total SM background uncertainty   & $18\%$   & $26\%$   & $18\%$   & $22\%$   & $25\%$  & $14\%$             \\
\noalign{\smallskip}\hline\noalign{\smallskip}
$VV$ theoretical uncertainties      & $8.0\%$  & $10\%$   & $1.0\%$  & $1.5\%$  & $3.6\%$ & $4.9\%$       \\
\ttbar ~theoretical uncertainties & $8.2\%$  & $6.6\%$  & $14\%$   & $8.6\%$  & $1.0\%$ & $6.3\%$       \\
\ttZ ~theoretical uncertainties   & $-$      & $-$      & $1.2\% $ & $2.0\%$  & $-$     & $-$       \\
\ttbar--\(Wt\) ~interference       & $-$      & $1.0\%$  & $-$      & $1.1\%$  & $-$     & $2.4\%$       \\
Other theoretical uncertainties   & $-$      & $-$      & $1.4\%$  & $1.6\%$  & $-$     & $-$       \\
\noalign{\smallskip}\hline\noalign{\smallskip}
MC statistical uncertainty        & $5.8\%$  & $7.4\%$  & $5.6\%$  & $6.7\%$  & $3.3\%$ & $2.7\%$       \\
\noalign{\smallskip}\hline\noalign{\smallskip}
$VV$  normalisation                 & $15\%$   & $20\%$   & $1.0\%$  & $2.0\%$  & $2.8\%$ & $8.6\%$       \\
\ttbar  ~normalisation            & $2.3\%$  & $1.9\%$  & $4.9\%$  & $3.3\%$  & $1.0\%$ & $6.1\%$       \\
\ttZ  ~normalisation              & $-$      & $-$      & $4.1\%$  & $4.5\%$  & $-$     & $-$       \\%
\noalign{\smallskip}\hline\noalign{\smallskip}
Jet energy scale                  & $5.5\%$  & $3.7\%$  & $3.8\%$  & $4.1\%$  & $1.0\%$ & $3.2\%$       \\
Jet energy resolution             & $2.3\%$  & $11\%$   & $9.0\%$  & $18\%$   & $1.3\%$ & $3.5\%$       \\
Lepton modelling                  & $1.3\%$  & $2.0\%$  & $1.0\%$  & $2.5\%$  & $1.3\%$ & $3.3\%$       \\
\MET modelling                    & $1.1\%$  & $2.2\%$  & $3.0\%$  & $1.8\%$  & $-$     & $1.0\%$       \\
Flavour tagging                    & $3.1\%$  & $2.9\%$  & $1.6\%$  & $1.0\%$  & $-$     & $1.3\%$       \\
Pile-up reweighting  and JVT      & $1.0\%$  & $1.0\%$  & $-$      & $-$      & $1.0\%$ & $-$       \\
Fake and non-prompt leptons       & $1.7\%$  & $-$      & $-$      & $4.6\%$  & $25\%$  & $-$       \\
\noalign{\smallskip}\hline\noalign{\smallskip}
\end{tabular}
\end{center}
\end{table}
\FloatBarrier
 
\section{Results}
\label{sec:results}
A set of simultaneous likelihood fits is performed, for each one of the three different selections, using standard minimisation software packages, HistFitter and pyhf~\cite{HistFitter, pyhf}. For the normalisation of the semi-data-driven backgrounds, only the CRs are considered in the background fit, while for the computation of the exclusion limits both the CRs and SRs are included as constraining channels. The likelihood is a product of Poisson probability density functions (pdf), describing the observed number of events in each CR/SR, and Gaussian pdf distributions   that describe  the nuisance parameters associated with  all the systematic uncertainties.  Systematic uncertainties that are correlated between different samples are accounted for in the fit configuration by using the same nuisance parameter. The uncertainties are applied in each of the CRs and SRs and their effect is correlated for events across all regions in the fit. 
 
The results of the background fit are shown in Figures~\ref{fig:pull_twob}--\ref{fig:pull_fourb} for each of the three analysis selections.
In general, good agreement, within about one standard deviation, is observed in all the SRs and VRs except in \SRTBWDF where the data fluctuates well below the fit.
 
\begin{figure}[htb!]
\begin{center}
\includegraphics[width=0.85\textwidth]{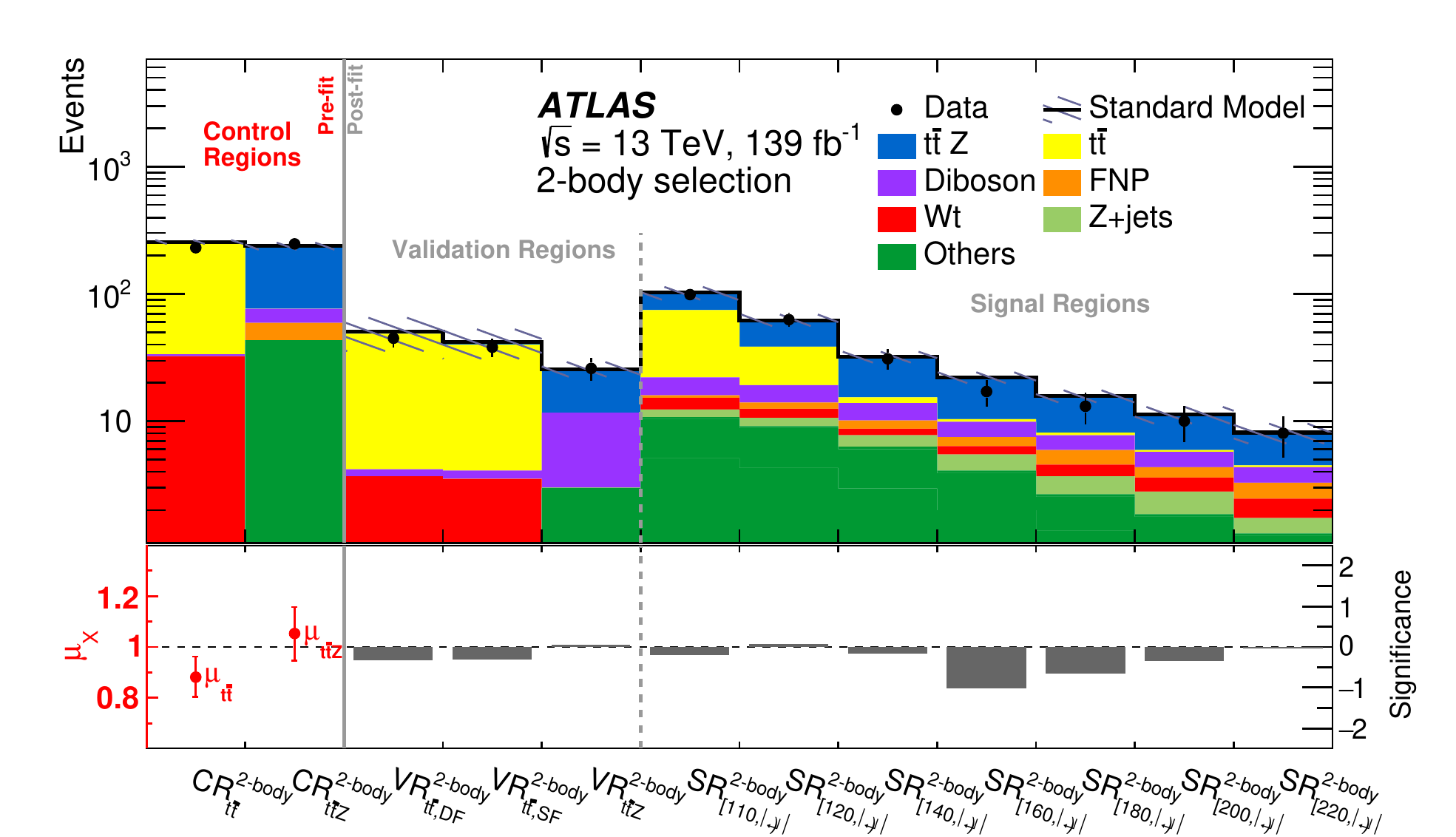}
\caption{Two-body selection. Expected and observed yields are shown. The upper panel shows the observed number of events in each of the  CRs, VRs and the inclusive SRs defined in the two-body selection, together with the expected SM backgrounds obtained  before the fit in the CRs and after the fit in the VRs and SRs.  ``Others'' includes contributions from $VVV$, $\ttbar t$, $\ttbar\ttbar$, $\ttbar W$, $\ttbar WW$, $\ttbar WZ$, $\ttbar H$, and $tZ$ processes. The shaded band represents the total uncertainty in the expected SM background. The lower panel shows the normalisation factors $\mu_X$ (left two bins) extracted in the CRs for the \ttbar and \ttZ processes, while, for the VRs and the inclusive SRs (right bins), the significance as defined in Ref.~\cite{Cousin:2008clt}. }
\label{fig:pull_twob}
\end{center}
\end{figure}
\begin{figure}[htb!]
\begin{center}
\includegraphics[width=0.85\textwidth]{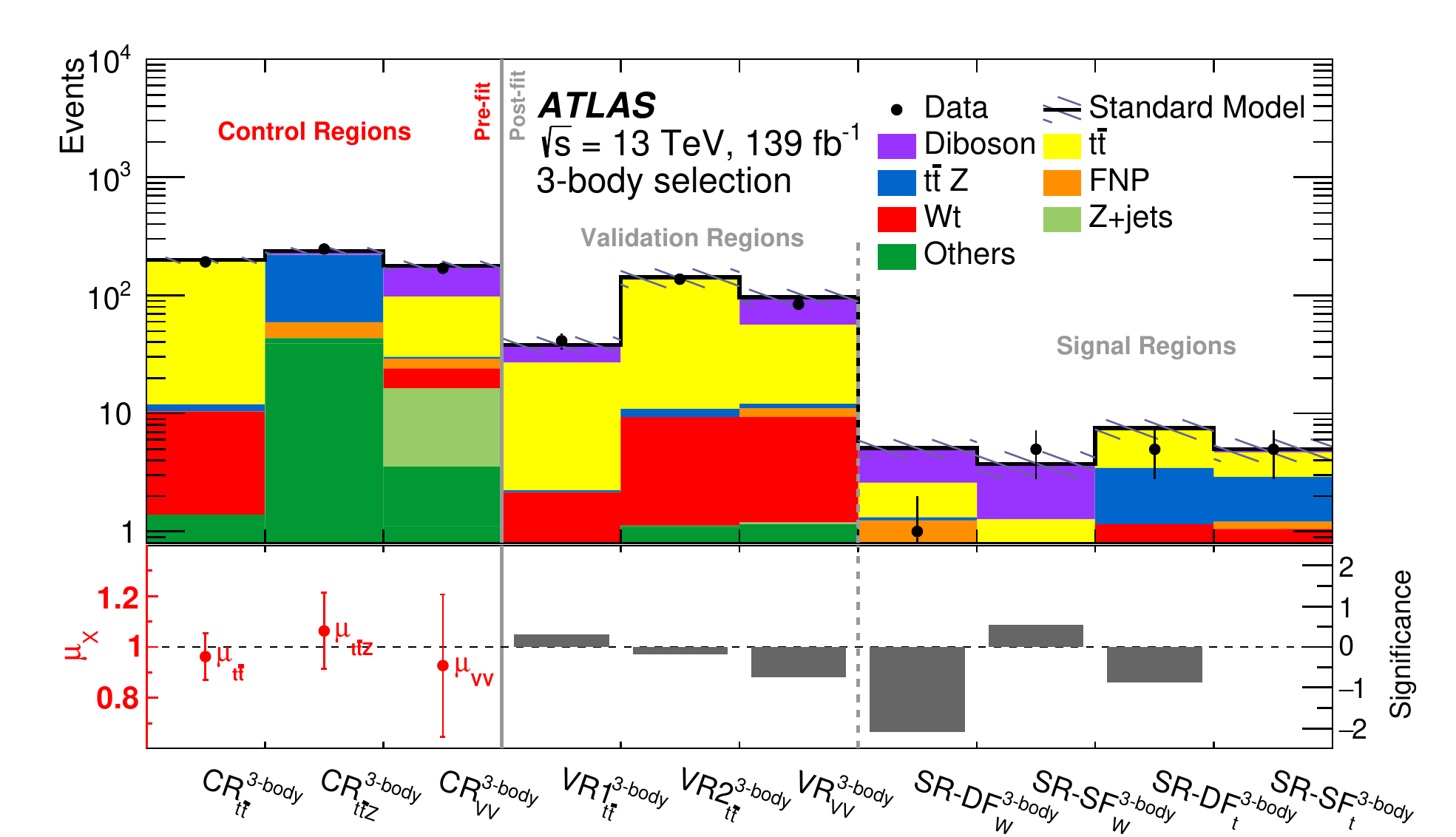}
\caption{Three-body selection. Expected and observed yields are shown. The upper panel shows the observed number of events in each of the CRs,VRs and SRs defined in the three-body selection, together with the expected SM backgrounds obtained  before the fit in the CRs and after the fit in the VRs and SRs. ``Others'' includes contributions from $VVV$, $\ttbar t$, $\ttbar\ttbar$, $\ttbar W$, $\ttbar WW$, $\ttbar WZ$, $\ttbar H$, and $tZ$ processes. The shaded band represents the total uncertainty in the expected SM background. The lower panel shows the normalisation factors $\mu_X$ (left three bins) extracted in the CRs for the \ttbar, \ttZ and diboson processes, while, for the VRs and the SRs (right bins), the significance as defined in Ref.~\cite{Cousin:2008clt}.  }
\label{fig:pull_threeb}
\end{center}
\end{figure}
 
\begin{figure}[htb!]
\begin{center}
\includegraphics[width=0.85\textwidth]{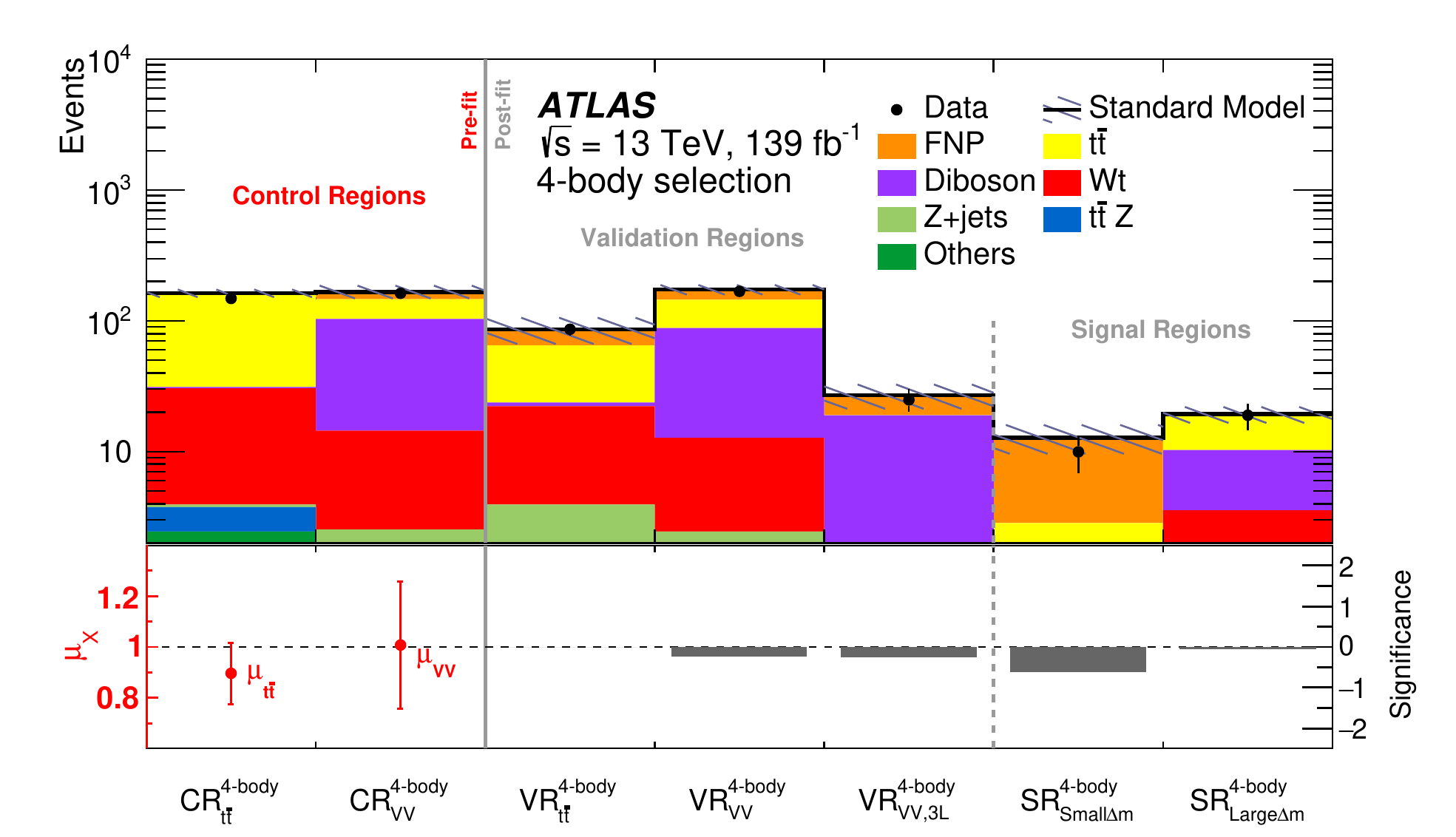}
\caption{Four-body selection. Expected and observed yields are shown. The upper panel shows the observed number of events in each of the  CRs, VRs and SRs defined in the four-body selection, together with the expected SM backgrounds obtained  before the fit in the CRs and after the fit in the VRs and SRs. ``Others'' includes contributions from $VVV$, $\ttbar t$, $\ttbar\ttbar$, $\ttbar W$, $\ttbar WW$, $\ttbar WZ$, $\ttbar H$, and $tZ$ processes. The shaded band represents the total uncertainty in the expected SM background. The lower panel shows the normalisation factors $\mu_X$ (left two bins) extracted in the CRs for the \ttbar and diboson processes, while, for the VRs and the SRs (right bins), the significance as defined in Ref.~\cite{Cousin:2008clt}.  }
\label{fig:pull_fourb}
\end{center}
\end{figure}

\FloatBarrier

\subsection{Two-body selection results}
\label{sec:twobody_results}
 
The estimated SM yields in the binned and inclusive SRs defined in the two-body selection are obtained  with a background fit  which simultaneously determines  the normalisations of the background contributions from \ttbar and \ttZ. Figure \ref{fig:twobody_SR} shows the \mttwoL\  
distribution  for events satisfying  all the selection criteria of the  $\SRtwoB_{110,\infty}$  (SF and DF) signal regions,
after the background fit. Each bin corresponds to one of the binned SRs. No significant excess over the SM prediction is observed, as can be seen from results shown in Tables~\ref{tab:twobody_SR_DF} and \ref{tab:twobody_SR_SF} for the binned SRs.

\begin{figure}[h!]
\begin{center}
\subfigure[]{
\includegraphics[width=0.48\textwidth]{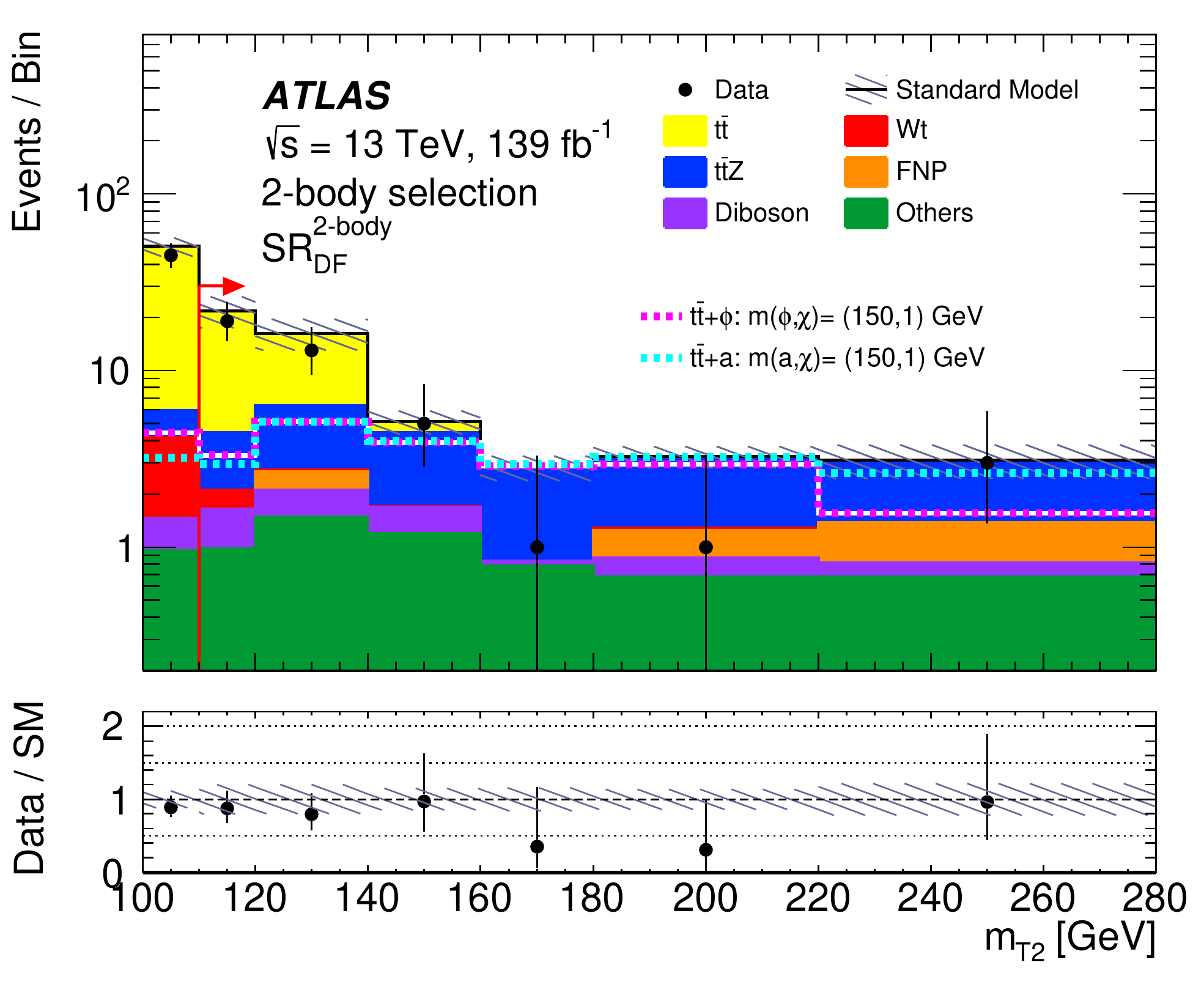}
}
\subfigure[]{
\includegraphics[width=0.48\textwidth]{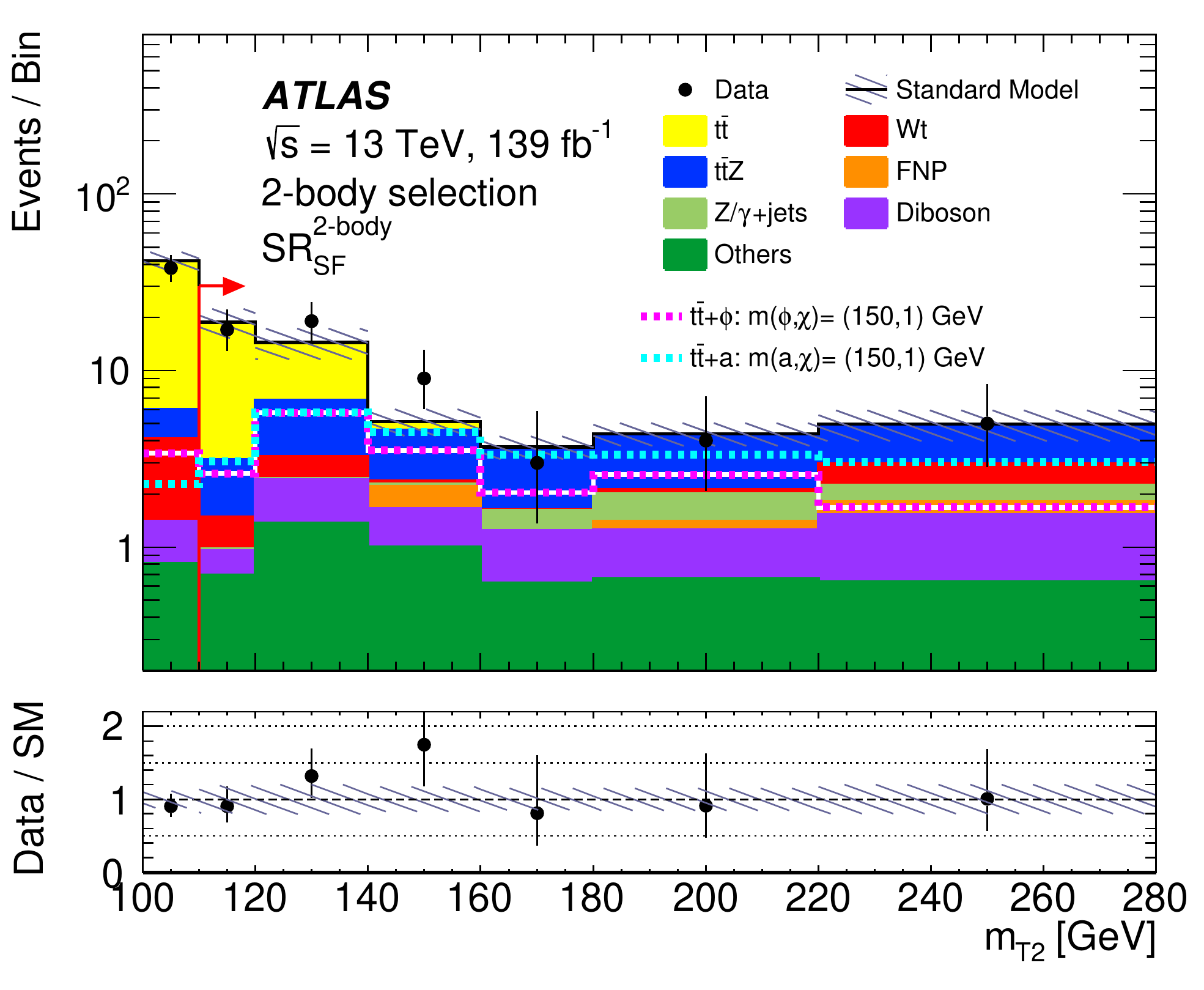}
}
\end{center}
\caption{Two-body selection. Distributions of \mttwoL\ in $\SRtwoB_{110,\infty}$ for (a) different-flavour and (b) same-flavour events satisfying the selection criteria of the given SR, except the one for the presented variable, after the background fit. The contributions from all SM backgrounds are shown as a histogram stack. ``Others'' includes contributions from $VVV$, $\ttbar t$, $\ttbar\ttbar$, $\ttbar W$, $\ttbar WW$, $\ttbar WZ$, $\ttbar H$, and $tZ$ processes. The hatched bands represent the total statistical and systematic uncertainty. The rightmost bin of each plot includes overflow events. Reference dark-matter signal models are overlayed  for comparison. Red arrows in the upper panels indicate the signal region selection criteria. The bottom panels show the ratio of the observed data to the total SM background prediction, with hatched bands representing the total uncertainty in the background prediction.}
\label{fig:twobody_SR}
\end{figure}
 
\begin{table}[h!]
\begin{center}
\caption{Two-body selection. Background fit results for the different-flavour leptons binned SRs. The `Others' category contains the contributions from $VVV$, $\ttbar t$, $\ttbar\ttbar$, $\ttbar W$, $\ttbar WW$, $\ttbar WZ$, $\ttbar H$, and $tZ$. Combined statistical and systematic uncertainties are given. Entries marked `--' indicate a negligible background contribution (less than 0.001 events). The individual uncertainties can be correlated, and do not necessarily add up in quadrature to the total background uncertainty.}
\label{tab:twobody_SR_DF}
\begin{adjustbox}{width=1.\textwidth}
\begin{tabular}{l|cccccc}
\noalign{\smallskip}\hline\noalign{\smallskip}
& \SRDFzero           & \SRDFone               & \SRDFtwo             & \SRDFthree             & \SRDFfour         & \SRDFfive               \\[-0.05cm]
\noalign{\smallskip}\hline\noalign{\smallskip}
Observed events          & $19$                & $13$                   & $5$                  & $1$                    & $1$               & $3$                    \\
\noalign{\smallskip}\hline\noalign{\smallskip}
 
Fitted bkg.\ events        & $22 \pm 4$~~          & $16.3 \pm 3.2$~~         & $5.1 \pm 0.8$        & $2.83 \pm 0.45$        & $3.25 \pm 0.45$   & $3.11 \pm 0.67$              \\
\noalign{\smallskip}\hline\noalign{\smallskip}
Post-fit, $t \bar t$     & $17 \pm 4$~~          & $10.0 \pm 3.2$~~         & $0.7 \pm 0.5$        & $0.01_{-0.01}^{+0.10}$~\, & $0.13 \pm 0.11$   & --              \\
Post-fit, $t \bar t +Z$  & $2.3 \pm 0.5$       & $3.5 \pm 0.7$          & $2.7 \pm 0.7$        & $2.0 \pm 0.4$          & $1.9 \pm 0.4$     & $1.7 \pm 0.6$              \\
$Wt$                     & $0.47 \pm 0.27$     & $0.05_{-0.05}^{+0.33}$~\, & $0.025 \pm 0.012$    & --                     & $0.033 \pm 0.013$ & --              \\
$Z/\gamma^*$ + jets      & --                  & --                     & --                   & --                     & --                & --              \\
Diboson                  & $0.67 \pm 0.27$     & $0.61 \pm 0.24$        & $0.49 \pm 0.16$      & $0.05_{-0.05}^{+0.07}$~\, & $0.19 \pm 0.13$   & $0.14 \pm 0.07$              \\
Others                   & $0.97 \pm 0.19$     & $1.48 \pm 0.28$        & $1.19 \pm 0.16$      & $0.78 \pm 0.12$        & $0.68 \pm 0.13$   & $0.67 \pm 0.11$              \\
Fake and non-prompt      & $0.0_{-0.0}^{+0.5}~ $ & $0.6 \pm 0.6$          & $0.0_{-0.0}^{+0.5}$~ & $0.0_{-0.0}^{+0.5}$~    & $0.37 \pm 0.23$   & $0.6 \pm 0.4$               \\
\noalign{\smallskip}\hline\noalign{\smallskip}
\end{tabular}
\end{adjustbox}
\end{center}
\end{table}
 
\begin{table}[h!]
\begin{center}
\caption{Two-body selection. Background fit results for the same-flavour leptons binned SRs. The `Others' category contains the contributions from $VVV$, $\ttbar t$, $\ttbar\ttbar$, $\ttbar W$, $\ttbar WW$, $\ttbar WZ$, $\ttbar H$, and $tZ$. Combined statistical and systematic uncertainties are given. The individual uncertainties can be correlated, and do not necessarily add up in quadrature to the total background uncertainty.}
\label{tab:twobody_SR_SF}
\begin{adjustbox}{width=1.\textwidth}
\begin{tabular}{l|cccccc}
\noalign{\smallskip}\hline\noalign{\smallskip}
& \SRSFzero           & \SRSFone            & \SRSFtwo               & \SRSFthree                & \SRSFfour       & \SRSFfive              \\[-0.05cm]
\noalign{\smallskip}\hline\noalign{\smallskip}
Observed events          & $17$                & $19$                & $9$                    & $3$                       & $4$             & $5$                    \\
\noalign{\smallskip}\hline\noalign{\smallskip}
 
Fitted bkg.\ events        & $18.8 \pm 3.5$~~      & $14.4 \pm 2.9$~~      & $5.1 \pm 0.9$          & $3.7 \pm 0.6$             & $4.4 \pm 0.7$   & $5 \pm 1$              \\
\noalign{\smallskip}\hline\noalign{\smallskip}
Post-fit, $t \bar t$     & $15.7 \pm 3.4$~~      & $7.6 \pm 2.3$       & $0.6 \pm 0.4$          & $0.007_{-0.007}^{+0.020}$~~ & $0.10 \pm 0.08$ & $0.16_{-0.16}^{+0.18}$~\,              \\
Post-fit, $t \bar t +Z$  & $1.65 \pm 0.35$     & $3.5 \pm 0.7$       & $2.2 \pm 0.5$          & $2.1 \pm 0.4$             & $2.18 \pm 0.45$ & $1.9 \pm 0.6$              \\
$Wt$                     & $0.5 \pm 0.5$       & $0.8 \pm 0.8$       & $0.10 \pm 0.04$        & $0.018_{-0.018}^{+0.019}$~~ & $0.12 \pm 0.06$ & $0.71 \pm 0.29$              \\
$Z/\gamma^*$ + jets      & $0.020 \pm 0.014$   & $0.044 \pm 0.003$   & $0.07_{-0.07}^{+0.17}$~\, & $0.38 \pm 0.13$           & $0.60 \pm 0.33$ & $0.4 \pm 0.4$              \\
Diboson                  & $0.27 \pm 0.20$     & $1.0 \pm 0.6$       & $0.65 \pm 0.24$        & $0.6 \pm 0.4$             & $0.59 \pm 0.28$ & $0.9 \pm 0.5$              \\
Others                   & $0.69 \pm 0.13$     & $1.37 \pm 0.21 $    & $0.99 \pm 0.16$        & $0.63 \pm 0.11$           & $0.67 \pm 0.14$ & $0.64 \pm 0.10$              \\
Fake and non-prompt      & $0.0_{-0.0}^{+0.4}$~ & $0.0_{-0.0}^{+0.4}$~ & $0.56 \pm 0.06$        & $0.0_{-0.0}^{+0.7}$~       & $0.15 \pm 0.12$ & $0.28 \pm 0.21$              \\
\noalign{\smallskip}\hline\noalign{\smallskip}
\end{tabular}
\end{adjustbox}
\end{center}
\end{table}

\FloatBarrier
\subsection{Three-body selection results}
\label{sec:threebody_results}
 
The  dominant background processes in the three-body selection are diboson, \ttbar and \ttZ production, and the yields are determined with a simultaneous fit. Figure~\ref{fig:threebodyNM1} shows the distributions of \MDR  in \SRTBW (top) and  in \SRTBT (bottom), for events satisfying all the selection criteria except the one for the presented variable, after the background fit.  Table~\ref{tab:threebodybkgonly_SRwt} shows the observed events in each signal region and the  SM background estimates.  No excess over the SM prediction is observed while a fluctuation of about $-2\sigma$ is observed in \SRTBWDF and is also visible in Figure~\ref{fig:threebodyNM1SRWDF}.
 
\begin{figure}[!htb]
\begin{center}
\subfigure[]{
\includegraphics[width=0.48\textwidth]{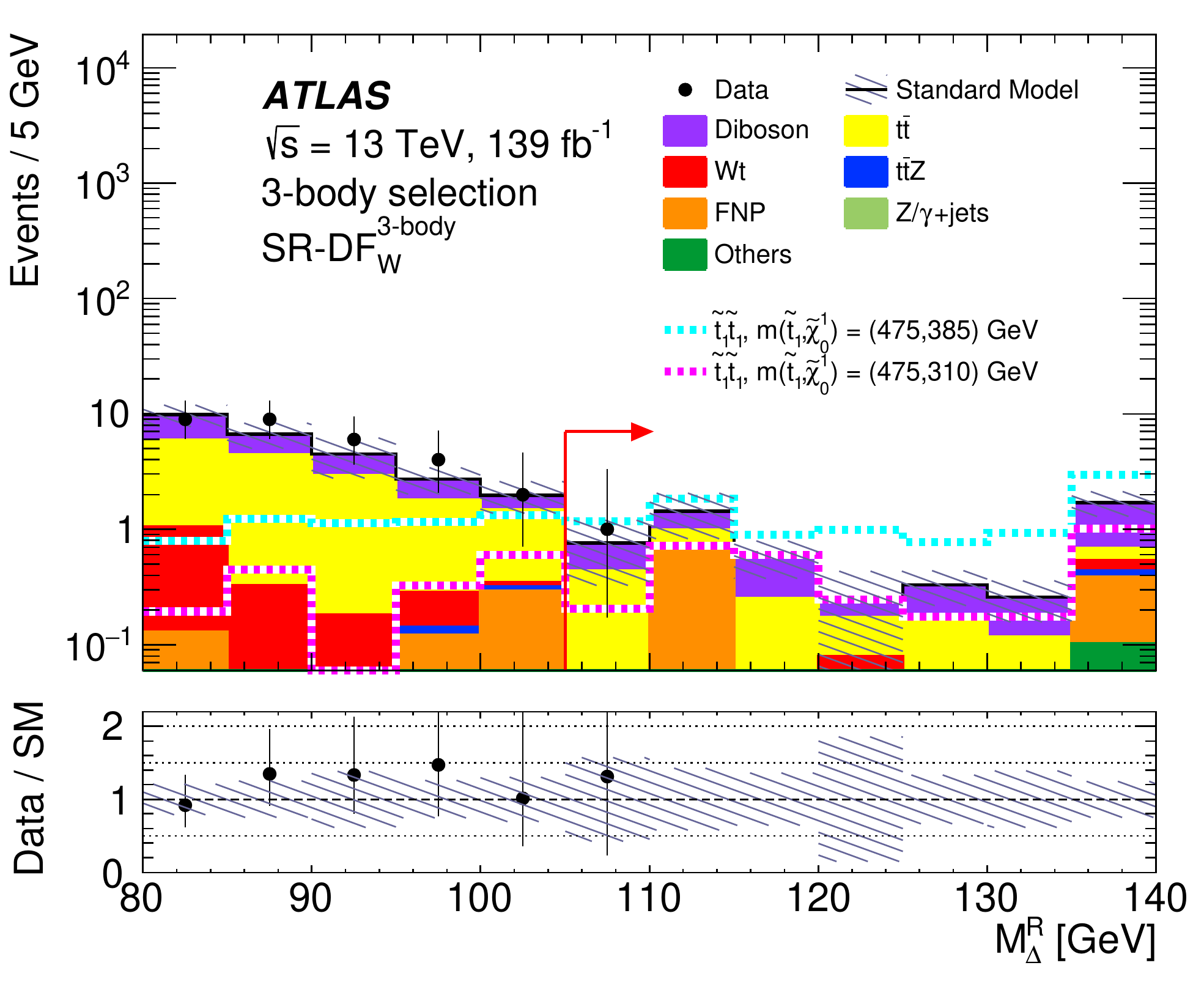}
\label{fig:threebodyNM1SRWDF}
}
\subfigure[]{
\includegraphics[width=0.48\textwidth]{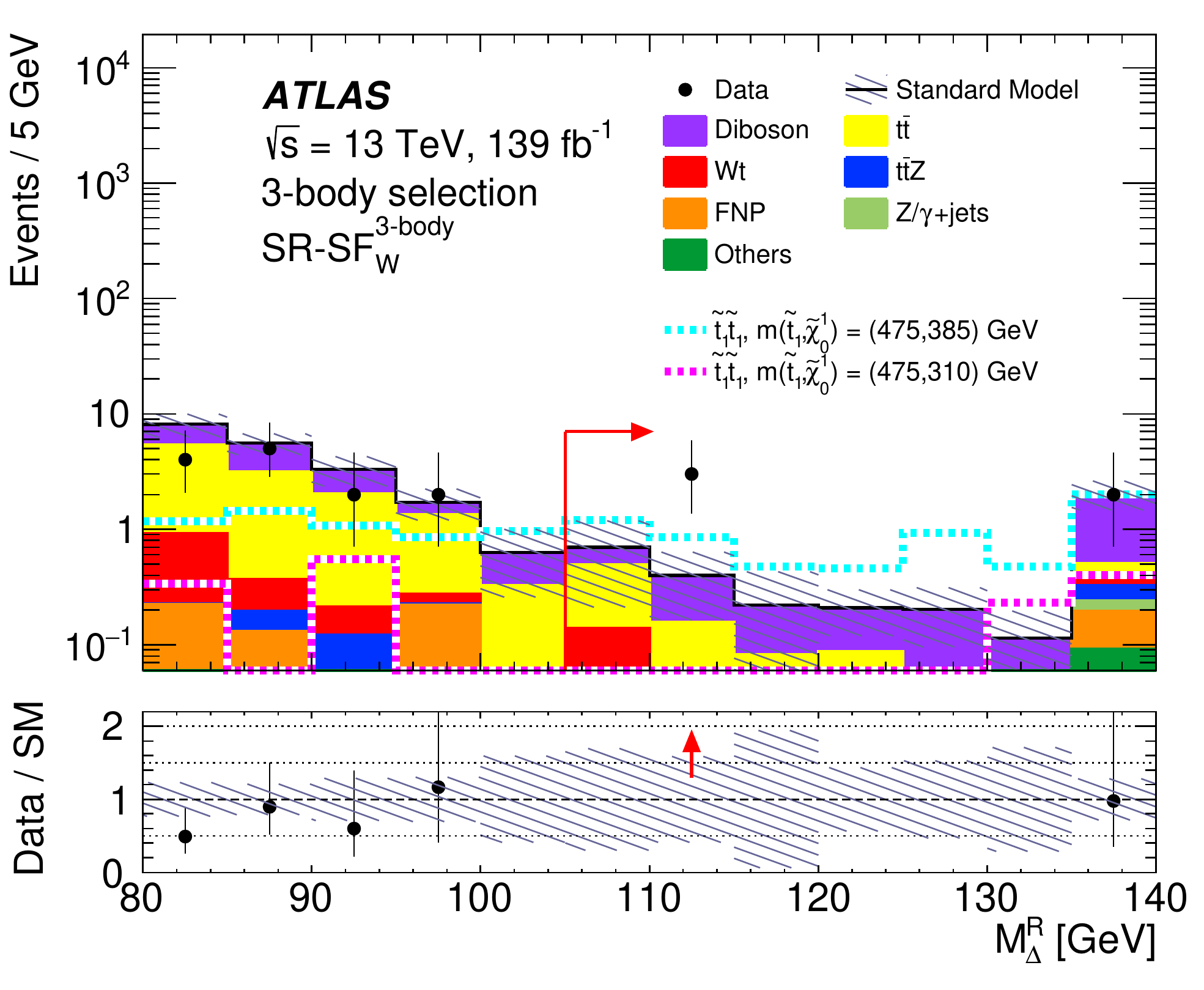}
}
\subfigure[]{
\includegraphics[width=0.48\textwidth]{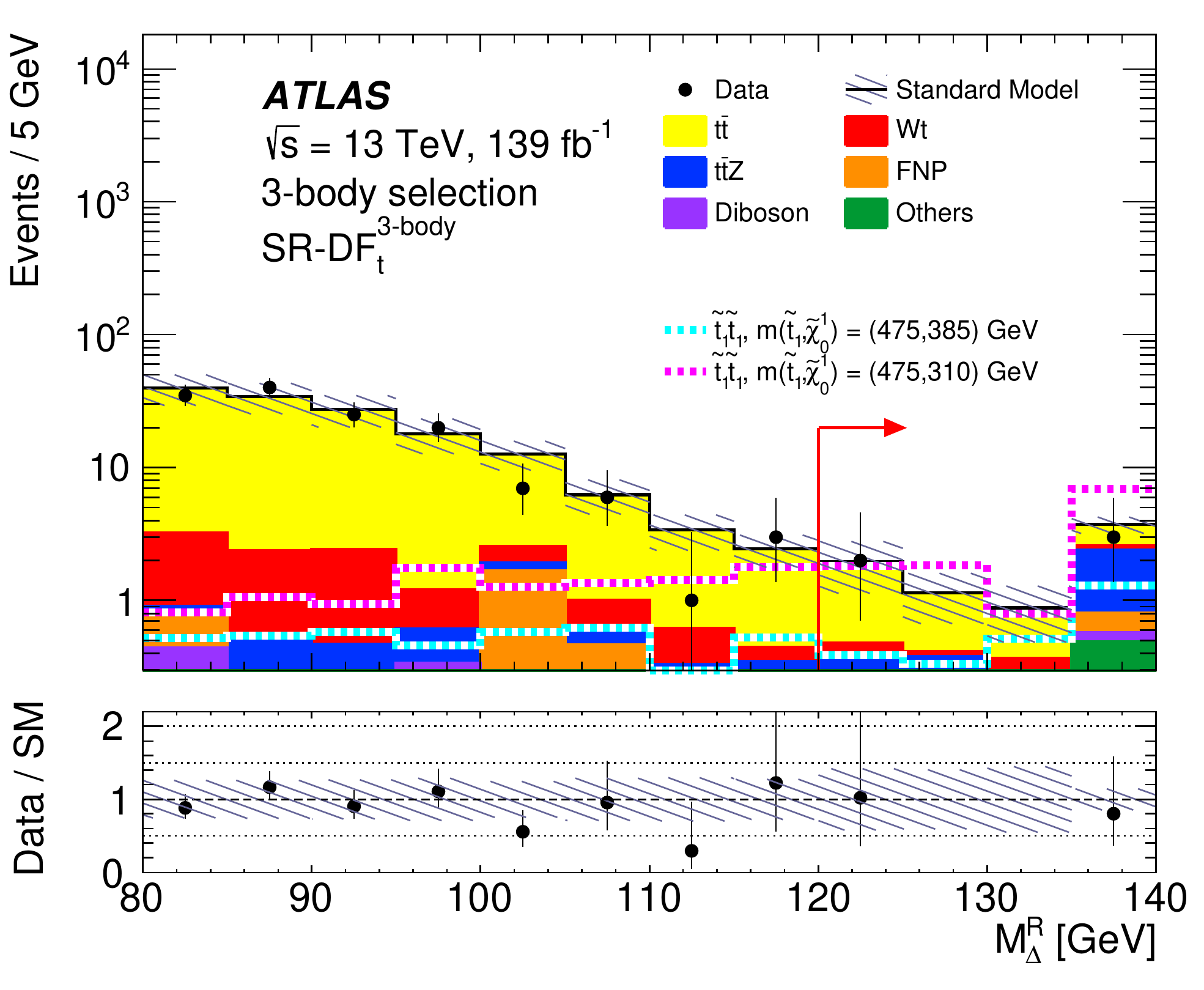}
}
\subfigure[]{
\includegraphics[width=0.48\textwidth]{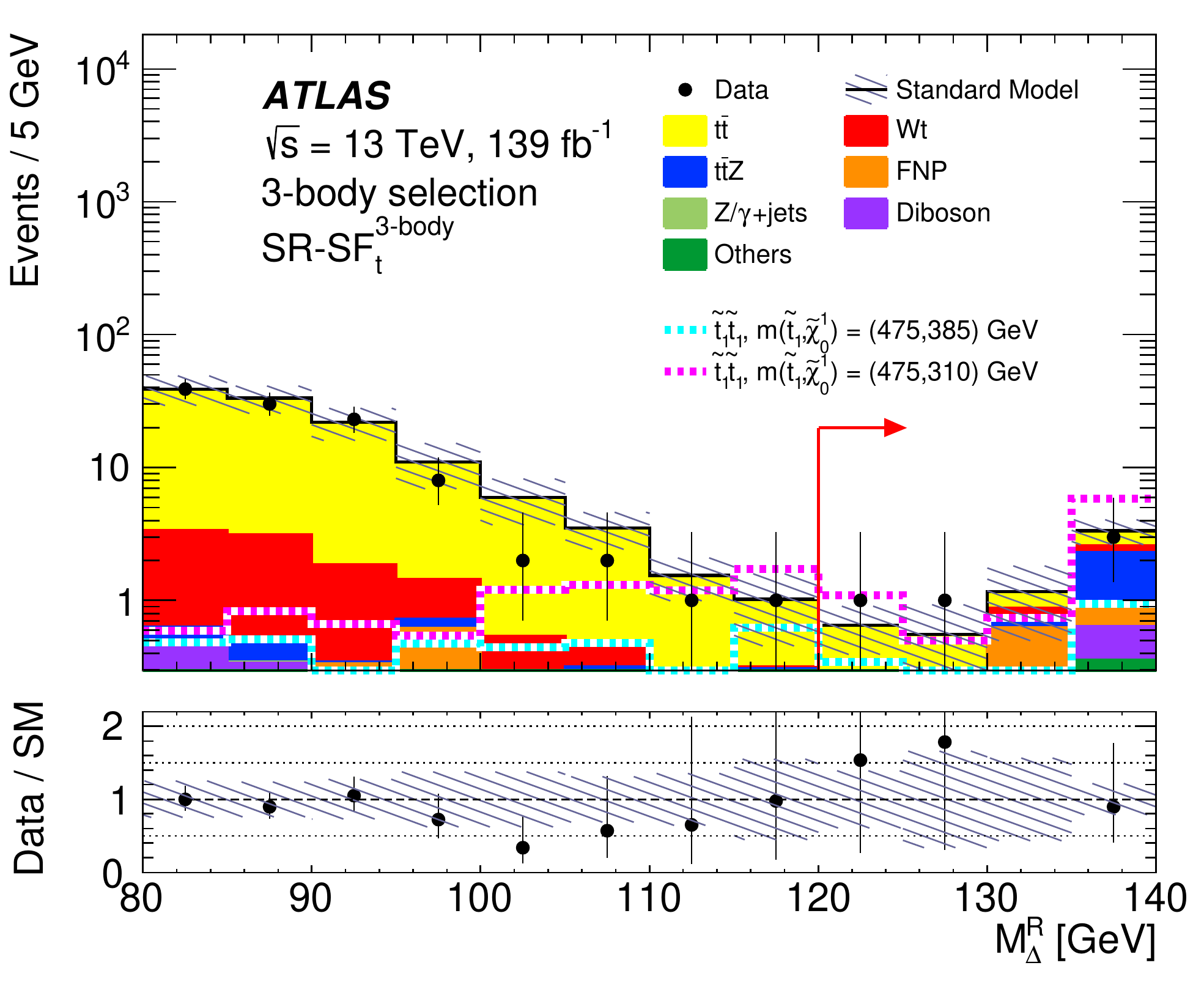}
}
\end{center}
\caption{Three-body selection. Distributions of $\MDR$ in (a,b) \SRTBW\ and (c,d) \SRTBT\ for (left) same-flavour and (right) different-flavour events satisfying the selection criteria of the given SR, except the one for the presented variable, after the background fit. The contributions from all SM backgrounds are shown as a histogram stack. ``Others'' includes contributions from $VVV$, $\ttbar t$, $\ttbar\ttbar$, $\ttbar W$, $\ttbar WW$, $\ttbar WZ$, $\ttbar H$, and $tZ$ processes.  The hatched bands represent the total statistical and systematic uncertainty. The rightmost bin of each plot includes overflow events. Reference top squark pair production signal models are overlayed for comparison. Red arrows in the upper panels indicate the signal region selection criteria. The bottom panels show the ratio of the observed data to the total SM background prediction, with hatched bands representing the total uncertainty in the background prediction; red arrows show data outside the vertical-axis range.}
\label{fig:threebodyNM1}
\end{figure}
 
\begin{table}[!htb]
\caption{Three-body selection. Observed event yields and background fit results for the three-body selection SRs. The `Others' category contains  contributions from $VVV$, $\ttbar t$, $\ttbar\ttbar$, $\ttbar W$, $\ttbar WW$, $\ttbar WZ$, $\ttbar H$, and $tZ$. Combined statistical and systematic uncertainties are given. Entries marked `--' indicate a negligible background contribution (less than 0.001 events). The individual uncertainties can be correlated, and do not necessarily add up in quadrature to the total background uncertainty.}
\label{tab:threebodybkgonly_SRwt}
\begin{center}
\begin{tabular}{l|cccc}
\noalign{\smallskip}\hline\noalign{\smallskip}
&\SRTBWDF           &\SRTBWSF                &\SRTBTDF                &\SRTBTSF\\[-0.05cm]
\noalign{\smallskip}\hline\noalign{\smallskip}
Observed events            & 1                 &5                       &5                       &5\\
\noalign{\smallskip}\hline\noalign{\smallskip}
Total (post-fit) SM events & $5.1 \pm 1.0 $   & $4.0 \pm 1.0$        & $7.5 \pm 1.4$          & $5.0 \pm 1.1$ \\
\noalign{\smallskip}\hline\noalign{\smallskip}
Post-fit, \ttbar           & $1.3 \pm 0.5$     & $0.76 \pm 0.32$        & $3.9 \pm 1.1$          & $1.8 \pm 0.7$\\
Post-fit, $t\bar t+Z$      & $0.085 \pm 0.034$ & $0.08 \pm 0.05$        & $2.3 \pm 0.4$          & $1.69 \pm 0.35$\\
Post-fit, diboson          & $2.5 \pm 1.0 $    & $2.5 \pm 1.0$          & $0.17 \pm 0.09$        & $0.34 \pm 0.14$\\
$Wt$                       & $0.30 \pm 0.05$   & $0.211 \pm 0.030$      & $0.4_{-0.4}^{+0.5}$~    & $0.54 \pm 0.19$\\
$Z/\gamma^{*}$ +jets       & --                & $0.044 \pm 0.019$      & --                     & $0.015_{-0.015}^{+0.027}$~~ \\
Others                     & $0.232 \pm 0.020$ & $0.25 \pm 0.05$        & $0.70 \pm 0.12$        & $0.49 \pm 0.08$ \\
Fake and non-prompt        & $0.70 \pm 0.09$   & $0.00_{-0.00}^{+0.25}$~\, & $0.00_{-0.00}^{+0.23}$~\, & $0.16_{-0.16}^{+0.23}$~\, \\
\noalign{\smallskip}\hline\noalign{\smallskip}
\end{tabular}
\end{center}
\end{table}

\FloatBarrier
\subsection{Four-body selection results}
\label{sec:fourbody_results}
 
The estimated SM yields in \SRFBA and \SRFBB are determined with a background fit that provides the normalisation factors for \ttbar and diboson production. Figure \ref{fig:fourbodySRplots} shows the distributions of (a) \MET in \SRFBA  and (b) \Rlj   in  \SRFBB for events satisfying the selection criteria of the given SR, except the one for the presented variable, after the background fit. The background fit results are shown in Table~\ref{tab:fourbodybkgonly_SR}. The observed yield in the SR is within one standard deviation of the background prediction.
 
\begin{figure}[!htb]
\begin{center}
\subfigure[]{
\includegraphics[width=0.48\textwidth]{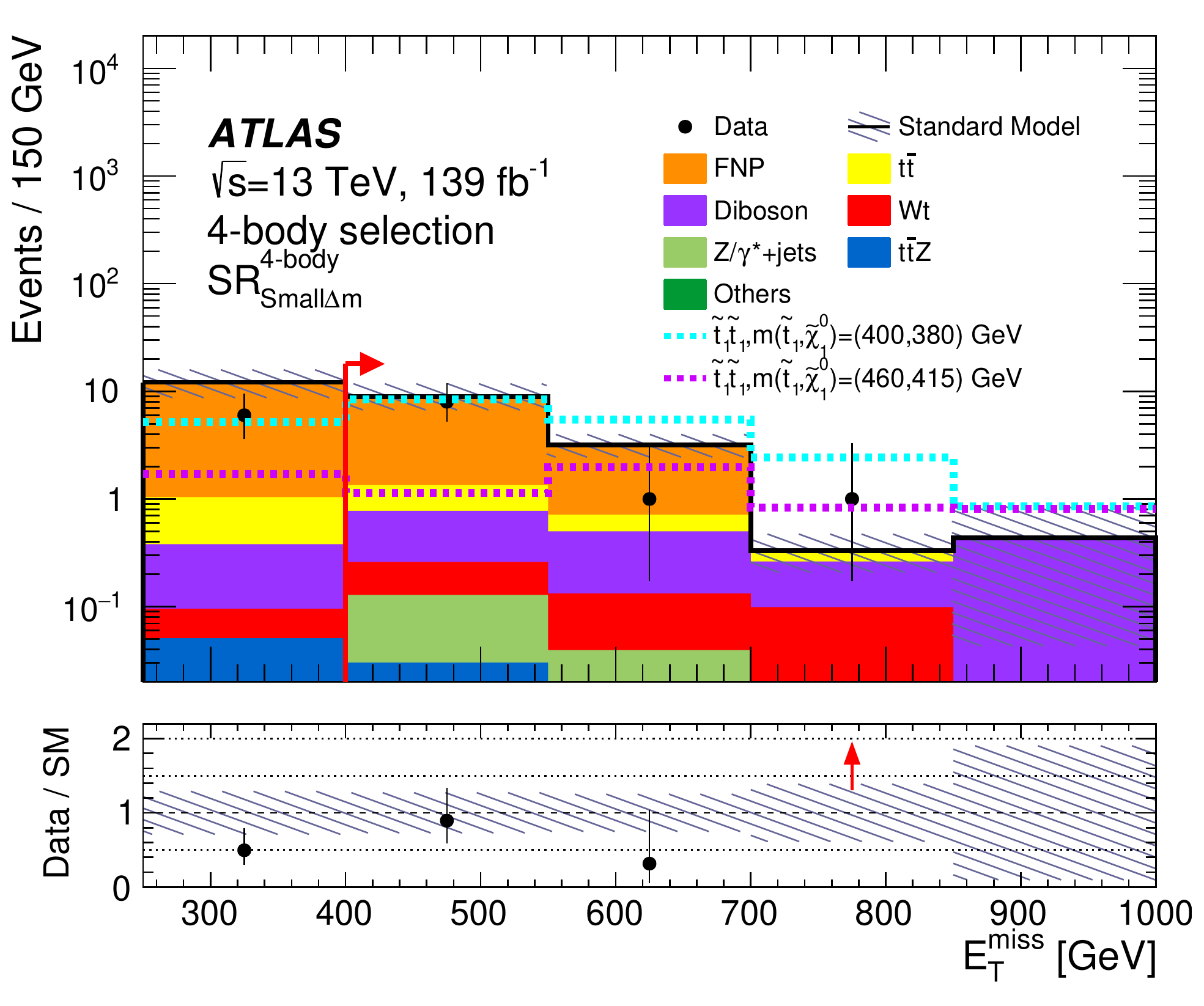}
}
\subfigure[]{
\includegraphics[width=0.48\textwidth]{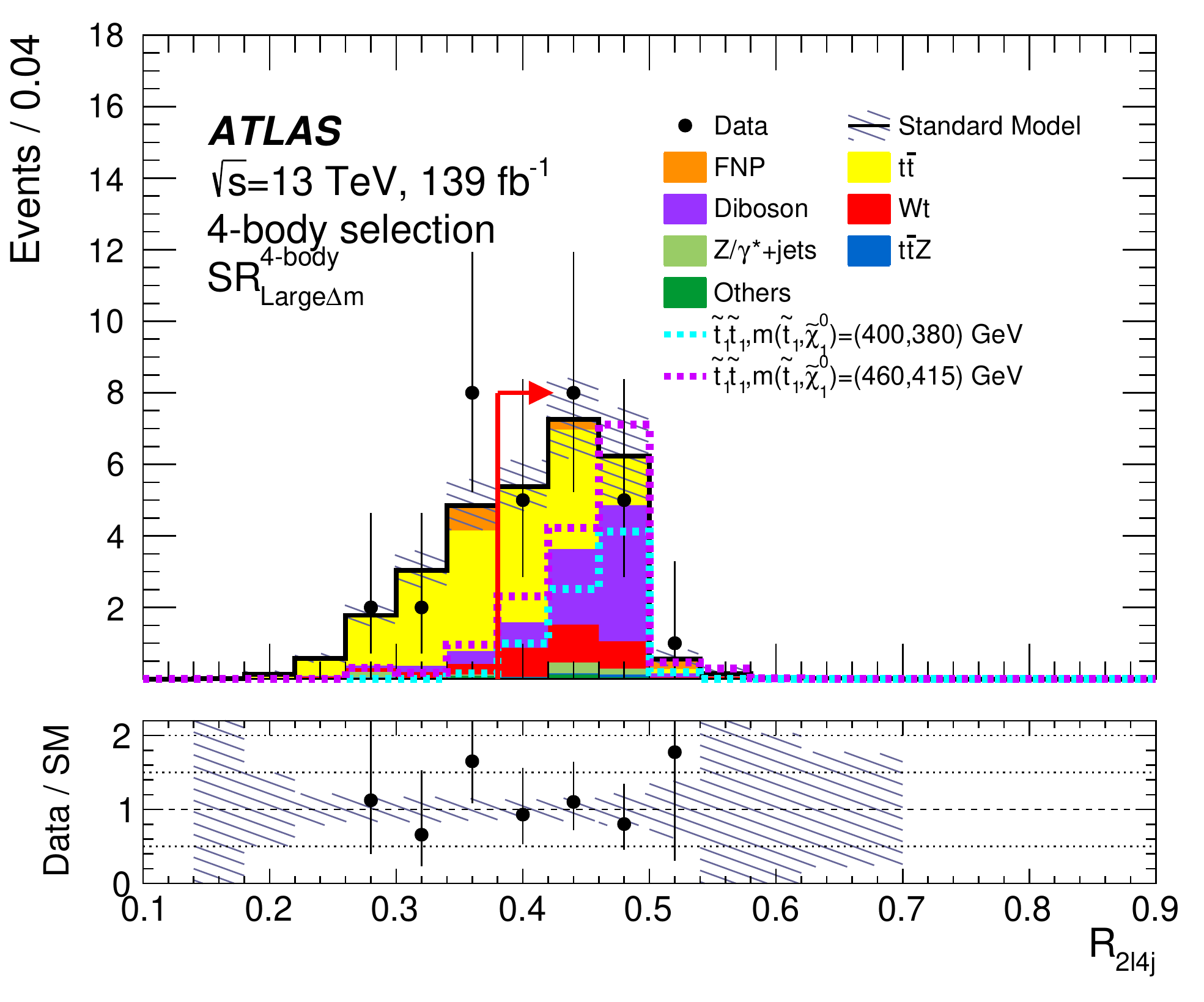}
}
\end{center}
\caption{Four-body selection. (a) distributions of  \MET in \SRFBA and (b) distribution of \Rlj  in \SRFBB for events satisfying the selection criteria of the given SR, except the one for the presented variable, after the background fit. The contributions from all SM backgrounds are shown as a histogram stack. ``Others'' includes contributions from $VVV$, $\ttbar t$, $\ttbar\ttbar$, $\ttbar W$, $\ttbar WW$, $\ttbar WZ$, $\ttbar H$, and $tZ$ processes.   The hatched bands represent the total statistical and systematic uncertainty. The rightmost bin of each plot includes overflow events. Reference top squark pair production signal models are overlayed for comparison. Red arrows in the upper panel indicate the signal region selection criteria.  The bottom panels show the ratio of the observed data to the total SM background prediction, with hatched bands representing the total uncertainty in the background prediction; red arrows show data outside the vertical-axis range.}
\label{fig:fourbodySRplots}
\end{figure}
 
\begin{table}[!htb]
\caption{Four-body selection. Observed event yields and  background fit results for \SRFBA and \SRFBB. The `Others' category contains the contributions from $VVV$, $\ttbar t$, $\ttbar\ttbar$, $\ttbar W$, $\ttbar WW$, $\ttbar WZ$, $\ttbar H$, and $tZ$. Combined statistical and systematic uncertainties are given. The individual uncertainties can be correlated, and do not necessarily add up in quadrature to the total background uncertainty.}
\label{tab:fourbodybkgonly_SR}
\begin{center}
\begin{tabular}{l|cc}
\noalign{\smallskip}\hline\noalign{\smallskip}
& \SRFBA                    &\SRFBB\\[-0.05cm]
\noalign{\smallskip}\hline\noalign{\smallskip}
Observed events            & $10$                      & $19$                    \\
\noalign{\smallskip}\hline\noalign{\smallskip}
Total (post-fit) SM events & $12.8 \pm 3.2$~~            & $19.3 \pm 2.7$~~              \\
\noalign{\smallskip}\hline\noalign{\smallskip}
Post-fit, \ttbar           & $0.87 \pm 0.26$           & $8.7 \pm 1.5$              \\
Post-fit, diboson          & $1.5 \pm 0.5$             & $6.8 \pm 2.3$              \\
$Wt$                       & $0.32 \pm 0.08$           & $2.7 \pm 0.5$              \\
$Z/\gamma^{*}$ +jets       & $0.128 \pm 0.023$         & $0.46 \pm 0.19$              \\
\ttZ                       & $0.047 \pm 0.010$         & $0.126 \pm 0.033$              \\
Others                     & $0.019_{-0.019}^{+0.021}$~~ & $0.26 \pm 0.07$              \\
Fake and non-prompt        & $10.0 \pm 3.1$~~            & $0.24 \pm 0.09$              \\
\noalign{\smallskip}\hline\noalign{\smallskip}
\end{tabular}
\end{center}
\end{table}

\FloatBarrier
\section{Interpretation}
\label{sec:interpretations}
 
No excess is observed in the data relative to the expected background.  The analysis results are therefore interpreted in terms of model-independent upper limits on the visible cross-section $(\sigma_{\mathrm{vis}})$ of new physics, defined as the 95\% confidence level (CL) upper limit on the number of signal events $(S^{95})$ divided by the integrated luminosity, and in terms of exclusion limits in the plane of the masses parameters of our simplified models.
For the two-body selection the upper limits are derived using the inclusive SRs.
 
The upper limits on $\sigma_{\mathrm{vis}}$ are derived, in each SR, by performing a model-independent hypothesis test, which introduces a free signal as an additional process to be constrained by the observed yield. The CL$_\mathrm{s}$ method~\cite{Read:2002hq} is used to derive all the exclusion confidence levels. Model-independent upper limits are presented in Table~\ref{tab:modindep}. These limits assume negligible signal contamination in the CRs, resulting in a more conservative result than from the model-dependent limits, where a small signal contamination is allowed in the CRs.
 
\begin{table}[!hbp]
\caption{\label{tab:modindep}
Model-independent 95\% CL upper limits on the visible cross-section ($\sigma_{\mathrm{vis}}$) of new physics, on the visible number of signal events ($S_{\mathrm{obs}}^{95}$), on the visible number of signal events ($S_{\mathrm{exp}}^{95}$) given the expected number of background events (and $\pm1\sigma$ excursions of the expected number), and the discovery $p$-value ($p(s = 0)$), all calculated with pseudo-experiments, are shown for each of the SRs. The $p$-value is reported as 0.5 if the observed yield is smaller than that predicted.}
\begin{center}
\begin{tabular}{llcccc}
\noalign{\smallskip}\hline\noalign{\smallskip}
Selection                   & {\textbf Signal Region} & $\sigma_{\mathrm{vis}}$ [fb] & $S_{\mathrm{obs}}^{95}$ & $S_{\mathrm{exp}}^{95}$         & $p(s=0)$\\
\noalign{\smallskip}\hline\noalign{\smallskip}
\multirow{7}{*}{Two-body}   & \SRdisczero & $0.21$ & $29.3$ & $ { 31 }^{ +11 }_{ -8 }$      & $ 0.5$\\
& \SRdiscone & $0.15$ & $21.4$ & $ { 21 }^{ +8 }_{ -6 }$~       & $ 0.4$\\
& \SRdisctwo & $0.10$ & $13.2$ & $ { 14 }^{ +5 }_{ -4 }$~       & $ 0.5$\\
& \SRdiscthree & $0.06$ & ~~$8.2$  & ~$ { 11 }^{ +5 }_{ -3.0 }$     & $ 0.5$\\
& \SRdiscfour & $0.06$ & ~~$7.9$  & $ { 9.6 }^{ +3.8 }_{ -2.8 }$  & $ 0.5$\\
& \SRdiscfive & $0.06$ & ~~$7.6$  & $ { 8.4 }^{ +3.6 }_{ -2.3 }$  & $ 0.5$\\
& \SRdiscsix & $0.05$ & ~~$7.6$  & $ { 7.5 }^{ +3.1 }_{ -2.0 }$  & $ 0.5$\\
\noalign{\smallskip}\hline\noalign{\smallskip}
\multirow{4}{*}{Three-body} & \SRTBWDF               & ~~$0.023$ & ~~$3.2$  & $ { 5.7 }^{ +2.3 }_{ -1.5 }$  & $ 0.5$ \\
& \SRTBWSF               & $0.05$  & ~~$7.0$  & $ { 5.6 }^{ +2.3 }_{ -1.5 }$  & ~~$ 0.27$\\
& \SRTBTDF               & $0.04$  & ~~$5.5$  & $ { 6.9 }^{ +2.9 }_{ -1.9 }$  & $ 0.5$\\
& \SRTBTSF               & $0.04$  & ~~$6.3$  & $ { 6.1 }^{ +2.6 }_{ -1.6 }$  & $ 0.5$\\
\noalign{\smallskip}\hline\noalign{\smallskip}
\multirow{2}{*}{Four-body}  & \SRFBA                 & $0.06$ & ~~$8.2$  & $ { 9.6 }^{ +3.8 }_{ -2.5 }$  & $ 0.5$\\
& \SRFBB                 & $0.08$ & $11.1$ & $ { 11.1 }^{ +4.5 }_{ -3.0 }$~~ & $ 0.5$\\
\noalign{\smallskip}\hline\noalign{\smallskip}
\end{tabular}
\end{center}
\end{table}
 
Model-dependent limits are computed for the various signal scenarios considered in the analysis. The hypothesis tests are performed including the expected signal yield and its associated uncertainties in the CRs and SRs. All limits are quoted at 95\% CL with the CL$_\mathrm{s}$ method. When setting limits, the two-body selection binned SRs $\SRtwoBDF_{[x,y)}$ and $\SRtwoBSF_{[x,y)}$ regions are combined. Similarly, the \SRTBWDF , \SRTBWSF , \SRTBTDF , and \SRTBTSF\ signal regions are combined for the three-body selection, and so are \SRFBA and \SRFBB for the four-body selection.
 
Limits for simplified models in which pair-produced \stopone decay with 100\% branching ratio into a top quark and \ninoone are shown in the \stopone--\ninoone mass plane in Figure~\ref{fig:tN1_exclusion_a} and in the $m(\stopone)$--$\Delta m(\stopone,\ninoone)$ plane in Figure~\ref{fig:tN1_exclusion_b}.
The exclusion contour is the envelope of the exclusion regions obtained separately for the three selections.
Top squark masses up to 1~\TeV\ are excluded for a massless lightest neutralino. Neutralino masses up to  500~\GeV\ are excluded for $m(\stopone)$  above the  top quark production kinematic limit.  In the three-body decay region, top squark masses are excluded up to  600~\GeV\ for $\DeltaM=120~\GeV$, up to $550~\GeV$ for $\DeltaM$ close to the top quark mass and up to 430~\GeV\ for $\DeltaM$ close to the \Wboson boson mass. In the four-body decay region, top squark masses are excluded up to 540~\GeV\ for $\DeltaM=40~\GeV$. Top squark decay around the \Wboson boson production kinematic limit is not fully excluded for $m(\stopone)$ above 400~\GeV\ because there the four-body and three-body decay exclusion regions do not overlap. The four-body selection loses sensitivity for \mwselgr\  due to the upper bound of the sub-leading lepton \pT while, for the three body selection, the \MDR requirement suppresses the sensitivity for \mwsells\  because of the smaller mass splitting. The three-body and two-body overlap in the sensitivity provides exclusion coverage  around the  top quark production kinematic limit up to $m(\stopone)$ of 540~\GeV.

\begin{figure}[!htb]
\begin{center}
\subfigure[]{
\includegraphics[width=0.8\textwidth]{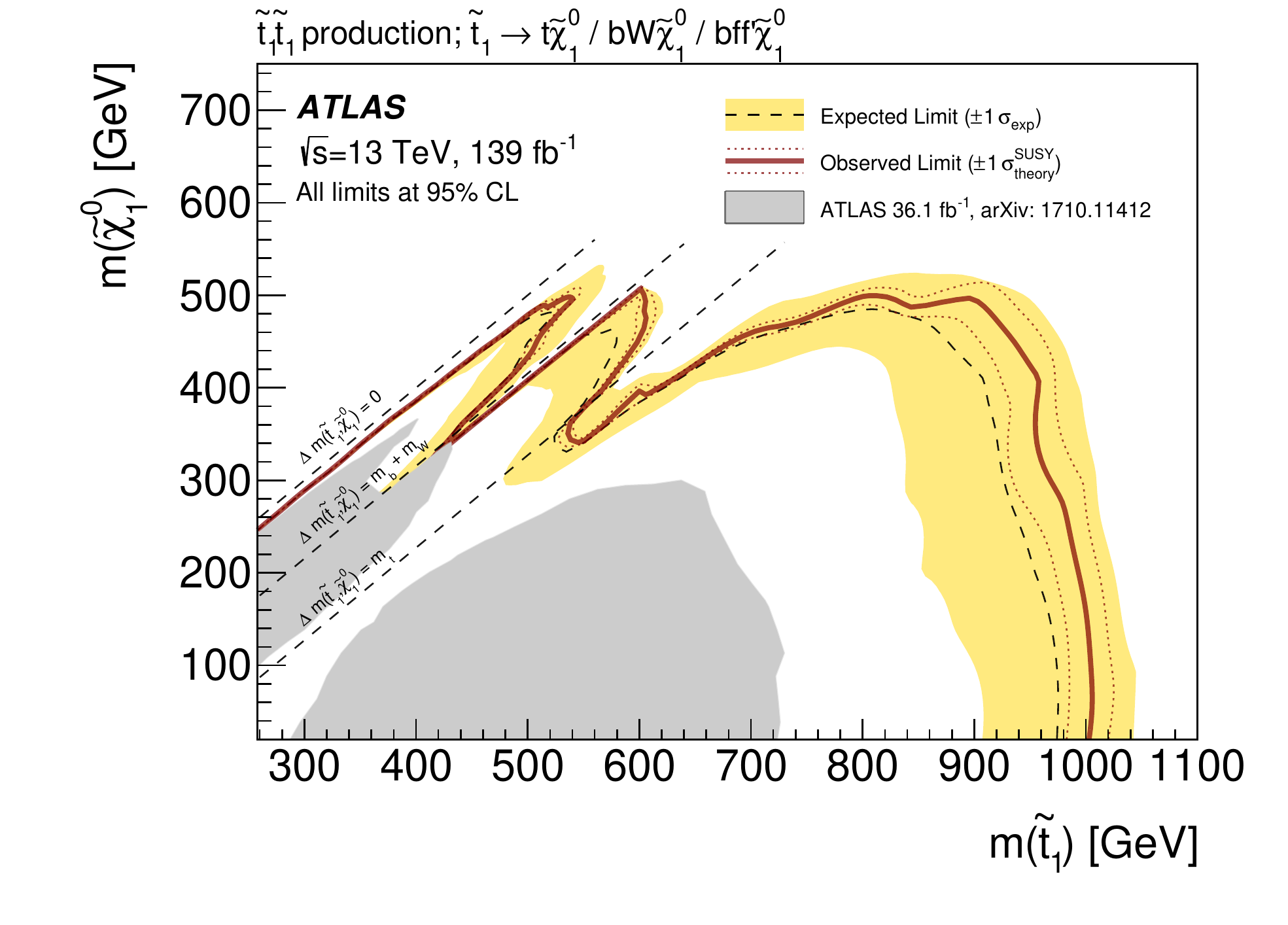}\label{fig:tN1_exclusion_a}
}
\subfigure[]{
\includegraphics[width=0.8\textwidth]{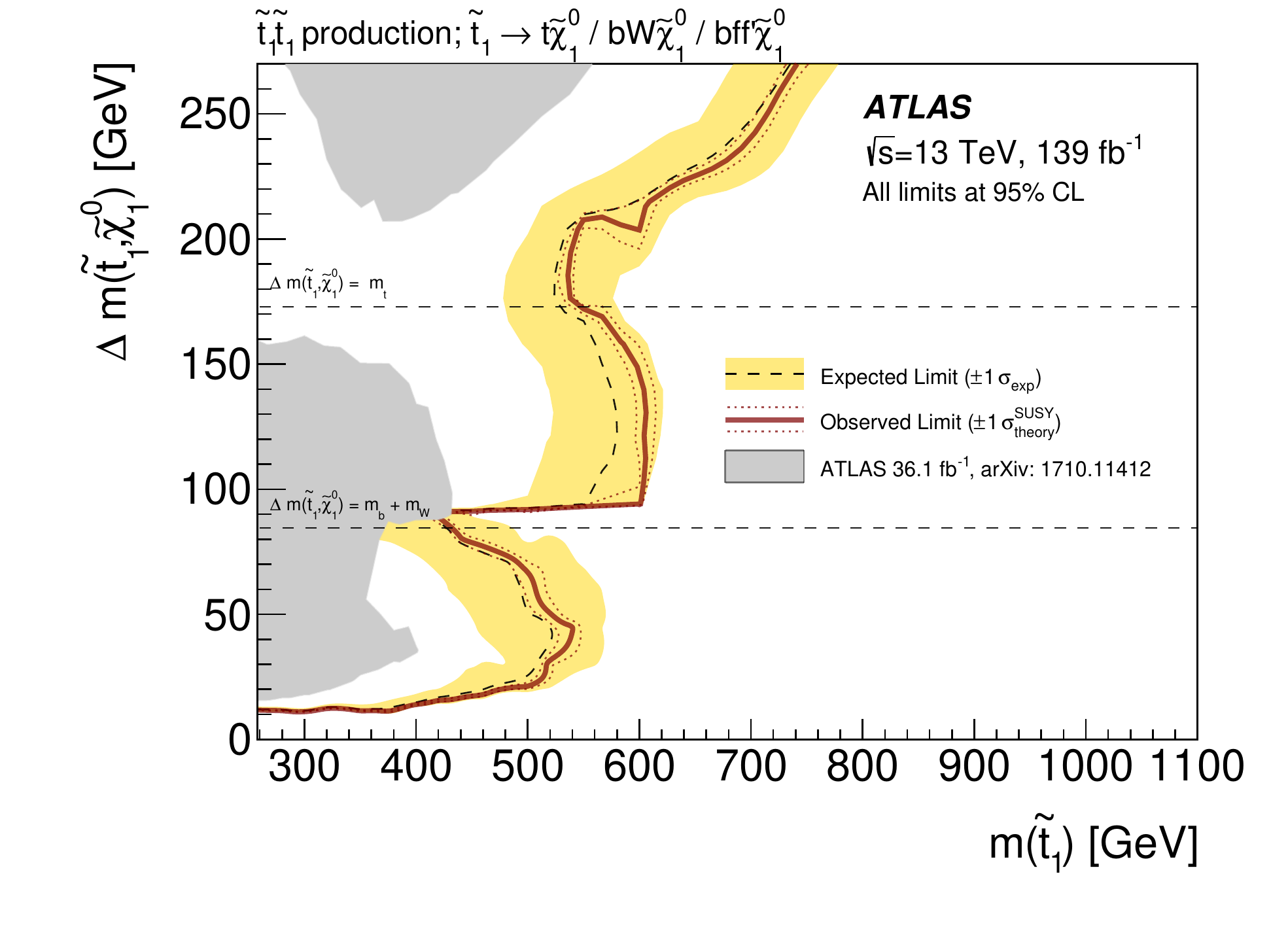}\label{fig:tN1_exclusion_b}
}
\end{center}
\caption{Exclusion limit contour (95\% CL) for a simplified model assuming \stopone pair production, decaying via $\stopone \rightarrow t^{(*)}\ninoone$ with 100\% branching ratio, in the (a) $m(\stopone)$--$m(\ninoone)$ and (b) $m(\stopone)$--$\Delta m(\stopone,\ninoone)$ planes. The dashed lines and the shaded bands are the expected limits and their $\pm1\sigma$ uncertainties.
The thick solid lines are the observed limits for the central value of the signal cross-section. The expected and observed limits do not include the effect of the theoretical uncertainties in the signal cross-section. The dotted lines show the effect on the observed limit when varying the signal cross-section by $\pm1\sigma$ of the theoretical uncertainty.
}
\end{figure}
\FloatBarrier
For the DM mediator models, Figure~\ref{fig:DM_g1_exclusion} shows upper limits at 95\% CL on the observed signal cross-section scaled to the theoretical signal cross-section for a coupling $g = g_q = g_{\chi} = 1$, denoted by $\sigma_{\mathrm{obs}}/\sigma_{\mathrm{Th}}(g=1.0)$.
These limits are obtained as a function of the mediator mass, assuming a specific DM particle mass of $1~\GeV$. Both the scalar and pseudoscalar mediator cases are considered.
The sensitivity is approximately constant for mediator masses below $100~\GeV$ and the models are excluded for scalar (pseudoscalar) mediator masses up to $250~(300)~\GeV$ when assuming $g = 1$.

\begin{figure}[!htb]
\begin{center}
\subfigure[]{
\includegraphics[width=0.48\textwidth]{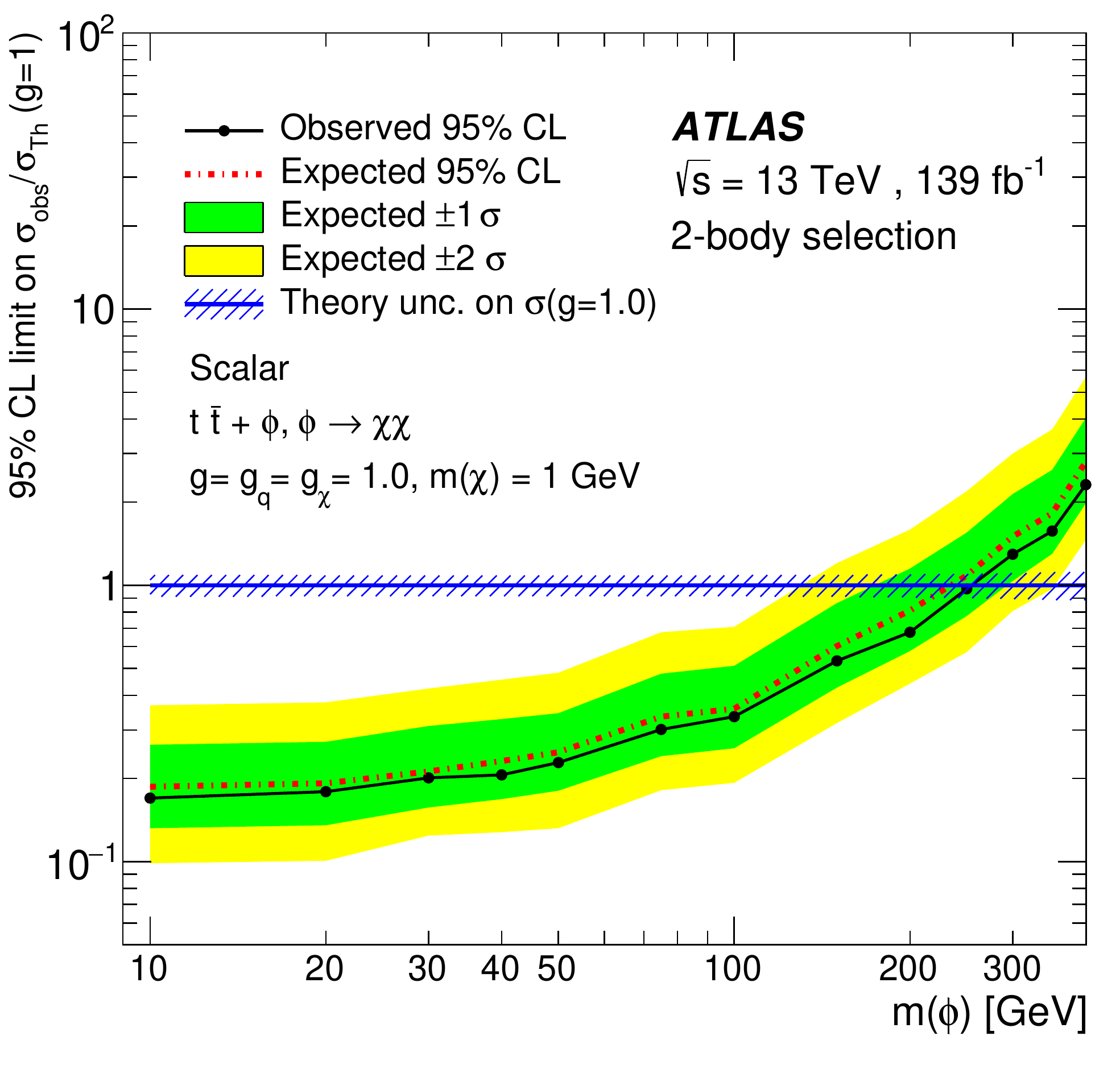}
}
\subfigure[]{
\includegraphics[width=0.48\textwidth]{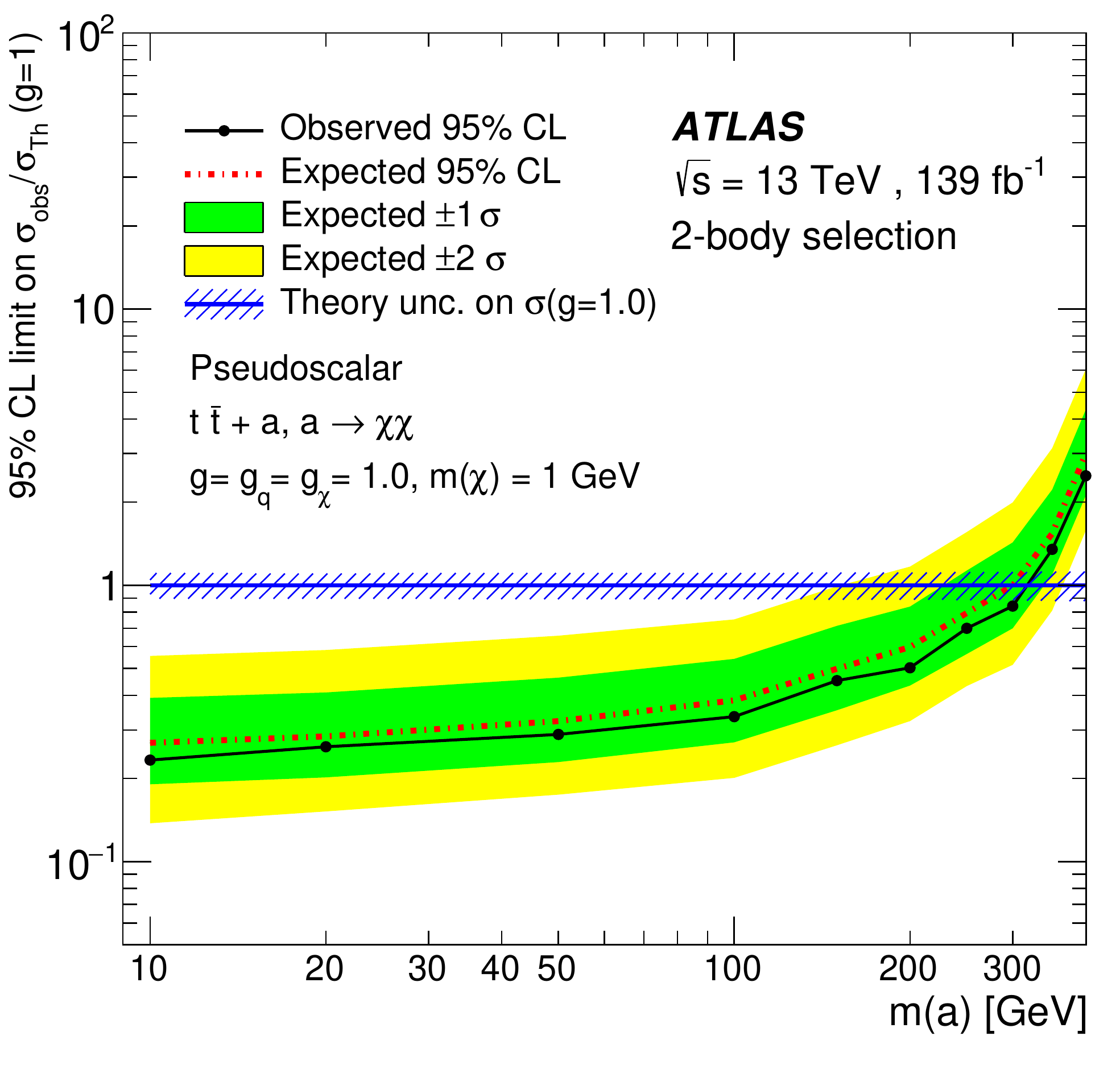}
}
\end{center}
\caption{Exclusion limits for (a) $\ttbar + \phi $ scalar and (b) $\ttbar + a $ pseudoscalar models as a function of the mediator mass for a DM particle mass of $m(\chi)=1~\GeV$. The limits are calculated at 95\% CL and are expressed in terms of the ratio of the excluded cross-section to the nominal cross-section for a coupling assumption of $g = g_q = g_{\chi} = 1$. The solid (dashed) lines shows the observed (expected) exclusion limits.}
\label{fig:DM_g1_exclusion}
\end{figure}
 
\FloatBarrier
 
\section{Conclusion}
\label{sec:conclusion}
This paper reports the results of a search for direct top squark pair production and for dark matter in a final state containing two leptons with opposite electric charge, jets and missing transverse momentum.
The search uses an integrated luminosity of $139~\ifb$ of proton--proton collisions at $\sqrt{s}=13~\TeV$, as collected by the ATLAS experiment at the Large Hadron Collider during Run~2 (2015--2018).
 
Compared to previous searches a significant improvement in sensitivity is obtained by using additional integrated luminosity and a new discriminating variable, the object-based \METsig.
Moreover, in the small-\DeltaM\ region, an important gain in sensitivity is also achieved by lowering the \pT threshold for lepton selection.
 
The data are found to be consistent with the Standard Model predictions. Assuming direct \stopone pair production
with both top squarks decaying in either the two-body channel $\stopone \rightarrow t \ninoone$, the three-body channel $\stopone \rightarrow b W \ninoone$,  or the four-body channel $\stopone \rightarrow b \ell \nu \ninoone$, constraints at 95\% confidence level are placed on the minimum \stopone and \ninoone masses up to about 1~\TeV\ and  $500~\GeV$ respectively. The results improve on the previous ATLAS limits obtained in a two-lepton final state and provide unique sensitivity among the ATLAS searches in the mass region where the decay $\stopone \rightarrow t \ninoone$ becomes kinematically allowed. For the dark-matter model, assuming spin-0 mediator production in association with a pair of top quarks and decay with 100\% branching ratio into a pair of dark-matter particles, scalar (pseudoscalar) mediator masses up to about $250~(300)~\GeV$ are excluded at 95\% confidence level for mediator couplings $g_q = g_\chi = 1$ to Standard Model and dark-matter particles.

\clearpage
\section*{Acknowledgements}
 

We thank CERN for the very successful operation of the LHC, as well as the
support staff from our institutions without whom ATLAS could not be
operated efficiently.
 
We acknowledge the support of ANPCyT, Argentina; YerPhI, Armenia; ARC, Australia; BMWFW and FWF, Austria; ANAS, Azerbaijan; SSTC, Belarus; CNPq and FAPESP, Brazil; NSERC, NRC and CFI, Canada; CERN; ANID, Chile; CAS, MOST and NSFC, China; COLCIENCIAS, Colombia; MSMT CR, MPO CR and VSC CR, Czech Republic; DNRF and DNSRC, Denmark; IN2P3-CNRS and CEA-DRF/IRFU, France; SRNSFG, Georgia; BMBF, HGF and MPG, Germany; GSRT, Greece; RGC and Hong Kong SAR, China; ISF and Benoziyo Center, Israel; INFN, Italy; MEXT and JSPS, Japan; CNRST, Morocco; NWO, Netherlands; RCN, Norway; MNiSW and NCN, Poland; FCT, Portugal; MNE/IFA, Romania; JINR; MES of Russia and NRC KI, Russian Federation; MESTD, Serbia; MSSR, Slovakia; ARRS and MIZ\v{S}, Slovenia; DST/NRF, South Africa; MICINN, Spain; SRC and Wallenberg Foundation, Sweden; SERI, SNSF and Cantons of Bern and Geneva, Switzerland; MOST, Taiwan; TAEK, Turkey; STFC, United Kingdom; DOE and NSF, United States of America. In addition, individual groups and members have received support from BCKDF, CANARIE, Compute Canada, CRC and IVADO, Canada; Beijing Municipal Science \& Technology Commission, China; COST, ERC, ERDF, Horizon 2020 and Marie Sk{\l}odowska-Curie Actions, European Union; Investissements d'Avenir Labex, Investissements d'Avenir Idex and ANR, France; DFG and AvH Foundation, Germany; Herakleitos, Thales and Aristeia programmes co-financed by EU-ESF and the Greek NSRF, Greece; BSF-NSF and GIF, Israel; La Caixa Banking Foundation, CERCA Programme Generalitat de Catalunya and PROMETEO and GenT Programmes Generalitat Valenciana, Spain; G\"{o}ran Gustafssons Stiftelse, Sweden; The Royal Society and Leverhulme Trust, United Kingdom.
 
The crucial computing support from all WLCG partners is acknowledged gratefully, in particular from CERN, the ATLAS Tier-1 facilities at TRIUMF (Canada), NDGF (Denmark, Norway, Sweden), CC-IN2P3 (France), KIT/GridKA (Germany), INFN-CNAF (Italy), NL-T1 (Netherlands), PIC (Spain), ASGC (Taiwan), RAL (UK) and BNL (USA), the Tier-2 facilities worldwide and large non-WLCG resource providers. Major contributors of computing resources are listed in Ref.~\cite{ATL-SOFT-PUB-2020-001}.
 

\clearpage

\printbibliography

\clearpage \input{atlas_authlist}

\end{document}

%% file: atlas_authlist.tex
 
\begin{flushleft}
\hypersetup{urlcolor=black}
{\Large The ATLAS Collaboration}

\bigskip

\AtlasOrcid[0000-0002-6665-4934]{G.~Aad}$^\textrm{\scriptsize 102}$,    
\AtlasOrcid[0000-0002-5888-2734]{B.~Abbott}$^\textrm{\scriptsize 128}$,    
\AtlasOrcid[0000-0002-7248-3203]{D.C.~Abbott}$^\textrm{\scriptsize 103}$,    
\AtlasOrcid[0000-0002-2788-3822]{A.~Abed~Abud}$^\textrm{\scriptsize 36}$,    
\AtlasOrcid[0000-0002-1002-1652]{K.~Abeling}$^\textrm{\scriptsize 53}$,    
\AtlasOrcid[0000-0002-2987-4006]{D.K.~Abhayasinghe}$^\textrm{\scriptsize 94}$,    
\AtlasOrcid[0000-0002-8496-9294]{S.H.~Abidi}$^\textrm{\scriptsize 167}$,    
\AtlasOrcid[0000-0002-8279-9324]{O.S.~AbouZeid}$^\textrm{\scriptsize 40}$,    
\AtlasOrcid{N.L.~Abraham}$^\textrm{\scriptsize 156}$,    
\AtlasOrcid[0000-0001-5329-6640]{H.~Abramowicz}$^\textrm{\scriptsize 161}$
% ...
\bigskip
\\

% ...
$^{36}$CERN, Geneva; Switzerland.\\
% ...
$^{40}$Niels Bohr Institute, University of Copenhagen, Copenhagen; Denmark.\\
% ...
$^{53}$II. Physikalisches Institut, Georg-August-Universit\"{a}t G\"ottingen, G\"ottingen; Germany.\\
% ...
$^{94}$Department of Physics, Royal Holloway University of London, Egham; United Kingdom.\\
% ...
$^{102}$CPPM, Aix-Marseille Universit\'e, CNRS/IN2P3, Marseille; France.\\
$^{103}$Department of Physics, University of Massachusetts, Amherst MA; United States of America.\\
% ...
$^{128}$Homer L. Dodge Department of Physics and Astronomy, University of Oklahoma, Norman OK; United States of America.\\
% ...
$^{156}$Department of Physics and Astronomy, University of Sussex, Brighton; United Kingdom.\\
% ...
$^{161}$Raymond and Beverly Sackler School of Physics and Astronomy, Tel Aviv University, Tel Aviv; Israel.\\
% ...
$^{167}$Department of Physics, University of Toronto, Toronto ON; Canada.\\
% ...

\end{flushleft}
